\documentclass{elsart}

\usepackage{amsfonts}
\usepackage{amsmath}
\usepackage{amssymb}
\usepackage{wasysym}
\usepackage{epsf,graphicx,psfrag}

\newtheorem{theorem}{Theorem}

\newenvironment{proof}
{\begin{trivlist} \item[]{\bf Proof. }}%
{\hspace*{\fill}$\rule{.3\baselineskip}{.35\baselineskip}$\end{trivlist}}

\begin{document}

\begin{frontmatter}

\title{Breathers in oscillator chains with Hertzian interactions}

\author[a]{Guillaume James\thanksref{cor}},
\author[b]{Panayotis G. Kevrekidis},
\author[c]{Jes\'us Cuevas}.
\thanks[cor]{Corresponding author.
             E-mail: guillaume.james@imag.fr}

\address[a]{Laboratoire Jean Kuntzmann, Universit\'e de Grenoble and CNRS, BP 53, 38041 Grenoble Cedex 9, France.}
\address[b]{Department of Mathematics and Statistics, University of Massachusetts, Amherst, Massachussets 01003-4515, USA.}
\address[c]{Grupo de F\'{\i}sica No Lineal.
Departamento de F\'{\i}sica Aplicada I. Escuela Polit\'{e}cnica Superior. Universidad de Sevilla. C/ Virgen de \'{A}frica, 7.
41011 Sevilla, Spain.}

\journal{Physica D}

\begin{abstract}
We prove nonexistence of breathers (spatially localized and time-periodic oscillations)
for a class of Fermi-Pasta-Ulam lattices representing an uncompressed chain of beads interacting
via Hertz's contact forces. We then consider the setting in which
an additional on-site potential is present, motivated by the 
Newton's cradle under the effect of gravity.
Using both direct numerical computations
and a simplified asymptotic model of the oscillator chain, the so-called discrete $p$-Schr\"odinger (DpS) equation,
we show the existence of discrete breathers and study their spectral properties and mobility.
Due to the fully nonlinear character of Hertzian interactions, 
breathers are found to be much 
more localized than in classical nonlinear lattices
and their motion occurs with less dispersion.
In addition, we study numerically the excitation of a traveling breather
after an impact at one end of a semi-infinite chain.
This case is well described by the DpS equation when local
oscillations are faster than binary collisions, a situation occuring e.g. in
chains of stiff cantilevers decorated by spherical beads.
When a hard anharmonic part is added to the local potential, 
a new type of traveling breather emerges, showing
spontaneous direction-reversing in a spatially homogeneous system.
Finally, the 
interaction of a moving breather with a point defect is also considered
in the cradle system. Almost total breather reflections are observed
at sufficiently high defect sizes, suggesting potential applications of
such systems as shock wave reflectors.
\end{abstract}


\end{frontmatter}

\section{Introduction}

The study of nonlinear waves in granular crystals is
the object of intensive research, both from a theoretical perspective and
for practical purposes, e.g. for the design of shock absorbers \cite{hong,sen,frater}, acoustic lenses \cite{spadoni}
or diodes~\cite{chiaranick}. Due to the nonlinear interactions between
grains, several interesting types of localized waves can be generated in
chains of beads in contact. Solitary waves are
the most studied type of excitations and can be
easily generated by an impact at one end of a chain
\cite{neste2,falcon,hinch,mackay,ap,porter,sen}.
These solitary waves, in the absence of an original compression
in the chain (the so-called precompression), 
differ from classical ones (i.e. KdV-type solitary 
waves \cite{pego})
due to the fully nonlinear character of Hertzian contact interactions. Indeed,
their decay is  super-exponential and their
width remains unchanged with amplitude \cite{english,atanas}.

Another interesting class of excitations consists of time-periodic and 
spatially
localized oscillations.
Such waves may correspond to Anderson modes \cite{hu} in the presence of spatial disorder,
or to defect modes localized at an impurity in a granular chain under precompression \cite{theo}.
A different class of spatially
localized oscillations that occur in the absence of defects consists of 
discrete breathers,
which originate from the combined effects of nonlinearity and spatial discreteness (see the reviews \cite{review,flg}).
These waves exist in diatomic granular chains
under precompression \cite{boe,theo2,jamesnoble},
with their frequency lying between the acoustic and optic phonon bands
and can be generated e.g. through modulational instabilities.
However, because precompression suppresses the fully nonlinear character
of Hertzian interactions, these excitations inherit the usual properties of
discrete breathers, i.e. their spatial decay is exponential
and their width diverges at vanishing amplitude, i.e. for frequencies close
to the bottom of the optic band \cite{jamesnoble}.

For granular systems without precompression, the above discussion 
raises the question of existence of spatially localized oscillations.
Defect modes induced by a mass impurity have been numerically observed on 
short transients
in unloaded granular chains \cite{hascoet,job}, but the existence of permanent
localized oscillations remains an open question. In this paper, we give a 
partial answer to this problem by showing the {\em nonexistence} of 
{\em time-periodic} spatially localized oscillations
in uncompressed granular chains.
This result seems surprising at a first glance, because
Hertzian models of granular chains fall within the class of 
Fermi-Pasta-Ulam (FPU) lattices,
which sustain discrete breathers under some general assumptions on the 
interaction
potentials and particle masses (see \cite{pankovbook} and references therein). 
However these conditions
do not hold for uncompressed granular chains. Using a simple averaging 
argument, we show that the non-attracting character of Hertzian interactions 
between grains (repulsive under contact, and vanishing in the absence of 
contact) precludes the existence
of time-periodic localized oscillations, both for spatially homogeneous 
systems and for inhomogeneous chains.

\begin{figure}[h]
\begin{center}
\includegraphics[scale=0.15]{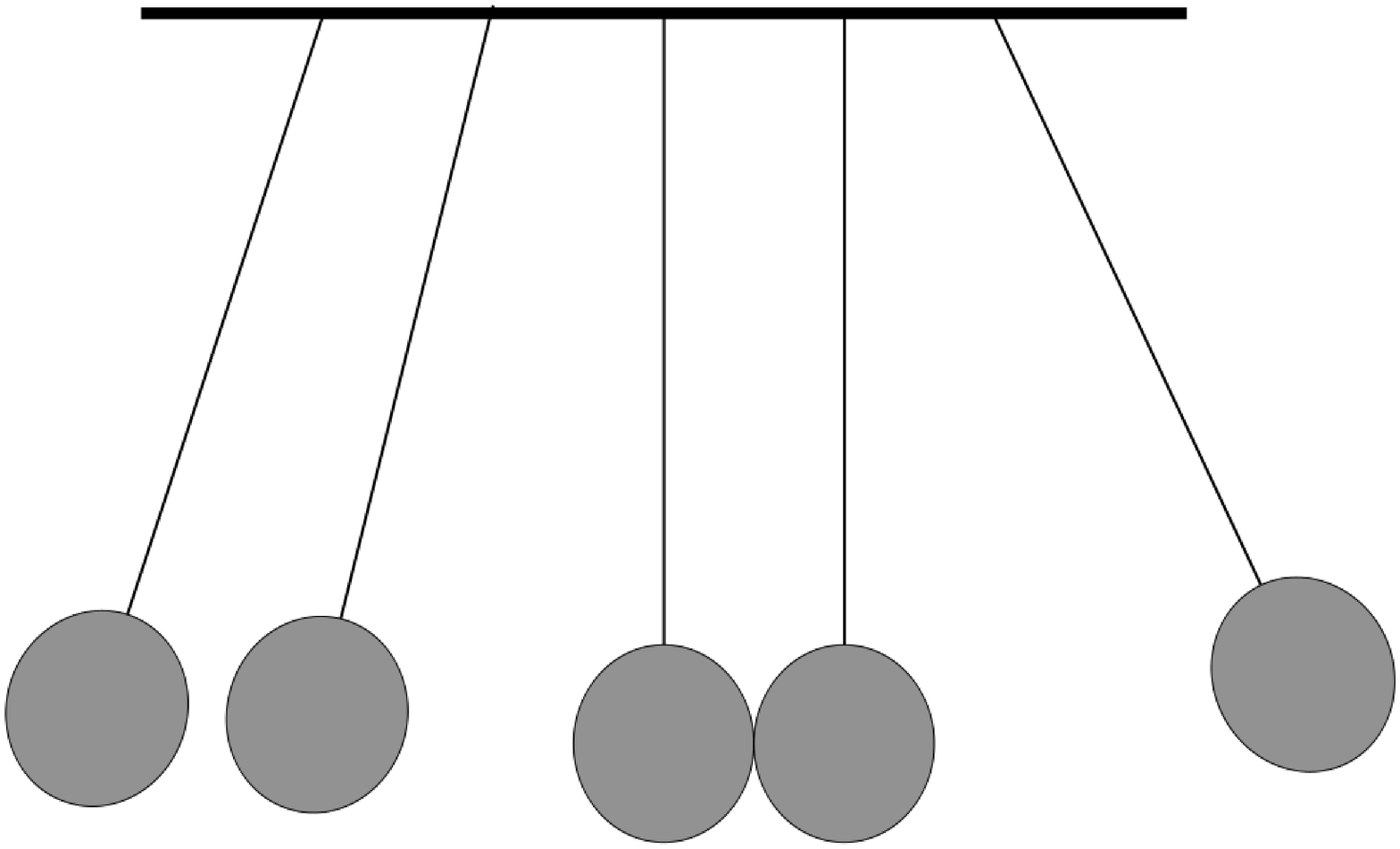}
\includegraphics[scale=0.15]{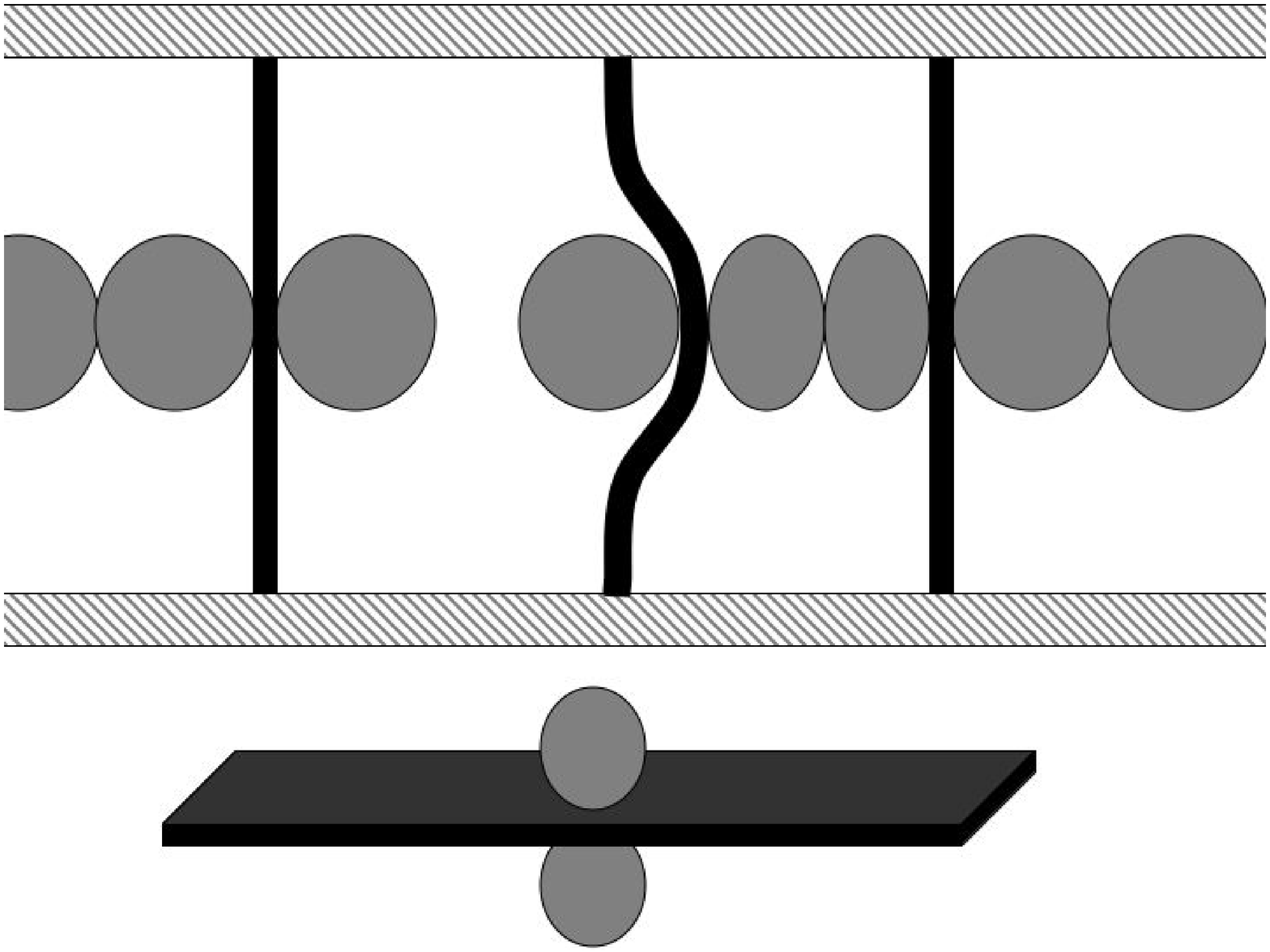}
\end{center}
\caption{\label{boules}
Left~: prototypical Newton's cradle. Right~:
array of clamped cantilevers decorated by spherical beads (displacements are
amplified for clarity).}
\end{figure}

However, in contrast to the above picture, the existence of
discrete breathers in a chain of linear oscillators coupled by Hertzian 
potentials has been recently reported~\cite{jamesc}. This model with an
on-site potential describes e.g. the small amplitude waves in
a Newton's cradle \cite{hutzler}, which consists of a chain of beads attached 
to pendula (see figure \ref{boules}, left).
In~\cite{jamesc}, static and moving breathers were numerically observed as a 
result of modulational instabilities of
periodic traveling waves, and extremely stable static breathers were generated from specific initial conditions.
In addition, a reduced model, the so-called discrete $p$-Schr{\"o}dinger 
(DpS) equation was derived as an asymptotic model for small
amplitude oscillations in the Newton's cradle, and successfully reproduced 
the above localization phenomena.
The discrete breathers possess special properties both in the original cradle 
model and the simplified DpS system,
i.e. their spatial decay is super-exponential and their width remains 
nearly constant at small amplitude.

In this paper we extend the above results in three directions.
The first one concerns traveling breather excitations, i.e.
localized waves displaying an internal oscillation in addition to their translational motion.
We show numerically that such waves can be excited
very simply in the cradle system by an impact at one end of a 
semi-infinite chain, and check that this dynamics is also reproduced by
the DpS model.
We also discuss some unusual properties of the moving breathers obtained in 
this way.
Due to the fully nonlinear Hertzian interactions, these breathers
display a strong localization (super-exponential decay)
and their dispersion remains very weak during propagation.
In contrast, introducing a precompression
attenuates spatial localization and enhances dispersive effects
(due to the fact that precompression
adds effectively a linear component to Hertzian interactions).
We illustrate this idea using both numerical simulations and
the discrete nonlinear Schr\"odinger equation, which allows us to approximate 
small amplitude traveling breathers under the effect of precompression.
In addition, we check that the whole phenomenology
remains valid at small amplitude when the linear local potential is
replaced by a smooth anharmonic potential.
However, such 
local nonlinearities yield additional phenomena, such as the excitation
of a surface mode after the impact for soft local potentials, 
and for hard potentials
the occurence of a direction-reversing traveling breather.
The latter is reminiscent of excitations known as ``boomerons''
(direction-reversing solitons) that were found
up to now only in particular
integrable models (see \cite{degas} and references therein).

Our second contribution concerns the computation of static breathers in the Newton's cradle
and their numerical continuation.
Using a modified Gauss-Newton method introduced in~\cite{crea}, we
obtain branches of site- and bond-centered breathers parametrized by their 
frequency
$\omega > \omega_0$ ($\omega_0$ being the frequency of the local linear oscillators).
These branches bifurcate from the trivial equilibrium 
when $\omega \rightarrow \omega_0$
and can be continued up to a strongly nonlinear regime.
Moreover, the Floquet spectra of these breathers display 
(in addition to the usual
double eigenvalue $+1$) an extra pair of eigenvalues very close to unity.
As an effect of this near-degeneracy, small perturbations of the breathers
along an associated pinning mode generate a
translational motion with negligible radiation, according to the
process analyzed in~\cite{aubryC}. This provides a connection
between these standing breathers and the traveling ones mentioned above.
In addition to these numerical computations, we obtain an analytical
quasi-continuum approximation of the breather profiles valid at small 
amplitude.
These approximate breathers have a compact support, which provides a
reasonable approximation to the super-exponential decay of the exact breathers.
This situation is analogous to what is known for the approximation of solitons in uncompressed granular chains \cite{neste2,english}.

Lastly, we examine
possible experimental realizations of these kinds of granular lattices
and the related observation of moving breathers after an impact, as well
as their potential practical usefulness (e.g., in the form of granular
protectors; see below).
In the usual Newton's cradle,
the period of local oscillations (of the order of a second) is much
larger than the collision time between two beads (typically of the order of $0.1$ ms \cite{lovett}),
so the propagation of compression
pulses and the oscillatory dynamics occur on two well-separated time scales.
Moreover, as shown in the present paper,
the DpS regime giving rise to discrete breathers is realized when
local oscillations are faster than binary collisions.
As a consequence, moving breathers would not be observable in practice
in a Newton's cradle.
However, we argue that reasonably simple
mechanical systems could be tailored so that the oscillation and
collision time scales become similar and the DpS regime takes place.
As a prototype for which this situation occurs,
we consider the chain of identical clamped cantilevers represented in
figure \ref{boules}.
Each cantilever is decorated by two spherical beads attached to its center, and the beads of two
successive cantilevers are tangent at the ground state.
Using a reduced oscillator chain model of this system (calibrated
for realistic material parameter values), we check that an
impact on the first cantilever generates a moving breather.
In addition, we argue that such devices may have potential applications
as granular protectors. Indeed, we observe that
moving breathers can be almost totally reflected by a localized impurity.

The outline of the paper is as follows. In section \ref{nonexist} we prove 
the non-existence
of time-periodic spatially localized oscillations in a broad class of 
Fermi-Pasta-Ulam lattices
with non-attracting interactions, including uncompressed granular chains. 
Section \ref{crdps}
studies the properties of static and moving breathers in granular chains with 
on-site
potentials and in the DpS equation, and highlights the relation between the
two models.
The new effects introduced by local anharmonic potentials are also discussed.
In addition we introduce the mixed granular-cantilever chain, and analyze an 
impact problem
in this system in connection with the previous findings.
Section \ref{caseprecomp} illustrates how precompression modifies the
properties of localized waves. Lastly, section \ref{conclu}
concludes this study and analyzes the results from a more general perspective,
in connection with possible experiments.

\section{\label{nonexist}Non-existence of breathers in FPU chains with repulsive interactions}

We consider an infinite chain of particles of masses $m_n >0$, interacting
with their nearest neighbors
via anharmonic potentials $V_n$. This type of
system (which can be thought of in general, i.e., for unequal masses
$m_n$, as a spatially inhomogeneous FPU lattice)
corresponds to the Hamiltonian
\begin{equation}
\label{ham1}
{\mathcal H}=
\sum_{n\in \mathbb{Z}}{\frac{m_n}{2}\, \dot{x}_n^2 + V_n (x_{n+1} - x_n)}
\end{equation}
where ${x}_n$ denotes the particle displacements from the ground state.
We consider
interaction potentials $V_n$ of the form
$$
V_n(x)=W_n [ (- x)_+ ],
$$
where $(a)_+ = \rm{max}(a,0)$, $W_n \in C^1(\mathbb{R}^+ , \mathbb{R}^+)$,
$W_n^\prime (0)=0$ and $W_n^\prime (x) > 0$ for all $x>0$. The form of
$V_n$ implies that particle interactions are repulsive under compression (i.e. for $x<0$) and
unilateral (interaction forces vanish under extension, i.e. for $x>0$).

Moreover we assume
\begin{equation}
\label{unif}
W_n^\prime (x)
 \leq f(x)  \ \ \ \forall x \in [0,r], \, \forall n \geq n_0,
\end{equation}
for some real constant $r >0$, integer $n_0$
and a monotone increasing function $f \in C^0([0,r])$ satisfying $f(0)=0$.
For example,
these assumptions are satisfied with $f(x)=\sup_{n \geq n_0} W_n^\prime (x)$
if the functions $W_n$ are convex in $[0,r]$ and belong to some finite set for $n \geq n_0$
(this is the case in particular for spatially periodic systems).
Another example is given by Hertzian interactions
$$
W_n (x)=\frac{1}{\alpha_n +1}\, \gamma_n\,  x^{\alpha_n +1},
$$
where the coefficients $\gamma_n , \alpha_n >0$ depend on
material properties and particle geometry. In that case
one can choose $f(x)=\gamma \, x^{\alpha}$ (and $r=1$)
provided $\gamma_n \leq \gamma$ and
$\alpha_n \geq \alpha >0$ for all $n\geq n_0$.

The Hamiltonian (\ref{ham1}) leads to the equations of motion
\begin{equation}
\label{eqm1}
m_n \, \ddot{x}_n  = V_n^\prime (x_{n+1} -x_n) - V_{n-1}^\prime (x_{n} -x_{n-1}), \ \ \ n \in \mathbb{Z}.
\end{equation}
In what follows we show that under the above assumptions, the only time-periodic breather solutions of
(\ref{eqm1}) are trivial equilibria. Due to the translational invariance of (\ref{ham1}),
breathers are defined as time-periodic solutions which converge (uniformly in time) towards
translations $x_{n}=c_{\pm}\in\mathbb{R}$ as
$n\rightarrow\pm\infty$. This implies that relative particle displacements vanish at infinity, i.e. one has
\begin{equation}
\label{loc}
\lim_{n\rightarrow \pm \infty}{\| x_n - x_{n-1}  \|_{L^\infty(0,T)}}=0
\end{equation}
for a $T$-periodic breather.
In what follows, we prove in fact a more general nonexistence
result of nontrivial periodic solutions vanishing as $n \rightarrow +\infty$.

\begin{theorem}
All time-periodic solutions of (\ref{eqm1}) satisfying
\begin{equation}
\label{loc2}
\lim_{n\rightarrow + \infty}{\| x_n - x_{n-1}  \|_{L^\infty}}=0
\end{equation}
are independent of $t$
and increasing with respect to $n$.
\end{theorem}

\begin{proof}
Let us consider a $T$-periodic solution of (\ref{eqm1}) and
integrate (\ref{eqm1}) over one period. This yields the equality
$$
\bar{F}_n = \bar{F}_{n+1},$$
where
$
\bar{F}_n = \frac{1}{T}\int_0^T{V_{n-1}^\prime (x_{n}(t) -x_{n-1}(t))\, dt}
$
is the average interaction force between masses $n-1$ and $n$.
Consequently $\bar{F}_n=\bar{F}$ is independent of $n$.

Now let us check that $\bar{F}$ vanishes thanks to the bound (\ref{unif}) uniform in $n$.
We have for all $n$
\begin{eqnarray*}
| \bar{F} | &=& \frac{1}{T}\int_0^T{W_{n-1}^\prime [ \, (x_{n-1}(t)-x_{n}(t) )_+ \, ] \, dt} \\
 & \leq    & \| W_{n-1}^\prime [ \, (x_{n-1}-x_{n} )_+ \, ]   \|_{L^\infty}. \\
\end{eqnarray*}
Taking into account (\ref{loc2}) and (\ref{unif}), the above inequality yields for $n$ large enough
$$
| \bar{F} | \leq \| f [ \, (x_{n-1}-x_{n} )_+ \, ]   \|_{L^\infty} = f[ \|  (x_{n-1}-x_{n} )_+    \|_{L^\infty} ]
$$
since $f$ is increasing. It follows that
$$
| \bar{F} | \leq f( \|  x_{n-1}-x_{n}     \|_{L^\infty} ) \rightarrow 0 \mbox{ as } n\rightarrow +\infty
$$
hence $\bar{F}=0$.

Now we use the fact that the interactions between particles are repulsive,
i.e. we have $-V_n^\prime (x)=  W_{n}^\prime [ \, (-x )_+ \, ] \geq 0$.
Since the $T$-periodic functions $F_n (t)=V_{n-1}^\prime (x_{n}(t) -x_{n-1}(t))$ are negative, continuous
and satisfy $\int_0^T{F_n(t)\, dt}=0$ as shown previously, we have consequently $F_n(t)=0$
for all $t$ and $n$. Using (\ref{eqm1}), this implies $\ddot{x}_n =0$ and thus
$x_n$ is an equilibrium solution (due to time-periodicity). Moreover one has
$x_{n} \geq x_{n-1}$ since $F_n=0$.
\end{proof}

\vspace{1ex}

We note that the above arguments do not work if an on-site potential is added to (\ref{ham1}), because
the average interaction forces are no more independent of $n$. In the next section, we numerically
show the existence of breathers for such type of nonlinear lattices.

\section{\label{crdps}Granular chains with local potentials and their correspondence to the DpS Equation}

We consider a nonlinear lattice with the Hamiltonian
\begin{equation}
\label{hamresc}
{\mathcal H}=
\sum_{n}{\frac{1}{2}\, \dot{y}_n^2+W( y_n ) + V(y_{n+1}-y_n)},
\end{equation}
where
\begin{equation}
\label{pothertz}
V(r)=\frac{2}{5}\,  ( -r )_+^{5/2}.
\end{equation}
The system (\ref{hamresc}) corresponds to a chain of identical particles in the local potential $W$,
coupled by the classical Hertz potential $V$ describing
contacts between smooth non-conforming surfaces.
Unless explicitly stated, the on-site potential $W$ will be chosen harmonic with
\begin{equation}
\label{lp}
W(y)=\frac{1}{2}\, y^2.
\end{equation}
In that case, the dynamical equations read
\begin{equation}
\label{eqm}
\ddot{y}_n + y_n  =(y_{n-1} - y_{n})_+^{3/2} - (y_n - y_{n+1})_+^{3/2}   .
\end{equation}
Figure \ref{boules} depicts
two examples of such systems.
In practical situations,
the assumption of a local harmonic potential
implies that the model will be valid for
small amplitude waves and suitable time scales
on which higher order terms can be neglected.
In order to capture higher order effects,
different parts of our analysis will be extended to
symmetric anharmonic local potentials
\begin{equation}
\label{anpot}
W(y)=\frac{1}{2}\, y^2+\frac{s}{4}\,  y^4,
\end{equation}
where the parameter $s$ measures the degree of anharmonicity.

In the work \cite{jamesc}, long-lived
static and traveling breather solutions of (\ref{eqm})
have been numerically observed, starting from suitably chosen
localized initial conditions, or from small perturbations of
unstable periodic traveling waves.
However, the classical result of MacKay and Aubry \cite{aubryMK}
proving the existence of static breathers near the anti-continuum limit
does not apply in that case. Indeed, if Hertzian interactions forces are cancelled
(or equivalently, if one considers breathers in the limit of vanishing amplitude), one obtains
an infinite lattice of identical linear oscillators, and the nonresonance assumption of reference \cite{aubryMK}
is not satisfied. Moreover, other existence proofs based on spatial dynamics and
the center manifold theorem \cite{jsc} do not apply, due to the fully-nonlinear character of interaction forces
(the same remark holds true in the case of traveling breathers \cite{ioossK,jamessire,sire}).
Variational tools \cite{aubkadel,pankovbook} may be suitable to obtain existence proofs in this
context, but this question is outside the scope of the present paper, 
where we shall resort chiefly to numerical and asymptotic methods.

In section \ref{dpsrev}, we recall the relation between (\ref{eqm}) and the DpS equation
derived in \cite{jamesc}. In section \ref{exist}, we numerically compute
breather solutions of (\ref{eqm}) by the Newton method, and compare them to
a quasi-continuum approximation deduced from the DpS equation.
Section \ref{stabmob} concerns the stability and mobility of breathers
in model (\ref{hamresc})-(\ref{anpot})
and the DpS equation, and the generation of traveling
breathers by an impact is studied in the same models in section \ref{impactpb}.
In section \ref{rescimpact}, we consider a general class of
granular chains with local potentials, and show that the
DpS regime occurs in the above impact problem when
local oscillations are faster than binary collisions.
Section \ref{canti} provides
an application of the above results to a chain of stiff cantilevers decorated by spherical beads.

\subsection{\label{dpsrev}The discrete $p$-Schr{\"o}dinger equation}

Small amplitude solutions of
system (\ref{hamresc})-(\ref{lp}) can be well approximated by an equation of the
nonlinear Schr{\"o}dinger type, namely the discrete $p$-Schr{\"o}dinger
(DpS) equation with $p=5/2$
\begin{equation}
\label{dps1}
i \, \dot{v}_n=(v_{n+1} - v_n) |v_{n+1}-v_n|^{p-2} 
- (v_n-v_{n-1})
|v_n-v_{n-1}|^{p-2}.
\end{equation}
The most standard model reminiscent of
this family of equations is the
so-called discrete nonlinear Schr{\"o}dinger (DNLS) equation,
studied in detail in a number of different contexts, including
nonlinear optics and atomic physics over the
past decade \cite{pgk,eilbeckj}.
However, the DpS equation is fundamentally different in that it contains a fully nonlinear inter-site coupling term,
corresponding to a discrete $p$-Laplacian.

\vspace{1ex}

To make the connection with the DpS equation
more precise, we sum up some basic elements of the analysis of \cite{jamesc}.
Let us consider the lattice model (\ref{eqm}) and the DpS equation
\begin{equation}
\label{dps}
2i \tau_0 \dot{A_n}=
(A_{n+1}-A_n)\, |A_{n+1}-A_n |^{1/2} 
-(A_{n}-A_{n-1})\, |A_{n}-A_{n-1} |^{1/2},
\end{equation}
where the time constant $\tau_0$ reads
$$
\tau_0 = \frac{5  (\Gamma{(\frac{1}{4})})^2 }{24 \sqrt{\pi}} \approx 1.545
$$
and $\Gamma$ denotes Euler's Gamma function.
Given a solution of (\ref{dps}) and $\epsilon >0$ small enough,
one obtains an approximate solution of (\ref{eqm})
\begin{equation}
\label{approx}
y_n^{\rm{app}} (t)=2\, \epsilon\, \mbox{Re }[\, A_n(\epsilon^{1/2} t )\, e^{it} \, ] .
\end{equation}
The approximate solution (\ref{approx}) and amplitude equation (\ref{dps})
have been derived in \cite{jamesc} using a multiple-scale expansion.
According to \cite{dumas}, for initial conditions of the form
$y_n (0) = 2\, \epsilon\, \mbox{Re }[\, A_n(0) \, ] + O(\epsilon^{3/2})$,
$\dot{y}_n (0) = -2\, \epsilon\, \mbox{Im }[\, A_n(0) \, ] + O(\epsilon^{3/2})$
with $\epsilon \approx 0$, this approximation is $O(\epsilon^{3/2})$-close
to the exact solution of (\ref{eqm}) at least up to times $t=O(\epsilon^{-1/2})$
(see also numerical results of \cite{jamesc} comparing
the DpS approximation and exact solutions of (\ref{eqm})).
Moreover, for some family of periodic traveling wave solutions of
the DpS equation, the ansatz (\ref{approx}) is $O(\epsilon^{3/2})$-close
to exact small amplitude periodic traveling waves of (\ref{eqm}) \cite{jamesc}.

Lastly, it is interesting to mention that the DpS equation depends on the terms
of (\ref{eqm}) up to order $O(|y|^{3/2})$ (see \cite{jamesc}, section 2.1). It follows that this equation remains unchanged
for smooth anharmonic on-site potentials $W(y)=\frac{1}{2}\, y^2 +O(|y|^3)$, because
the associated extra nonlinearity is at least quadratic. Consequently, the addition
of a local anharmonicity doesn't change the dynamics of (\ref{eqm}) for small
amplitude waves, on the timescales governed by the DpS equation.

\subsection{\label{exist}Computation of static breathers}

\begin{figure}[h]
\psfrag{n}[0.9]{ $n$}
\psfrag{x_n(0)}[1][Bl]{ $y_n (0)$}
\psfrag{u_n}[1][Bl]{ $u_n^{(1)}$}
\begin{center}
\includegraphics[scale=0.35]{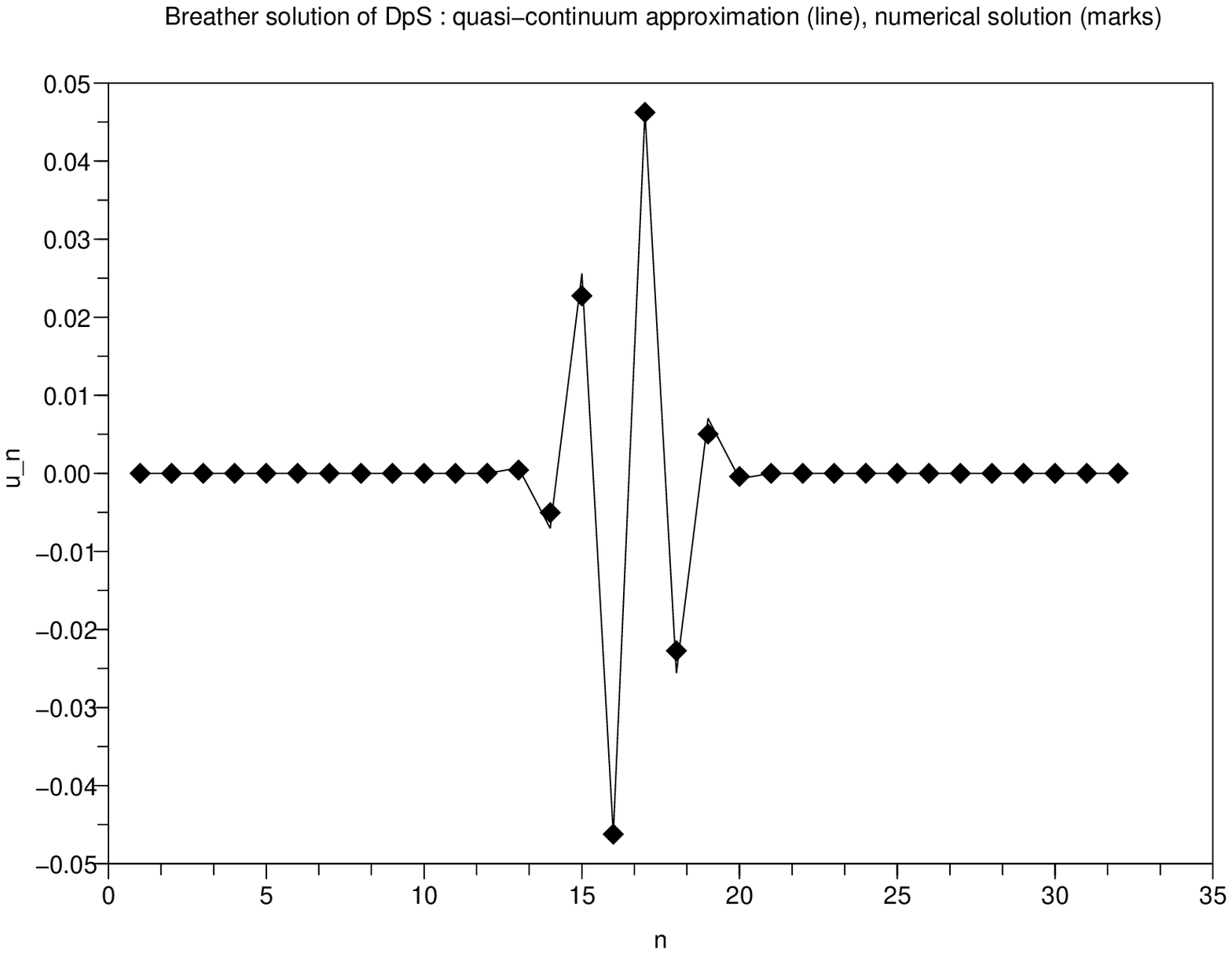}
\includegraphics[scale=0.35]{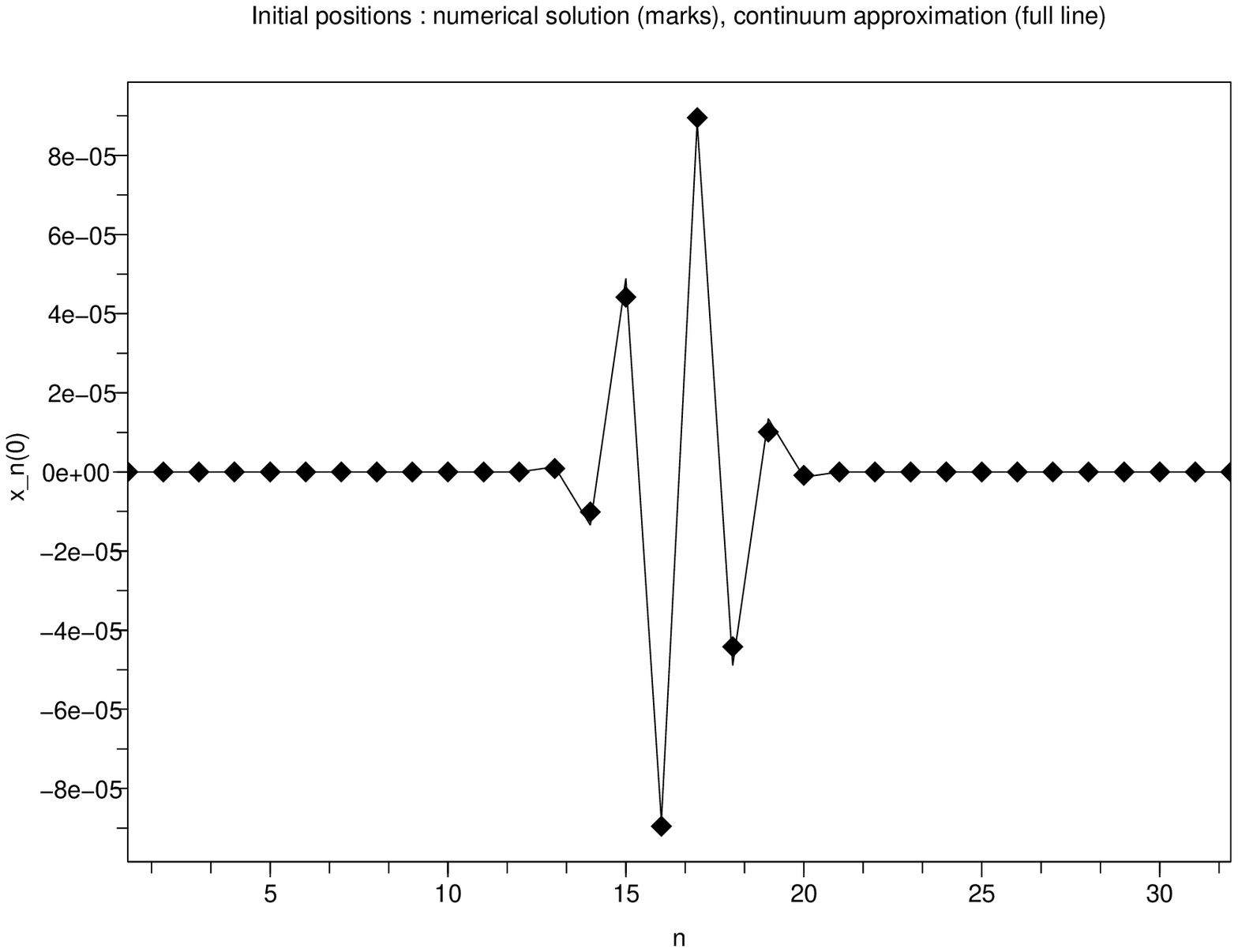}
\includegraphics[scale=0.35]{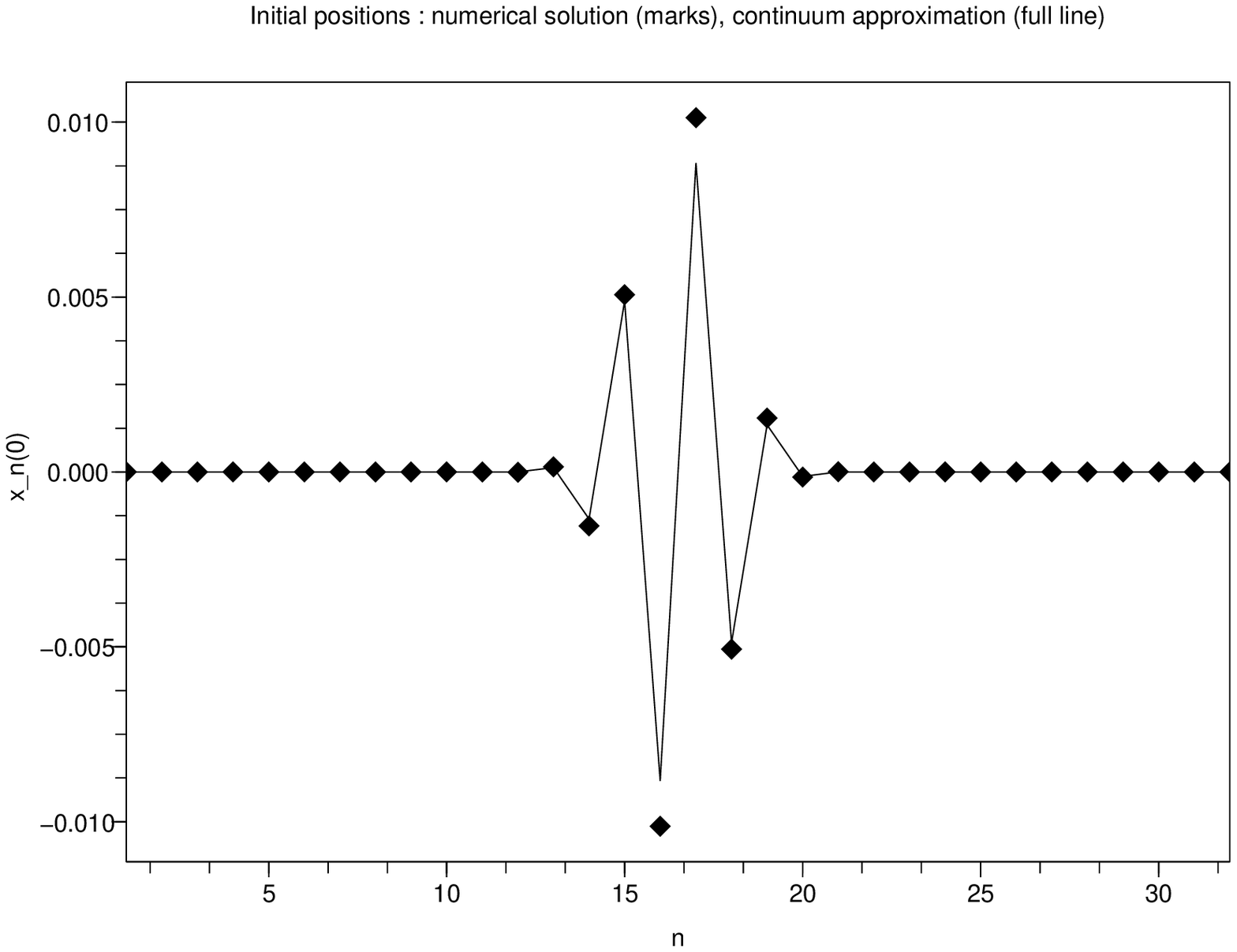}
\end{center}
\caption{\label{bcbreathers}
Top panel: spatially antisymmetric solution of the stationary DpS equation (\ref{dpsstat}), computed
numerically for $\mu=1$ (marks). This solution is compared to the quasi-continuum approximation
$u_n^{(1)}$ defined by equation (\ref{un1}) (continuous line).
The other graphs compare a bond-centered breather solution of (\ref{eqm}) computed numerically
(marks) and its quasi-continuum approximation $y_n^{(1)}$ (continuous line).
The middle plot corresponds to a small amplitude breather
($\omega_b=1.01$), and the bottom plot to a more strongly
nonlinear regime ($\omega_b=1.1$).}
\end{figure}

\begin{figure}[h]
\psfrag{n}[0.9]{ $n$}
\psfrag{x_n(0)}[1][Bl]{ $y_n (0)$}
\psfrag{u_n}[1][Bl]{ $u_n^{(2)}$}
\begin{center}
\includegraphics[scale=0.35]{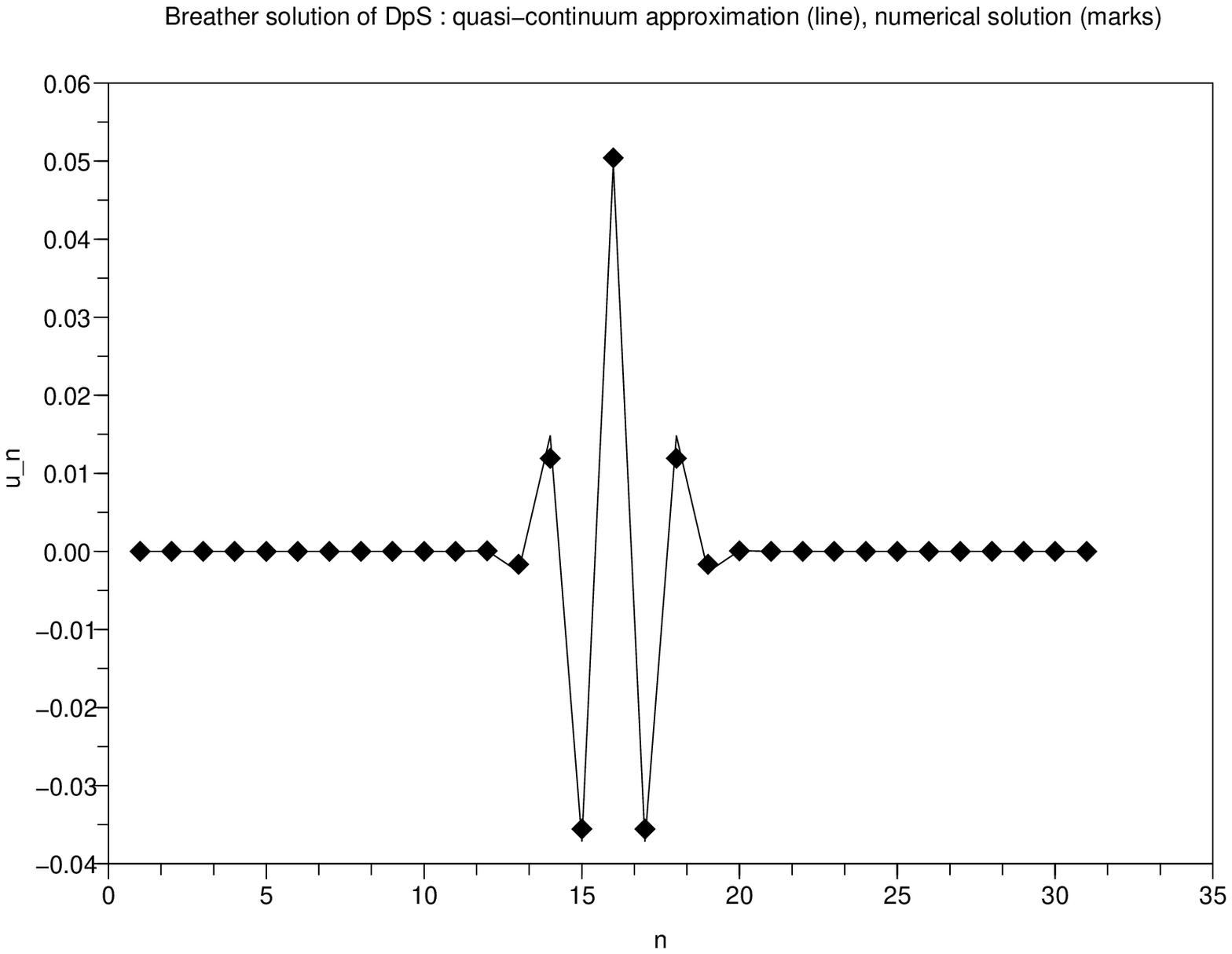}
\includegraphics[scale=0.35]{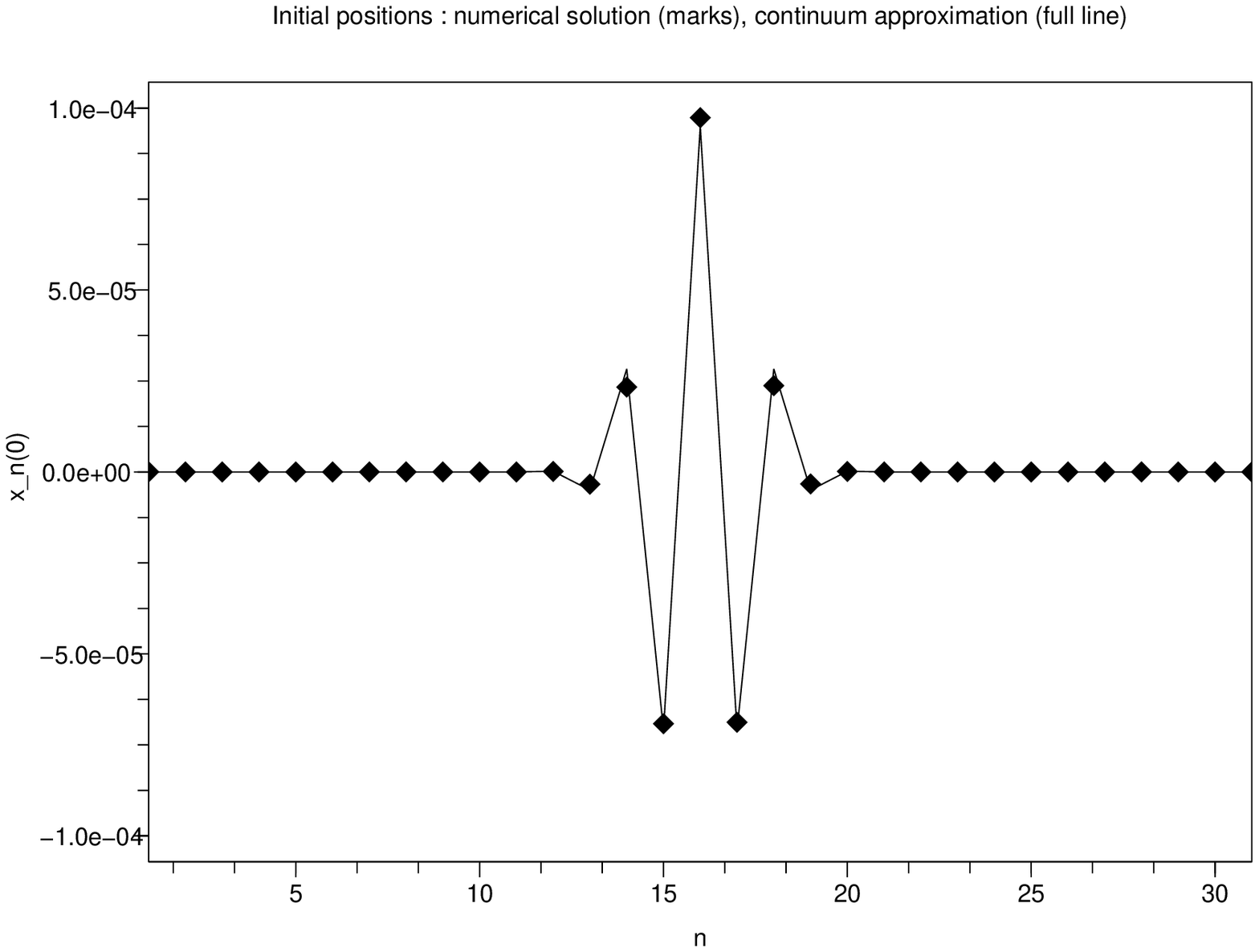}
\includegraphics[scale=0.35]{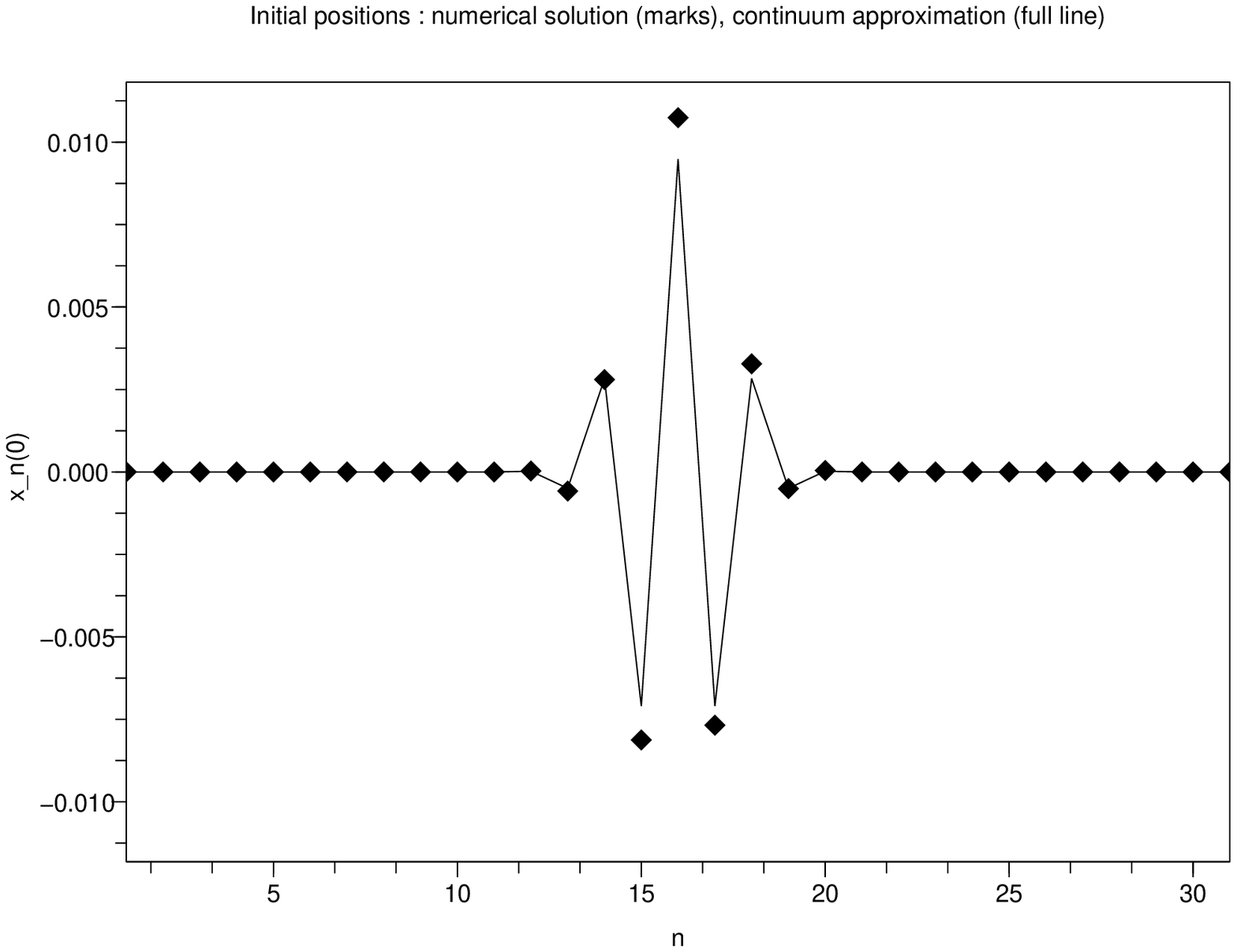}
\end{center}
\caption{\label{ssbreathers}
Top panel: spatially symmetric solution of the stationary DpS equation (\ref{dpsstat}), computed
numerically for $\mu=1$ (marks). This solution is compared to the quasi-continuum approximation
$u_n^{(2)}$ defined by equation (\ref{un2}) (continuous line).
The other graphs compare a site-centered breather solution of (\ref{eqm}) computed numerically
(marks) and its quasi-continuum approximation $y_n^{(2)}$ (continuous line).
The middle plot corresponds to a small amplitude breather
($\omega_b=1.01$), and the bottom plot to a more strongly
nonlinear regime ($\omega_b=1.1$).
}
\end{figure}

The work of \cite{jamesc} illustrated the existence of time-periodic
and spatially localized
solutions of the DpS equation.
Figures \ref{bcbreathers} and \ref{ssbreathers} (top panels)
display the profiles of
spatially antisymmetric or symmetric breather solutions of the DpS equation (\ref{dps1}). These
are sought by using the standard stationary ansatz for DNLS
type equations of the form $v_n=\exp(i \mu t) \, u_n$ with $\mu >0$ and
$u_n \in \mathbb{R}$.
The resulting coupled nonlinear algebraic equations read
\begin{equation}
\label{dpsstat}
-\mu {u}_n=(u_{n+1} - u_n) |u_{n+1}-u_n|^{1/2} 
- (u_n-u_{n-1})
|u_n-u_{n-1}|^{1/2}
\end{equation}
and are solved via a fixed point iteration of the Newton-Raphson type,
for free end boundary conditions.

Note that equation (\ref{dps1}) has
a scale invariance, since any
solution $v_n (t )$ generates a one-parameter family of solutions
$a\, v_n (|a|^{1/2}\, t )$, $a\in \mathbb{R}$. Thanks to this
scale invariance, the whole families of antisymmetric and symmetric breathers
can be reconstructed from the case $\mu=1$ of (\ref{dpsstat}).
In particular, breather amplitudes are $\propto \mu^2$ and the breather
width remains unchanged when $\mu \rightarrow 0$,
a property that strongly differs from the broadening of
DNLS breathers at small amplitude (see e.g. \cite{cjkms}, section 3).

\vspace{1ex}

In what follows we approach the two breather profiles using a
quasi-continuum approximation. Fixing $\mu=1$ and introducing
$$
w_n = (u_{n+1}-u_{n})\, |u_{n+1}-u_{n} |^{1/2},
$$
equation (\ref{dpsstat}) becomes
\begin{equation}
\label{dpsstat2}
w_{n+1}-2w_{n}+w_{n-1}+
w_{n} \, |w_{n}|^{-1/3} =0,
\end{equation}
where the nonlinear coupling has been linearized
(at the expense of having an on-site nonlinearity
non-differentiable at the origin).
The spatial profiles of figures \ref{bcbreathers} and \ref{ssbreathers}
suggest to use the so-called 
staggering transformation $w_n=(-1)^n\, f(n)$, which yields
\begin{equation}
\label{map}
f(n+1)-2f(n)+f(n-1)=-4f(n) + f(n) \, |f(n)|^{-1/3}.
\end{equation}
Now we look for an {\em approximate} solution $F$ of (\ref{map}).
For this purpose we use the formal approximation
$F(n\pm 1)\approx F(n) \pm F^\prime(n)
+ \frac{1}{2}F^{\prime\prime}(n)$, in same the spirit as
the approximations of soliton profiles performed in reference \cite{neste2}
(the accuracy of
this approximation will be checked a posteriori by numerical computations)\footnote{Note that $w_n$ corresponds to a spatially modulated binary oscillation, 
and a continuum approximation is obtained for its envelope, whereas
the continuum approximation of \cite{neste2} was performed on the full soliton profiles.}.
This leads to the differential equation
\begin{equation}
\label{diff}
F^{\prime\prime}=-4F + F\, |F|^{-1/3},
\end{equation}
which possesses a family of compactly supported solutions
$F(x)=\pm g(x+\phi)$, where
$$
g(x)=\big(\frac{3}{10}\big)^3\, \cos^6{\big(  \frac{x}{3} \big)} \mbox{ for } |x|\leq \frac{3\pi}{2}, \ \ \
g=0 \mbox{ elsewhere.}
$$
Replacing $f$ by its approximation $F$ and performing
appropriate choices of sign and spatial shifts in $F$, one obtains the symmetric approximate solutions
of (\ref{dpsstat2})
$$
w_n^{(1)}=(-1)^{n+1}g(n) , \ \ \
w_n^{(2)}=(-1)^{n+1}g(n+\frac{1}{2}).
$$
The case $\mu=1$ of (\ref{dpsstat}) yields
$u_n = w_{n-1}-w_n$, therefore we get the following
quasi-continuum approximations
of the antisymmetric and symmetric breather profiles
\begin{equation}
\label{un1}
u_n^{(1)}=(-1)^{n}\, [g(n)+g(n-1)] ,
\end{equation}
\begin{equation}
\label{un2}
u_n^{(2)}=(-1)^{n}\, [g(n+\frac{1}{2})+g(n-\frac{1}{2})].
\end{equation}
The first graphs of figures \ref{bcbreathers} and \ref{ssbreathers}
show the excellent agreement of these
approximations with the numerical solutions of the stationary
DpS equation.
Returning to the ansatz (\ref{approx}) and the
time-dependent (non-renormalized) DpS equation (\ref{dps}),
we obtain approximate breather solutions of (\ref{eqm}) taking the form
\begin{equation}
\label{ansatzapproxst2}
y_n^{(s)} (t)= 2 \epsilon\, {u}_n^{(s)}\, \cos{(\omega_{\rm{b}} t )},
\ \ \
\omega_{\rm{b}} =1+ \frac{\epsilon^{1/2}}{2\tau_0}, \ \ \ s=1,2.
\end{equation}
It is interesting to observe that approximation (\ref{ansatzapproxst2})
is unaffected by smooth on-site nonlinear terms for $\epsilon \approx 0$,
since we have noticed that the DpS equation remains unchanged.

In what follows we compare
the above approximations with breather solutions of (\ref{eqm})
computed numerically for free end boundary conditions.
Let us note $Y_n = (y_n,\dot{y}_{n})$.
We use a method described in~\cite{crea}
to compute zeros $Y_n (0)= (y_n(0),0)$ of the period map of the flow of (\ref{eqm})
(these initial conditions correspond to breathers even in time).
The method of  \cite{crea} is based on an adapted Gauss-Newton scheme
and path-following.

\begin{figure}[h]
\psfrag{n}[0.9]{ $n$}
\psfrag{x_n}[1][Bl]{ $y_n $}
\begin{center}
\includegraphics[scale=0.4]{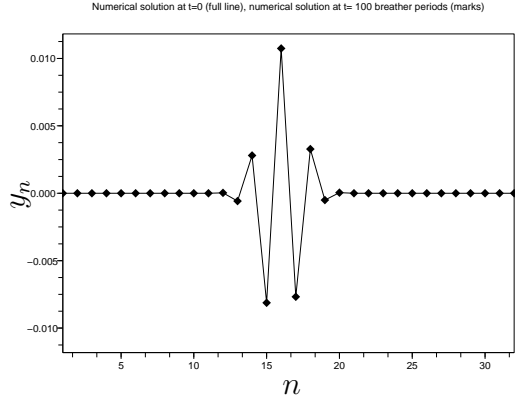}
\end{center}
\caption{\label{initialvsfinal}
Initial breather positions computed by the Newton method
for a chain of $31$ particles
(full line), compared to
their evolution at $t=100\, T_b$ (marks). Computations are performed for a site-centered
breather with frequency $\omega_b=1.1$.}
\end{figure}

\begin{figure}[h]
\psfrag{n}[0.9]{ $n$}
\psfrag{|x_n(0)|}[1][Bl]{ $|y_n(0)| $}
\begin{center}
\includegraphics[scale=0.4]{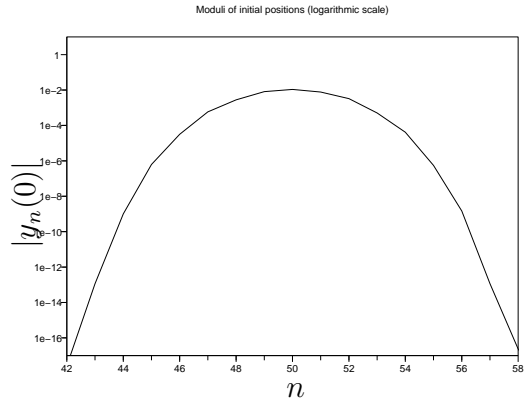}
\end{center}
\caption{\label{supexpdecay}
Moduli of the initial breather positions computed by the Newton method, plotted
in semi-logarithmic scale. Computations are performed for
a chain of $99$ particles and
a site-centered
breather with frequency $\omega_b=1.1$.}
\end{figure}

An example of computation of a breather with frequency $\omega_b = 1.1$
is shown in figure \ref{initialvsfinal}.
The initial guess used for the Newton method
is the site-centered approximate breather solution derived from the DpS equation.
After 5 iterations, the relative residual error
$$
E=\frac{{\| {\{ Y_{n}( {T_b}) - Y_{n}(0) \}}_n  \|}_{\infty}}{{\| {\{ Y_{n}(0) \}}_n  \|}_{\infty}},
\ \ \ T_b = \frac{2\pi}{\omega_b},
$$
reaches $1.5. 10^{-9}$ and the relative variation (measured using sup norms)
of particle positions between successive iterates drops to $1.7. 10^{-11}$.
Figure \ref{initialvsfinal} compares the initial breather positions
computed by the Newton method and their evolution at $t=100\, T_b$, which shows that
the breather oscillations are extremely stable. The super-exponential spatial decay
of the breather is shown in figure \ref{supexpdecay}.

Using the above numerical scheme,
we obtain two branches of breather solutions of (\ref{eqm}) with different symmetries,
parametrized by their frequency $\omega_b >1$. They consist of
bond-centered breathers, i.e. spatially antisymmetric
solutions satisfying $y_{-n+1}=-y_n$
(figure \ref{bcbreathers}) and site-centered breathers (figure \ref{ssbreathers}).
The latter possess subtle symmetry properties.
Since the Hertz potential is non-even, equation (\ref{eqm}) is not 
invariant by the symmetry
$S y_n := y_{-n}$. However, the set of $T_b$-periodic solutions of
(\ref{eqm}) is invariant under the transformation
$S'y_n(t)= -y_{-n}(t+T_b/2)$.
The site-centered breathers of (\ref{eqm}) are left invariant by $S'$ and not by $S$
(their asymmetry under $S$ increases with $\omega_b$ and becomes
visible in the bottom panel of figure \ref{ssbreathers}). 
On the contrary, the DpS equation admits both symmetries $S$ and $S'$, which both leave the site-centered DpS breathers invariant.

These different types of symmetries are illustrated by
figures \ref{bcbreathers} and \ref{ssbreathers}, which compare
the approximations (\ref{ansatzapproxst2}) with breather solutions of (\ref{eqm})
computed by the Newton method (middle and bottom plots).
While approximation (\ref{ansatzapproxst2}) is excellent at
small amplitude (case $\omega_b=1.01$), its accuracy deteriorates
in a more strongly nonlinear regime (case $\omega_b=1.1$).

\begin{figure}[h]
\begin{center}
\includegraphics[scale=0.35]{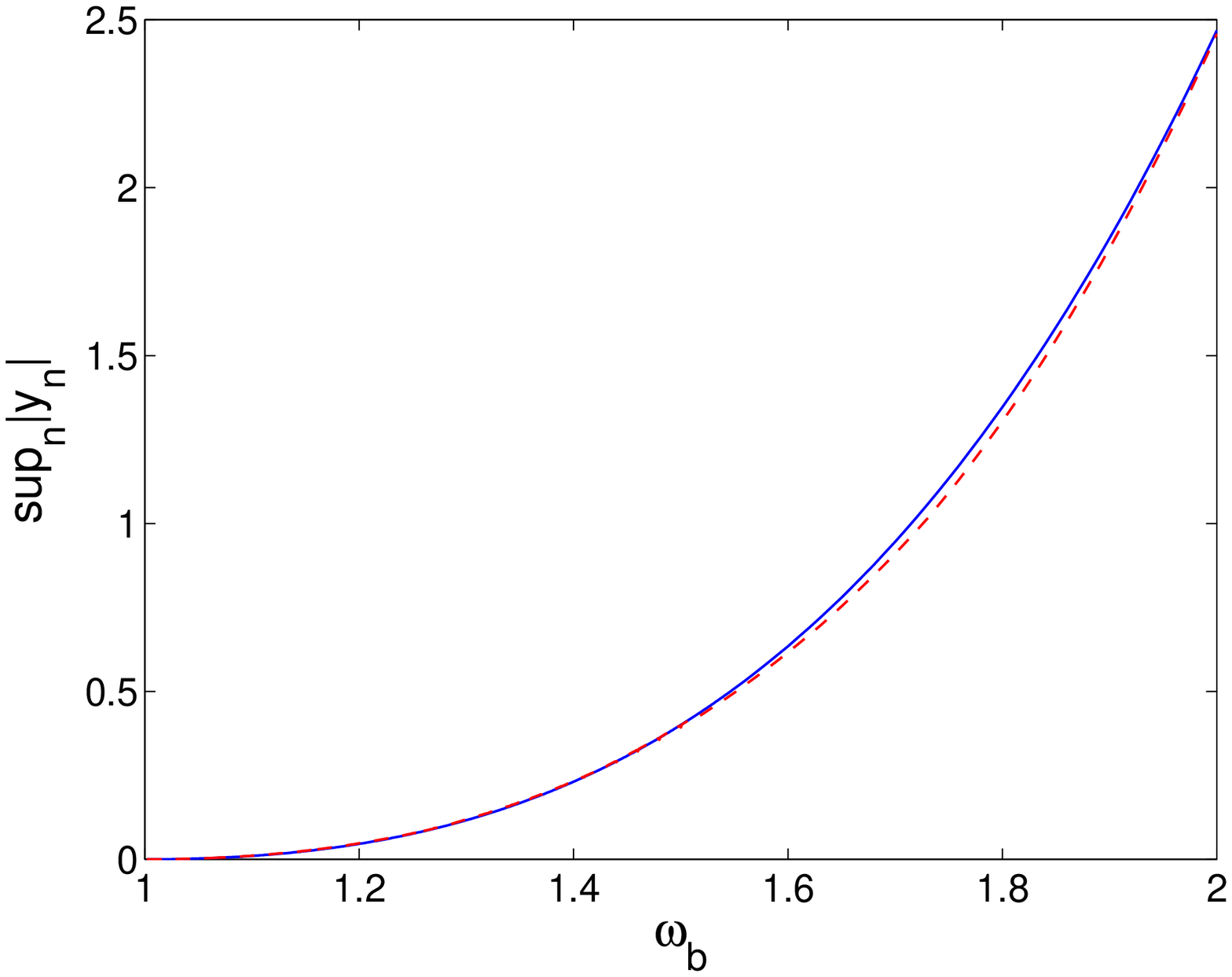}
\includegraphics[scale=0.35]{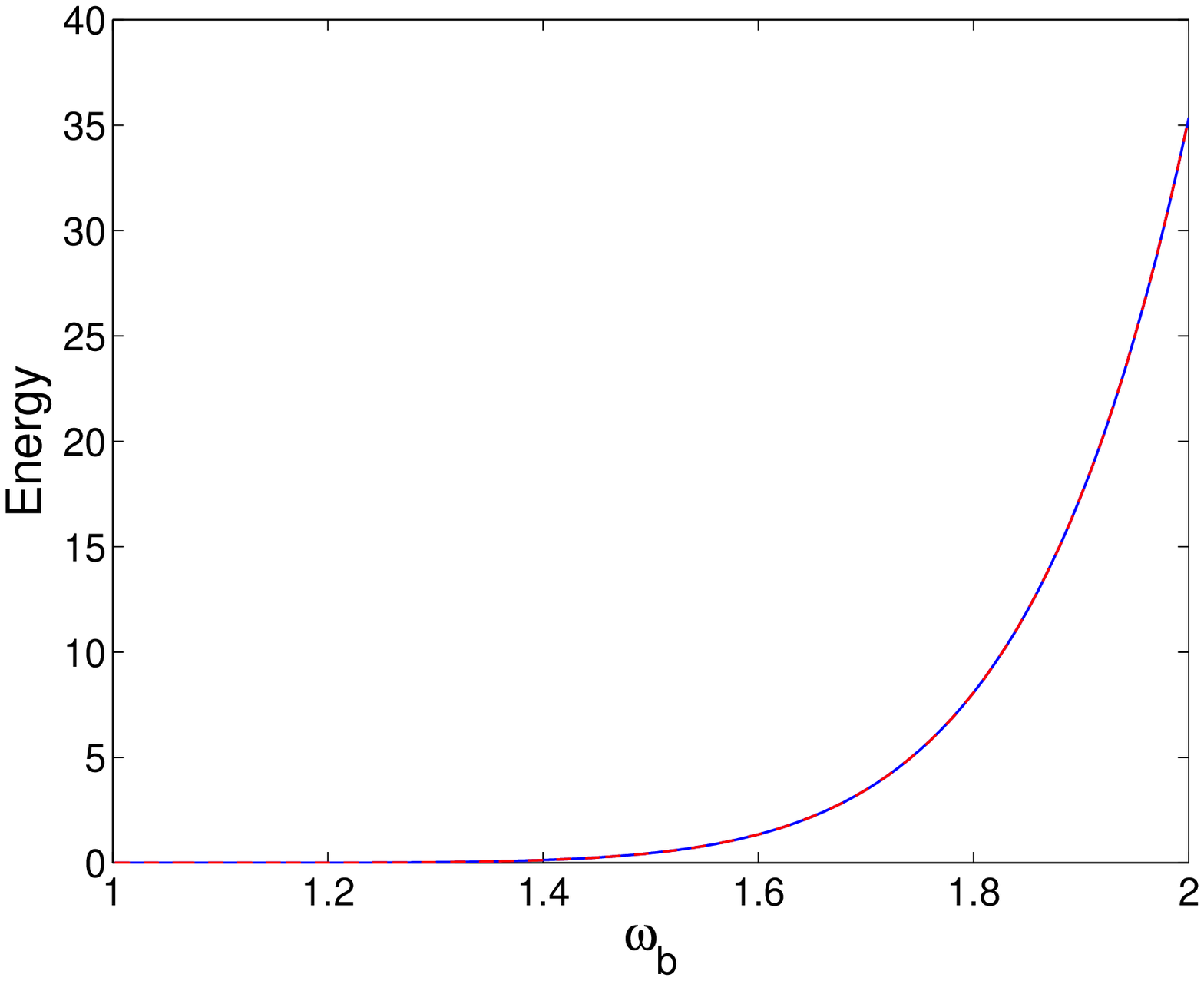}
\end{center}
\caption{\label{breathersnewton}
Maximal amplitude at $t=0$ (left plot) and energy (\ref{hamresc}) (right plot)
of breather solutions of (\ref{eqm}) as a function of their frequency $\omega_b$.
The continuous line corresponds to bond-centered breathers, and
the dashed line to site-centered breathers. Notice that the energy curves are indistinguishable.
}
\end{figure}

More details on the continuation of discrete breathers in $\omega_b$ are
shown in figure \ref{breathersnewton}, which compares
the maximal amplitude and the energy of the bond-centered and site-centered
breather solutions of (\ref{eqm}) when $\omega_b$ is varied
(the continuation is performed for $\omega_b \in (1,2]$).
Both solutions bifurcate from $y_n=0$ when $\omega_b \rightarrow 1^+$,
and their amplitude and energy increases with $\omega_b$.
While minor differences between the breather
amplitudes are visible, one observes that 
the respective energy curves are indistinguishable.

More generally, considering system (\ref{hamresc}) with
the local anharmonic potential (\ref{anpot}) and
choosing
$s \in [-1,1]$, we obtain branches of site-centered and bond-centered breathers
bifurcating from the origin when $\omega_b \rightarrow 1^+$ (results not shown).
The persistence of both types of symmetries is due to the evenness of $W$.

\vspace{1ex}

In what follows we study in more detail the energy barrier separating site-centered and bond-centered
breathers.
As illustrated below in section \ref{stabmob}, this allows us to approximate
the so-called Peierls-Nabarro energy barrier, which corresponds to the
amount of energy required to put a stable static breather into motion
under a momentum perturbation.

A notion of energy barrier separating discrete breathers
is usually defined as follows (cf. also~\cite{msep}).
From (\ref{un1})-(\ref{ansatzapproxst2}), one can deduce a family of
{\em approximate} static breather solutions of (\ref{hamresc})-(\ref{anpot})
\begin{equation}
\label{approxb}
y_n(t)= 2 \epsilon\,
[g(n+\frac{1}{2}-Q)+g(n-\frac{1}{2}-Q)]
(-1)^{n} \, \cos{(\omega_{\rm{b}} t )},
\end{equation}
where $\omega_{\rm{b}} =1+ \frac{\epsilon^{1/2}}{2\tau_0}$ and
$Q \in \mathbb{R}$
(the cases $Q=0$ and $Q=1/2$ corresponding respectively to site-centered and bond-centered breathers).
According to the works of \cite{msep,kastner,sepulchre},
approximate traveling breather solutions of (\ref{hamresc})-(\ref{anpot})
can be obtained from (\ref{approxb}). Their dynamics is described by an
effective Hamiltonian, whose critical points correspond to
site-centered and bond-centered breathers having the same area
$$
\mathcal{A}=\int_{0}^{2\pi /\omega_b}{\sum_{n}{\dot{y}_n^2}\, dt}.
$$
The absolute energy difference $\tilde{E}_{PN}$ between the two
breather solutions provides an approximation of the Peierls-Nabarro barrier. 
However, because the latter appears to be very small in system (\ref{hamresc})-(\ref{anpot})
(a phenomenon that will be illustrated in section \ref{stabmob}),
a very precise computation of $\tilde{E}_{PN}$ would be necessary. This 
yields additional numerical difficulties, due to the fact that
the two breather frequencies have to be retrieved from a given area $\mathcal{A}$.
Due to these difficulties, we shall use a more straightforward approach and
define (following ref. \cite{dauxois}) 
the approximate Peierls-Nabarro barrier ${E}_{PN}$ as the absolute energy difference 
between site-centered and bond-centered breathers having the same frequency $\omega_b$.

\begin{figure}[h]
\begin{center}
\includegraphics[scale=0.35]{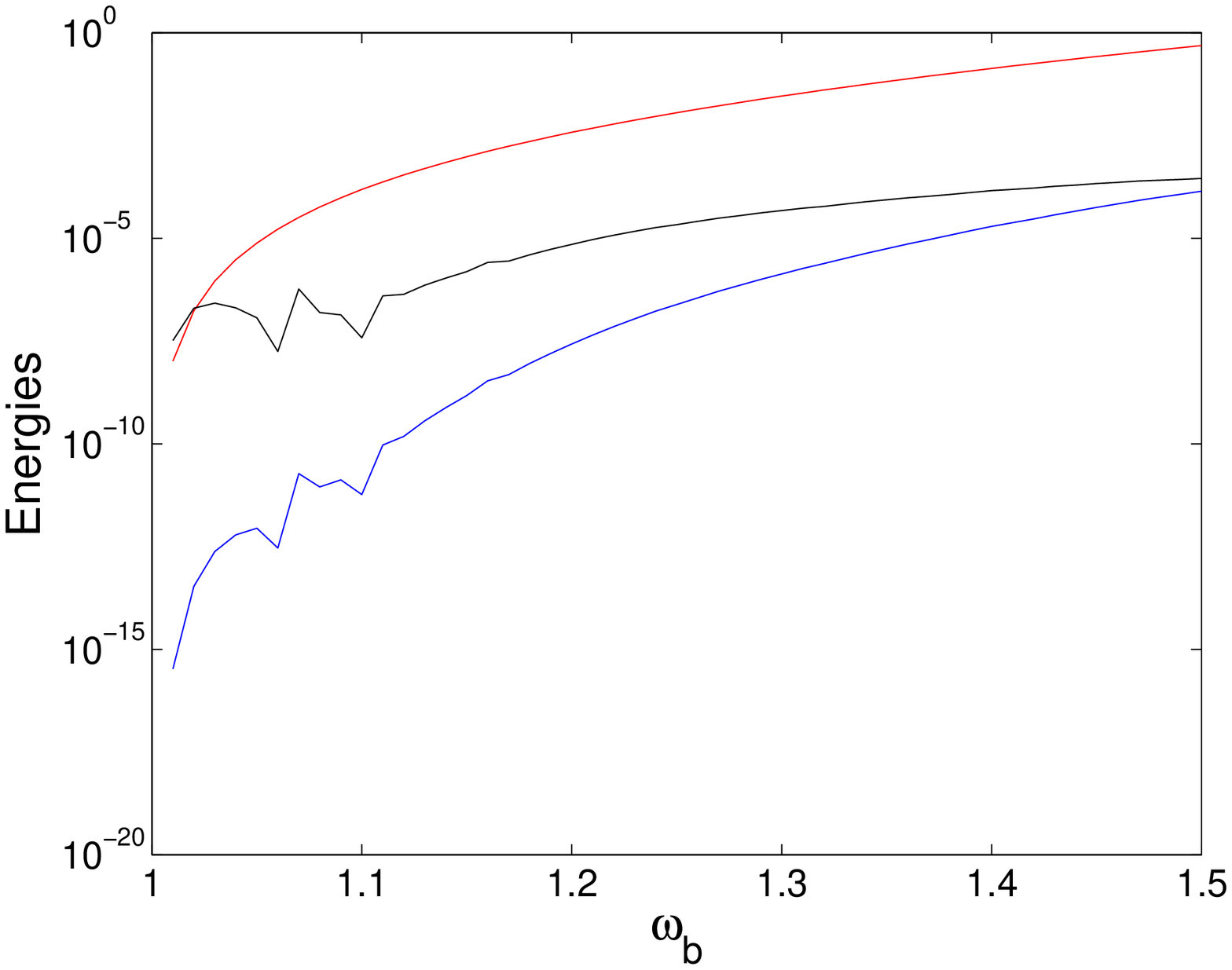}
\includegraphics[scale=0.35]{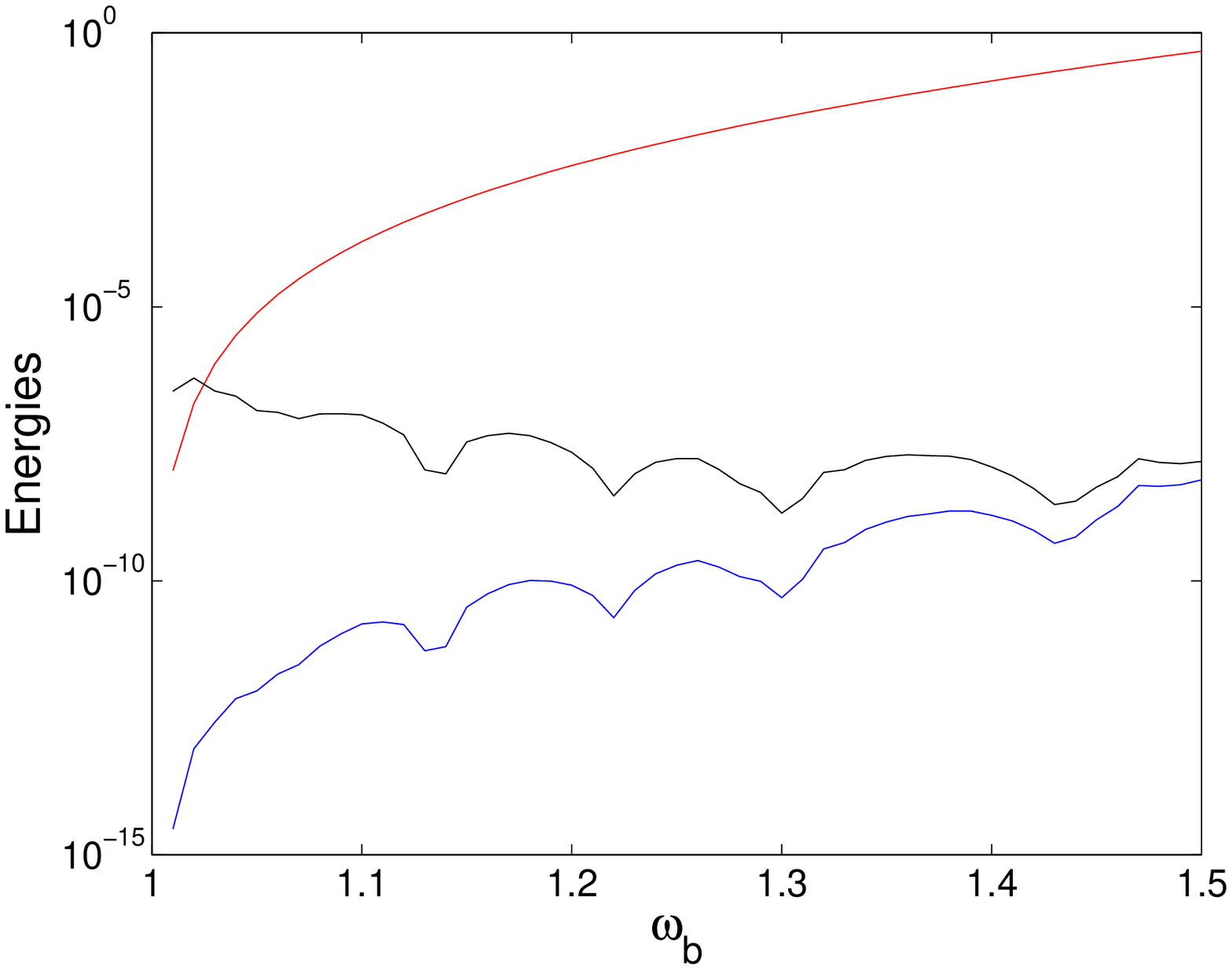}
\includegraphics[scale=0.35]{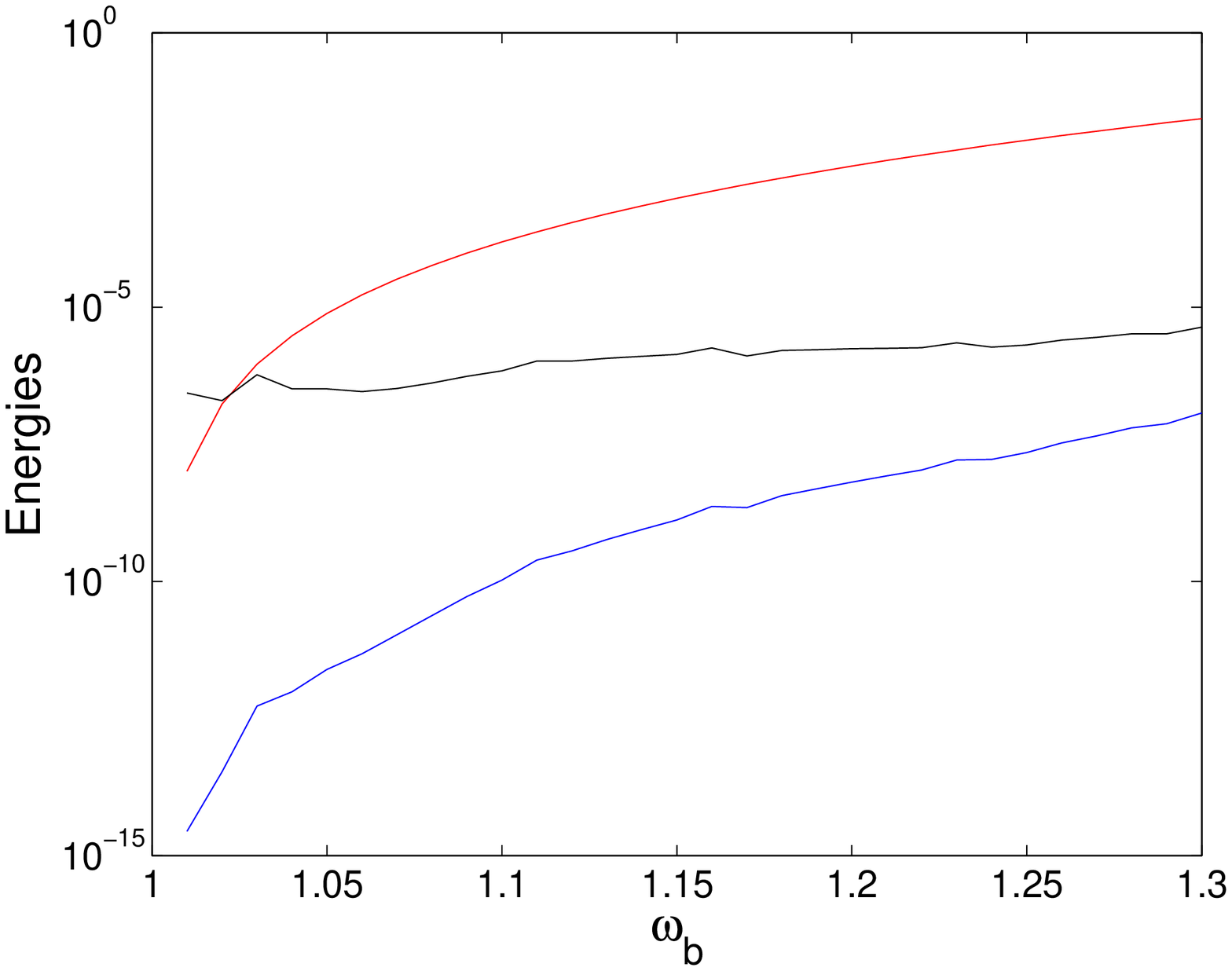}
\end{center}
\caption{\label{pnb}
Approximate Peierls-Nabarro barrier computed as a function of breather frequency,
for different degrees of anharmonicity of the on-site potential
(top left plot : $s=-1/6$, top right plot : $s=0$, bottom plot : $s=1$).
The red curves give the energy $E_{bc}$ of bond-centered breathers
defined by (\ref{hamresc}).
The blue curves correspond to the approximate Peierls-Nabarro barriers $E_{PN}$ (see text),
and the black curve to the relative energy ratio $E_{PN}/E_{bc}$ between
the energy barrier and the bond-centered breather energy.}
\end{figure}

We obtain extremely small values of ${E}_{PN}$
both for harmonic and anharmonic on-site potentials, even quite far from the small amplitude regime.
This result is illustrated by figure \ref{pnb} for $s=-1/6$, $s=0$ and $s=1$.
For small amplitude breathers ($\omega_b \approx 1.01$ in our computations),
the different values of $s$ yield comparable values of ${E}_{PN}$,
of the order of $10^{-14}-10^{-15}$. We find that ${E}_{PN}$ increases with the
breather amplitude but remains very small in our parameter range
(e.g. ${E}_{PN}$ is close to $10^{-4}$ for $\omega_b =1.5$ and $s=-1/6$).
The harmonic case yields even much smaller barriers
(by $3-4$ orders of magnitude for $\omega_b=1.3$).
As shown by figure \ref{pnb}, the smaller relative energy difference between 
site-centered and bond-centered breathers is also achieved in the harmonic case.
These results indicate that extremely small perturbations of the
breathers are capable of putting them into motion
(even more critically for harmonic on-site potentials), 
a phenomenon that will be illustrated in the next section.

\subsection{\label{stabmob}Breather stability and mobility}

In this section we examine the stability properties
of spatially antisymmetric and symmetric breather solutions of (\ref{dps1}) and (\ref{hamresc}),
and link these properties with the existence of traveling breather
solutions.
The linear (spectral) stability of breather solutions of (\ref{dps1})
is investigated by means of the perturbation~\cite{pgk}:
\begin{eqnarray}
v_n(t)=\exp(i {\mu} t) \left[u_n + \left(a_n \exp(\lambda t)
+ b_n^{\star} \exp(\lambda^{\star} t) \right) \right]
\label{dsp2}
\end{eqnarray}
where $u_n$ is a spatially symmetric or antisymmetric solution
of (\ref{dpsstat}) homoclinic to $0$.
The resulting linear problem for the eigenvalue $\lambda$ and
the eigenvector $(a_n, b_n)^T$ (where $^T$ denotes transpose) is
solved by standard numerical linear algebra solvers and the
results are depicted by means of the spectral plane
$(\lambda_r, \lambda_i)$ of the eigenvalues $\lambda= \lambda_r
+ i \lambda_i$.

\begin{figure}[h]
\begin{center}
\includegraphics[scale=0.4]{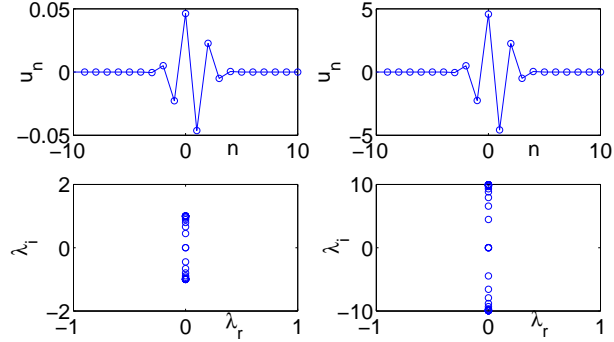}
\end{center}
\caption{\label{dps_fig2} The profiles (top panels) and the linear
stability (bottom panels) of the spatially antisymmetric solution
of the DpS equation are shown for the values of $\mu=1$ (left panels)
and $\mu=10$ (right panels). This inter-site solution is linearly stable.}
\end{figure}

From the bottom panels of Fig. \ref{dps_fig2},
we can infer that spatially antisymmetric solutions are spectrally stable
and therefore should be structurally robust
(a result confirmed by our direct numerical simulations --data
not shown here--). This is due to the absence of eigenvalues of
non-vanishing real part in this Hamiltonian system (in which
whenever $\lambda$ is an eigenvalue, so are $\lambda^{\star}$,
$-\lambda$ and $-\lambda^{\star}$).

\begin{figure}[h]
\begin{center}
\includegraphics[scale=0.4]{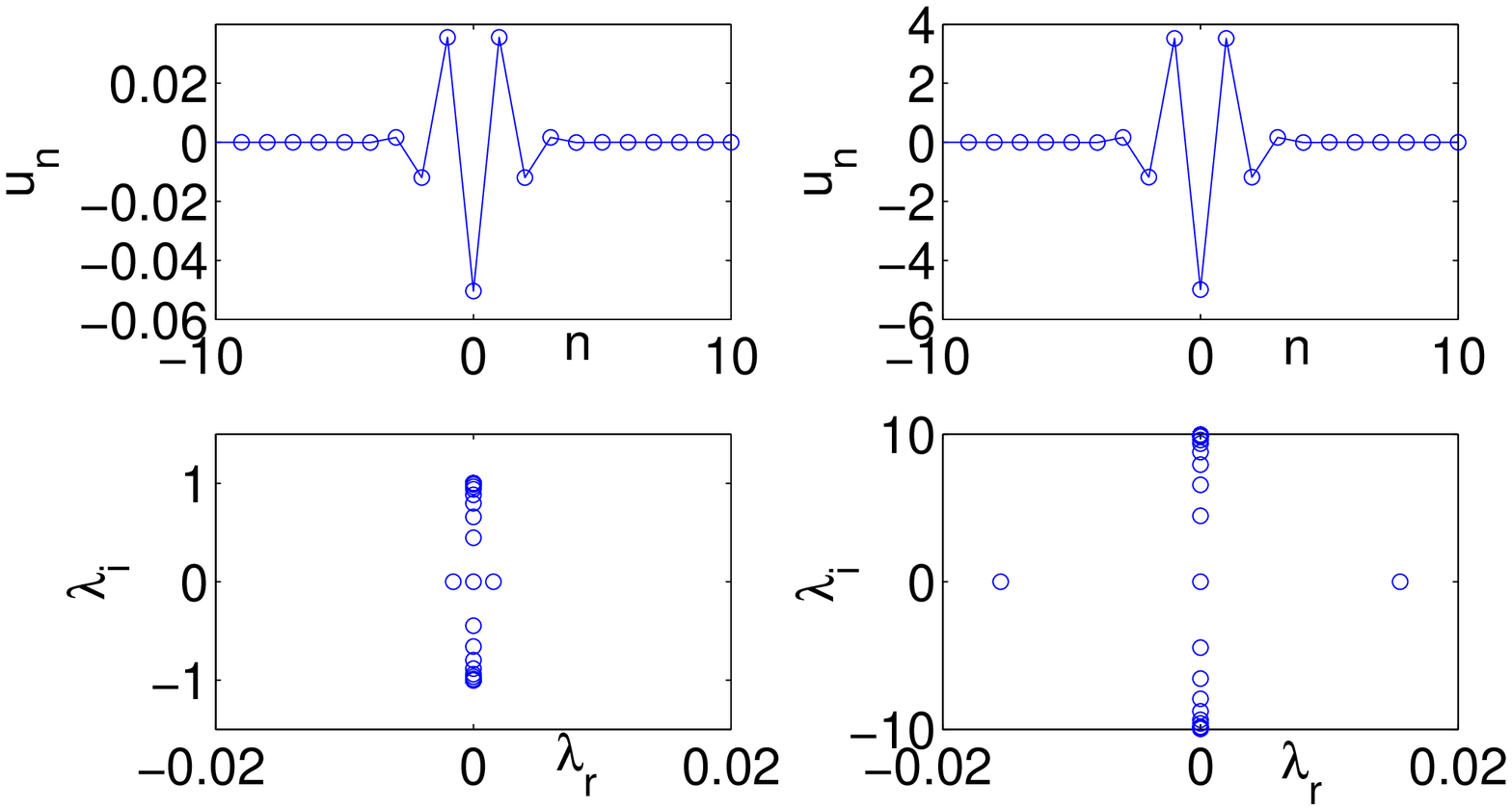}
\includegraphics[scale=0.4]{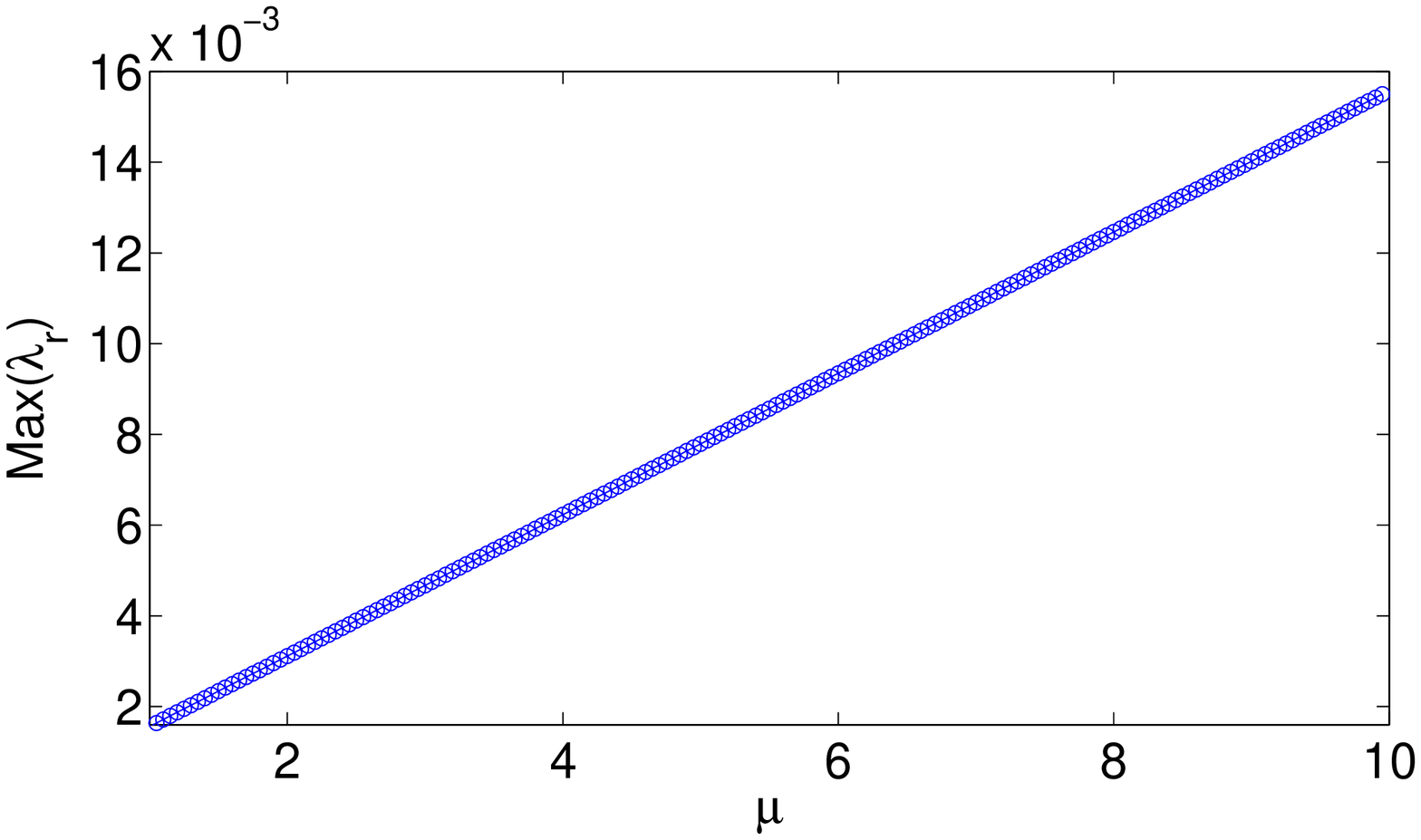}
\end{center}
\caption{\label{dps_fig1} The top panel is directly analogous
to the results of Fig. \ref{dps_fig2}, but for the case of the
site-centered solution. The presence of a real eigenvalue pair of
linearly growing magnitude as $\mu$ increases can be observed in the spectral
plane and is more clearly highlighted in the figure of the bottom
panel.}
\end{figure}

\begin{figure}[h]
\begin{center}
\includegraphics[scale=0.35]{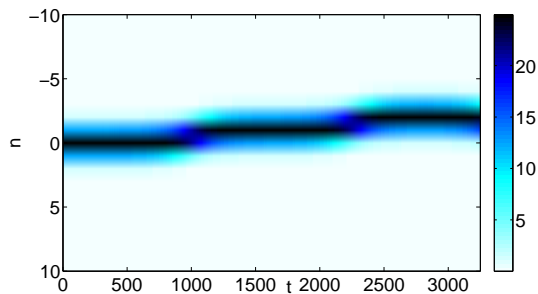}
\end{center}
\caption{\label{dps_fig3} The figure shows the space-time
contour plot of the square modulus of the field $|v_n(t)|^2$
for equation (\ref{dps1}).
The initial condition is a site-centered localized mode
(from the right panel of Fig. \ref{dps_fig1}), perturbed
by a uniformly distributed random perturbation of amplitude
$0.01$. The perturbation leads to the manifestation of the
instability of the site-centered mode which, in turn,
leads to its mobility.}
\end{figure}

On the other hand, the stability and associated dynamical properties
are more interesting in the case of the site-centered solution of
Fig. \ref{dps_fig1}. In this case, we can observe the presence
of a real eigenvalue pair. As can be seen in the bottom panel
of Fig. \ref{dps_fig1}, the real part of the relevant eigenvalue
pair (which corresponds to the instability growth rate) grows
linearly with the eigenvalue parameter $\mu$, inducing a progressively
stronger instability for larger amplitude solutions.
The dynamical manifestation of this instability is
illustrated in Figure \ref{dps_fig3}. Here we perturb
the dynamically unstable solution of the right panel
of Fig. \ref{dps_fig1} by a uniformly distributed
random perturbation (of amplitude $0.01$). The projection
of this random field on the unstable eigenvector of the
site centered mode excites the manifestation of the
dynamical instability of this mode which is, in turn,
illustrated in the space-time evolution (where the
colorbar corresponds to the field $|v_n(t)|^2$) of Fig. \ref{dps_fig3}.
Clearly, the instability of the site-centered mode is associated
with a ``translational'' eigenmode of the linearization problem,
whose excitation induces the motion of the localized mode.

In the above analysis, the breather stability properties remain qualitatively
unchanged for all values of $\mu$. This follows from
the scale invariance of (\ref{dps1}) pointed out in section \ref{exist},
which also implies the linear dependence of the eigenvalues $\lambda$ on $\mu$.
However, we note in passing that this 
simplification is obviously not valid for the 
model (\ref{eqm}).

\vspace{1ex}

Having determined the spectral stability of bond-centered and site-centered
breather 
solutions in the DpS equation, we now consider the same problem for
their analogues in the original lattice (\ref{hamresc}), including in our
analysis the effect of a possible addition of a 
local anharmonic potential (\ref{anpot}).

We have computed the Floquet spectrum of (\ref{hamresc})-(\ref{anpot})
linearized at the bond-centered breather and
the site-centered breather, for different values of the breather frequency $\omega_b  \in (1,2 ]$
and the anharmonicity parameter $s \in [-1,1]$. The Floquet spectrum
includes a quadruplet of eigenvalues close to $+1$ and eigenvalues on the unit circle
accumulating near $e^{\pm i 2\pi/\omega_b}$.
The spectral properties of these discrete breathers differ from usual ones \cite{marinstab} for several reasons.
Firstly, no bands of continuous spectrum are present on the unit circle for the infinite chain.
This is due to the fact that system (\ref{eqm}) linearized at $y_n =0$
(the limit of a breather solution at infinity) consists of an infinite chain of uncoupled identical linear oscillators,
and thus the phonon band reduces to a single frequency, equal to unity in the present case.
Secondly, another nonstandard property originates from the quadruplet of eigenvalues close to $+1$.
Due to the Hamiltonian character of (\ref{eqm}), $+1$ is always at least a double eigenvalue of the Floquet matrix.
In addition, we always find an extra pair of eigenvalues in the immediate
vicinity of $+1$ corresponding to a
pinning mode (see below). This contrasts with the case of Klein-Gordon lattices, where this situation
is a codimension-one phenomenon, occuring
near critical values of the coupling constant and for particular classes of on-site potentials
\cite{aubryC,chen,aubryrev}.

\begin{figure}[h]
\begin{center}
\includegraphics[scale=0.4]{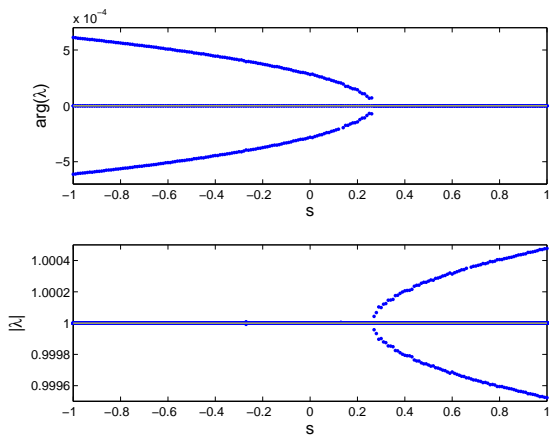}
\end{center}
\caption{\label{floqcradleis}
Arguments (upper plot) and moduli (lower plot) of the quadruplet of
Floquet eigenvalues $\lambda$ close to $+1$, corresponding to
system (\ref{eqm})-(\ref{anpot})
linearized at the bond-centered breather. Computations are performed
for $\omega_b = 1.1$, and eigenvalues are plotted
as a function of the anharmonicity parameter $s \in [-1,1]$.
}
\end{figure}

\begin{figure}[h]
\begin{center}
\includegraphics[scale=0.4]{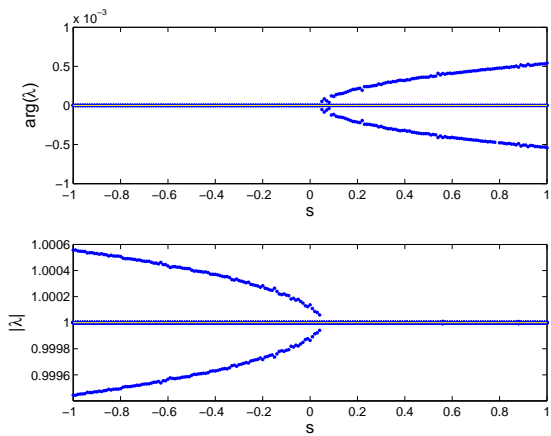}
\end{center}
\caption{\label{floqcradleos}
Same plot as in figure \ref{floqcradleis}, for
the site-centered breather with $\omega_b = 1.1$.}
\end{figure}

In what follows we describe the evolution of the quadruplet of eigenvalues close to $+1$
for $\omega_b = 1.1$ and $s \in [-1,1]$.
The following figures display the moduli and arguments of these eigenvalues
for the bond-centered breather (figure \ref{floqcradleis}) and
the site-centered breather (figure \ref{floqcradleos}).
For the bond-centered breather, a pair of Floquet multipliers $\lambda , \lambda^{-1}$ emerges 
from the unit circle after
a collision at $+1$, for $s > s_0^{b} \approx 0.26$. For the site-centered breather, 
a pair of multipliers $\lambda , \lambda^{-1}$
(with $\lambda >1$) exists for $s < s_0^{s}\approx 0.05$, and enters the unit circle 
for $s > s_0^{s}$ after a collision at $+1$.

From the above spectral study, 
one can infer that for harmonic
on-site potentials (i.e. $s=0$) and $\omega=1.1$,
the site-centered breather
is weakly unstable and the bond-centered breather is spectrally stable.
These results agree with the above results
obtained for the DpS equation. This provides a consistent picture, given
that the DpS equation correctly approximates breather profiles
of amplitudes $\epsilon = O((\omega_b - 1)^2)$ for
$\omega_b \approx 1$ (section \ref{exist}).
The DpS admits weakly unstable site-centered and stable bond-centered 
breather solutions,
and approximates the dynamics of (\ref{eqm})
for $O(\epsilon)$ initial data
on times of order $O(\epsilon^{-1/2})$ \cite{dumas}. Hence, we expect
a parallel to the instability of site-centered modes of the DpS dynamics
in Eq.~(\ref{eqm}). 
Note that these instabilities are extremely small
for $\omega_b$ close to $1$, because the instability of the site-centered breather
is already very weak at the renormalized (slow) time-scale of the DpS equation
(see figure \ref{dps_fig1}), and becomes $O(\epsilon^{1/2})$ times weaker
at the level of (\ref{eqm}) for a breather with amplitude $\epsilon$.

The above picture persists for $s\approx 0$, but the site-centered and bond-centered
breathers display a change of stability at the two different critical values
$s=s_0^{b,s}>0$ ($s_0^{s}$ being quite small),
after which their dynamical stability differs from the stability of the DpS 
breathers.
It would be interesting to analyze the bifurcations of new types of
discrete breathers near these critical values of $s$, and this problem
will be considered in a future work.

\begin{figure}[h]
\begin{center}
\includegraphics[scale=0.35]{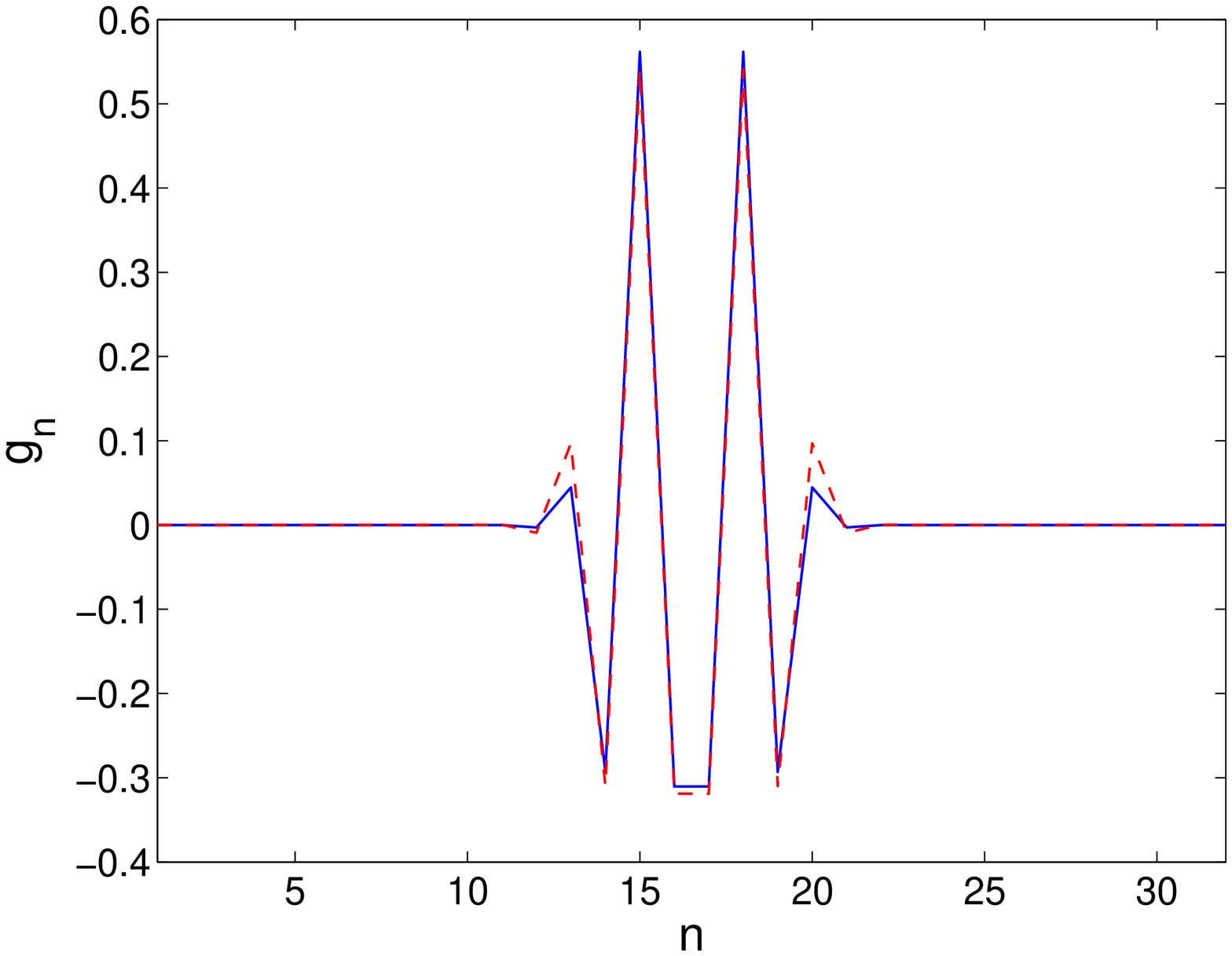}
\includegraphics[scale=0.35]{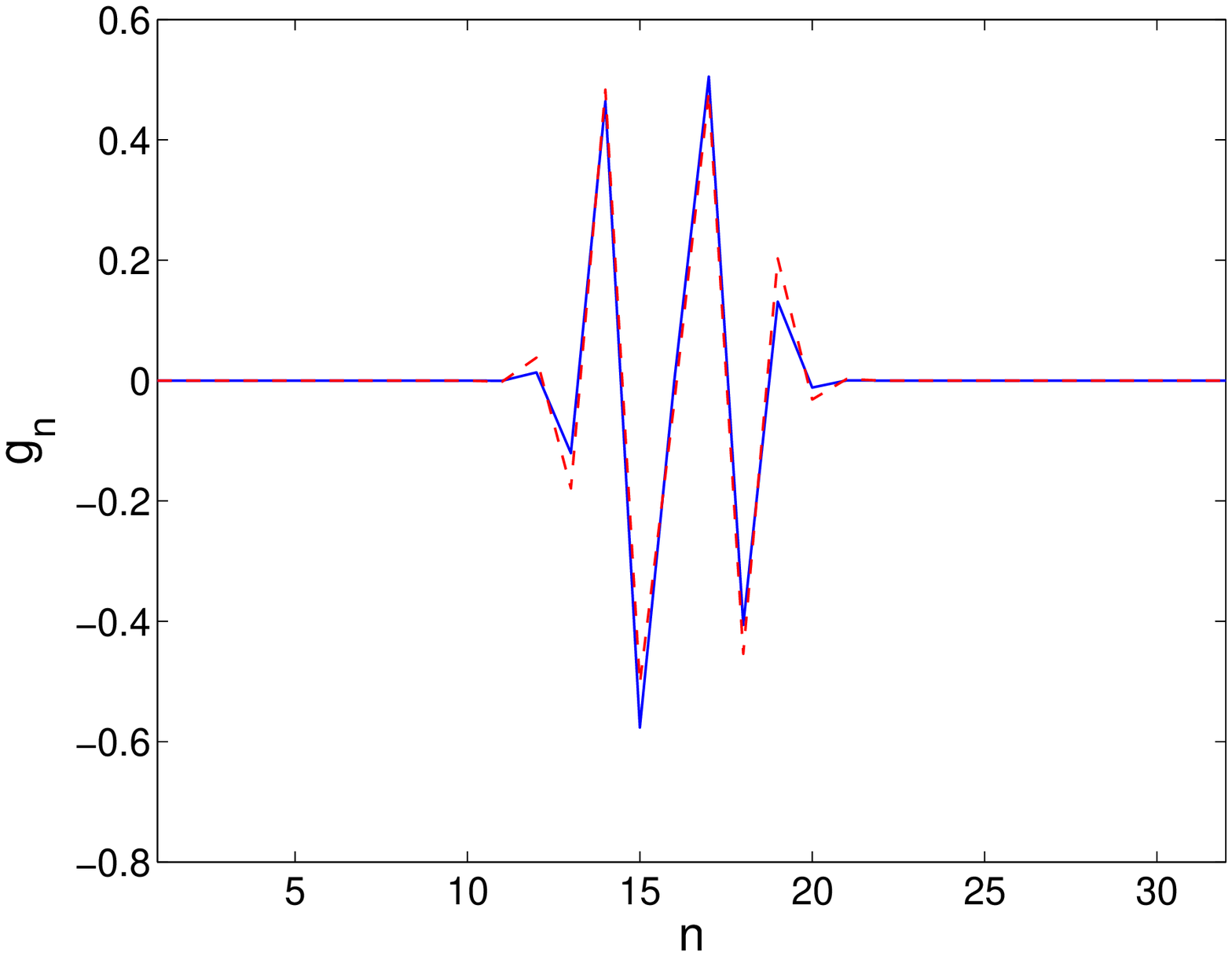}
\end{center}
\caption{\label{pinning} Pinning mode (full line) and discrete gradient (dashed line) corresponding to a bond-centered (left plot) and site-centered (right plot) stationary breather in system (\ref{eqm}),
for the breather frequency $\omega_b=1.1$. The components of the pinning mode correspond
to particle positions at $t=0$ (initial particle velocities vanish).}
\end{figure}

\begin{figure}[h]
\begin{center}
\includegraphics[scale=0.35]{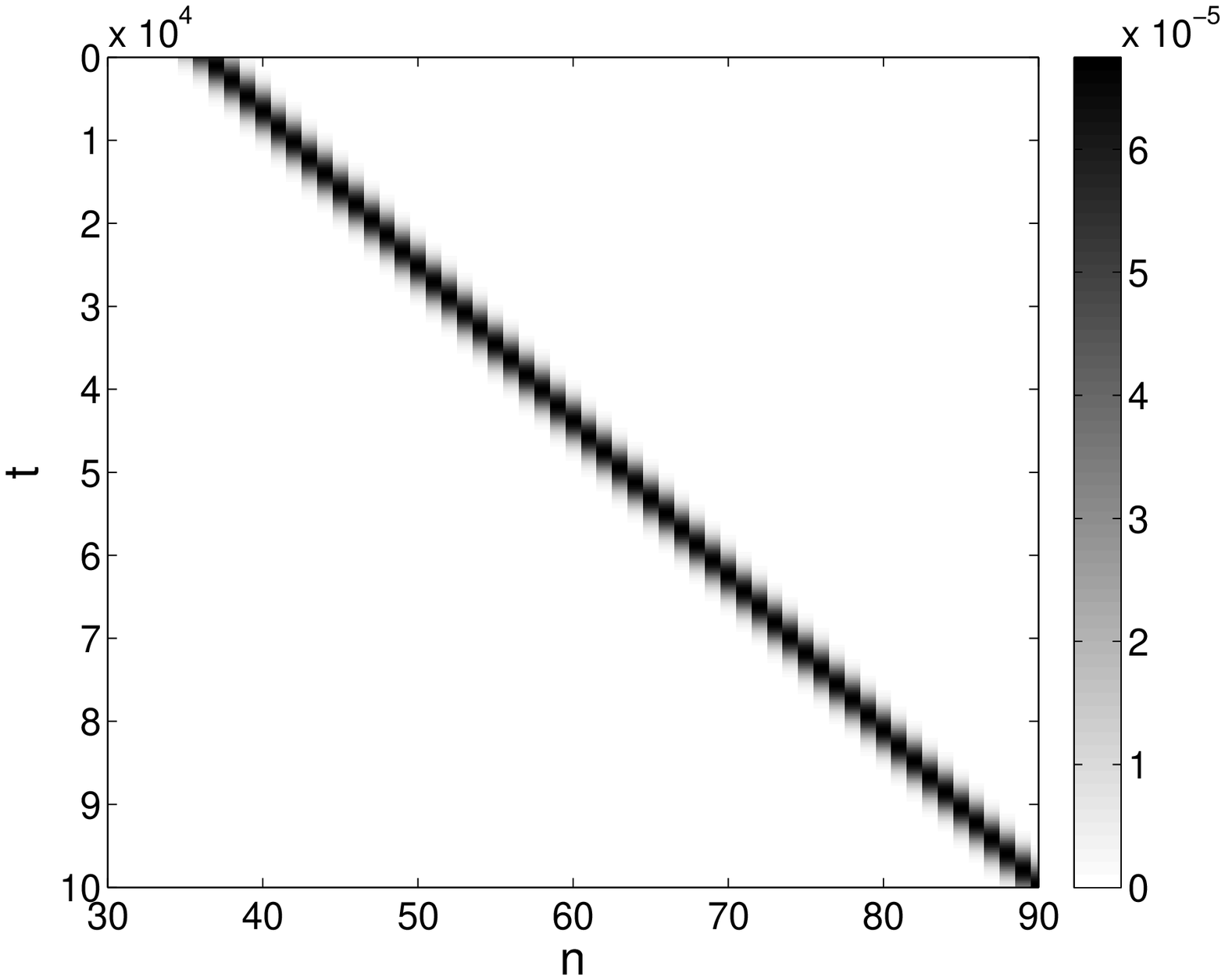}
\includegraphics[scale=0.35]{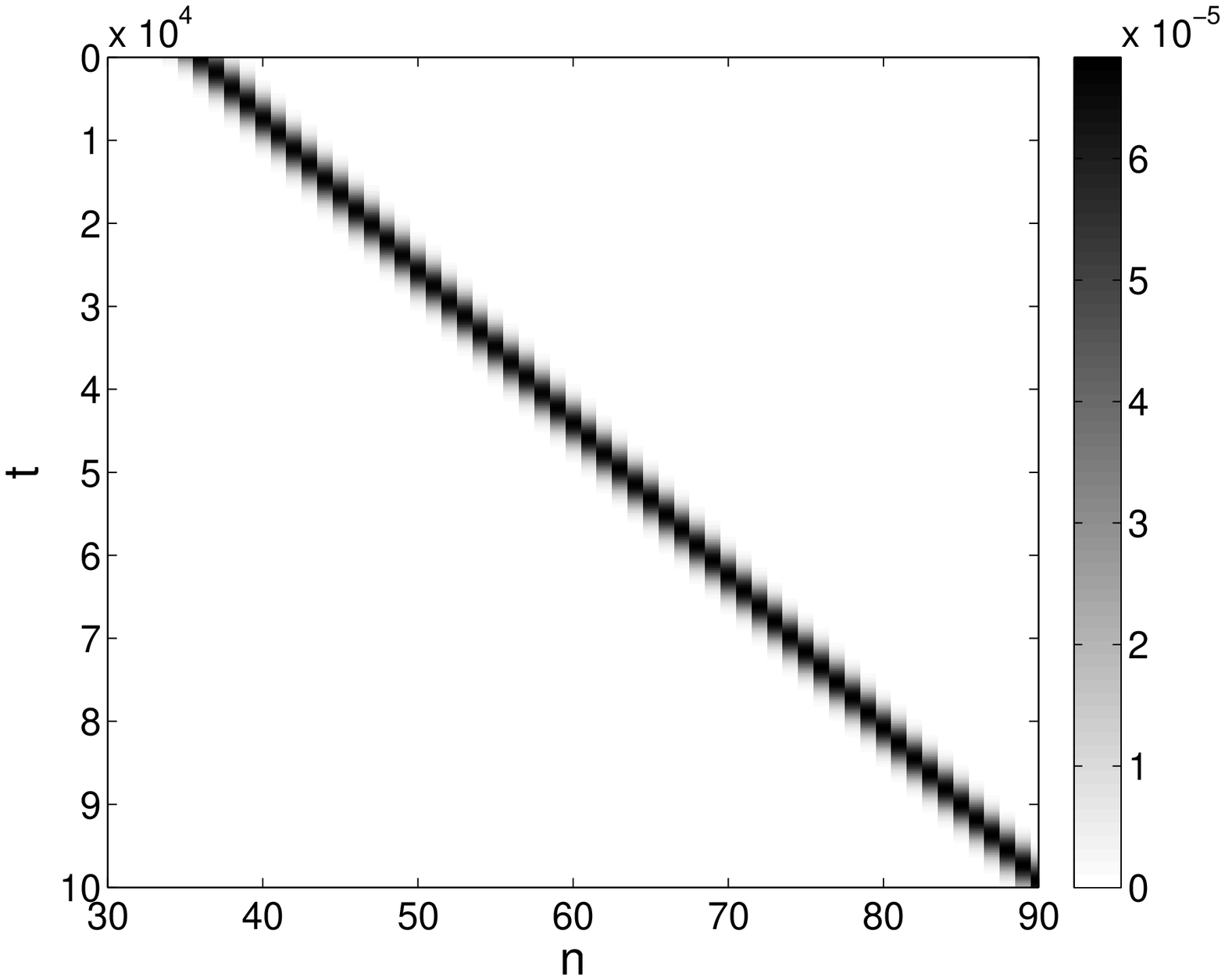}
\end{center}
\caption{\label{moving} Energy density plot of a moving breather in system (\ref{hamresc})-(\ref{lp}), obtained by perturbing 
along the pinning mode
a bond-centered (left plot) and site-centered (right plot) stationary breather with frequency $\omega_b=1.1$. The initial velocity perturbation has a magnitude $c=2 \times10^{-4}$.
The velocities $v$ of resulting traveling breathers are very close, i.e.
$v \approx 5.446.10^{-4}$ for the site-centered case and
$v \approx 5.364.10^{-4}$ for the bond-centered case, resulting in nearly identical figures.}
\end{figure}

In what follows we illustrate the effect of the additional Floquet
eigenvalues close to $+1$ on the breather dynamics,
again considering the case $\omega_b = 1.1$.
Figure \ref{pinning} compares an eigenvector associated with
one of these eigenvalues and the renormalized discrete gradient
$$
g_n=\frac{y_{n+1}(0)-y_{n-1}(0)}{\sum_n|y_{n+1}(0)-y_{n-1}(0)|^2},
$$
which reveals that the two profiles are very close.
The associated mode will thus be referred to as a translation mode or
pinning mode, and
the effect of a perturbation along its direction is to shift the breather 
center~\cite{chen}.
The existence of this mode has the effect of enhancing the breather mobility.
To illustrate this, we perturb at $t=0$ the velocity components of
a stationary breather, adding
the discrete gradient $g_n$ multiplied by a velocity factor $c$.
The kinetic energy imprinted to the lattice is then $c^2/2$.
We consider below the energy density at the $n$-th site, which is defined from (\ref{hamresc}):
\begin{equation}
\label{endens}
e_n=\frac{1}{2}\, \dot{y}_n^2+W( y_n ) + \frac{2}{5}\, \gamma (y_n - y_{n+1})_+^{5/2}.
\end{equation}
Fig. \ref{moving} shows the energy density plot in the system 
of Eqs. (\ref{hamresc})-(\ref{lp}),
for a site-centered and a bond-centered breather perturbed with $c=2
\times 10^{-4}$.
This perturbation results in a translational motion of the breather at an 
almost constant velocity
with negligible dispersion, which illustrates the strong mobility of
discrete breathers in the present model.
These results are consistent with the approximation $E_{PN}$ of the Peierls-Nabarro barrier
computed previously, since we found $E_{PN} \approx 1.77.10^{-11}$ for  
$s=0$ and $\omega_b = 1.1$ (see figure \ref{pnb}).
The above momentum perturbation increases the kinetic energy of the
bond-centered breather by $c^2/2=2 \times10^{-8}$, which is well-above $E_{PN}$.

\begin{figure}[h]
\begin{center}
\includegraphics[scale=0.35]{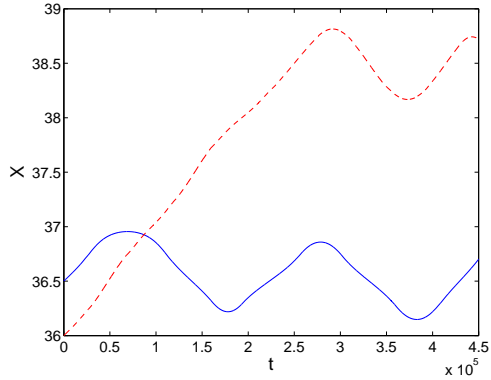}
\end{center}
\caption{\label{Xcomp} Time-evolution of the breather energy center after a momentum perturbation of a 
bond-centered (full line) and site-centered (dashed line) breather. All parameters are the same as
in figure \ref{moving}, except the initial velocity perturbation $c$. We fix $c=3.10^{-6}$, which corresponds to
an increase of kinetic energy below the Peierls-Nabarro barrier.}
\end{figure}

To describe the effect of breather perturbations below the Peierls-Nabarro barrier, it is convenient to 
consider the breather energy center
\begin{equation}
\label{encenter}
X=\frac{\sum_{n=n'-m}^{n'+m} ne_n}{\sum_{n=n'-m}^{n'+m} e_n}
\end{equation}
with $n'$ being the location of the maximum energy density of the breather and $m>0$ an integer which accounts for the width of the breather (we have fixed $m=5$). Figure \ref{Xcomp} displays $X(t)$
for $c=3.10^{-6}$, i.e. $c^2/2=4.5 \times10^{-12}$ lying below $E_{PN}$. In that case,
only the unstable site-centered breather is able to move along the lattice (it is able to jump $2$ sites but gets pinned subsequently).
For the stable bond-centered breather, a transition from pinning to mobility is obtained for $c>c_{c} \approx 6.19.10^{-6}$. 
The value of the Peierls-Nabarro barrier resulting from dynamical simulations is thus
$c_c^2/2 \approx 1.92.10^{-11}$, which is quite close to the approximation $E_{PN}$ computed previously.

\subsection{\label{impactpb}Study of an impact problem}

Having demonstrated the mobility of breather modes in the DpS equation,
in direct analogy with the dynamics of the full oscillator
model, we attempt the excitation of the first site of a
Newton's cradle and the associated
DpS chain, and observe the ensuing space-time evolution.

\vspace{1ex}

\begin{figure}[h]
\begin{center}
\includegraphics[scale=0.35]{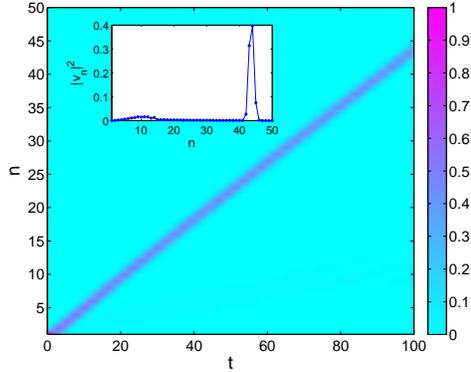}
\end{center}
\caption{\label{dps_fig4} The space-time evolution of the square modulus of
the field, similarly to Fig. \ref{dps_fig3}, under the DpS equation is shown
for an initial
excitation of the domain boundary site with $A_n(0)=-i\, \delta_{n,1}$
($\delta_{i,j}$ denotes the usual Kronecker symbol).
The corresponding spatial profile is
depicted in the inset for the final
shown simulation time.
Notice the robust propagating localized mode (traveling breather), as well
as the presence of a weak residual (fairly extended) excitation near the
boundary.}
\end{figure}

Consider the equation (\ref{dps}) on a semi-infinite lattice with $n\geq 1$ and
a free end boundary condition at $n=1$.
We numerically compute the solution of (\ref{dps}) with the 
initial condition
\begin{equation}
\label{ci}
A_1 (0)= -i, A_n(0)=0 \mbox{ for }n \geq 2.
\end{equation}
It can be clearly seen in Fig. \ref{dps_fig4} that the
result is the formation
of a localized excitation which is traveling robustly through
the chain. This is the traveling breather resulting from
the mobility of the discrete breathers that we considered
before. In addition to this strongly localized nonlinear
excitation, we can observe a weak residual excitation at the
end of the chain (this is somewhat reminiscent of the
phenomenology described in \cite{hinch}). Strictly
speaking, we cannot consider this mode to be a surface mode of
the chain \cite{molina},
as our observations indicate that its profile
is fairly extended and non-stationary (or periodic). 

\vspace{1ex}

\begin{figure}[h]
\begin{center}
\includegraphics[scale=0.4]{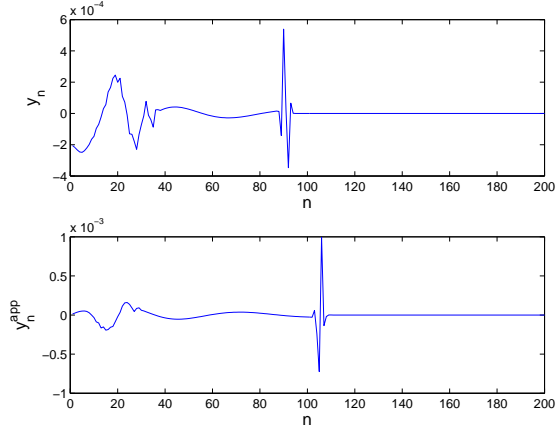}
\end{center}
\caption{\label{scomp}
Comparison between the solution of (\ref{eqm})-(\ref{cic}) (upper plot) and
its approximation given by (\ref{dps})-(\ref{approx})-(\ref{ci}) (lower plot),
at a fixed time $t \approx 8090$ corresponding to $\tau = 250$.
Computations are performed for $\epsilon=0.9548 .10^{-3}$.
For this value of $\epsilon$, the static breathers given by (\ref{ansatzapproxst2})
have frequency $1.01$, and the DpS equation yields very good approximations
of the static breathers.
}
\end{figure}

\begin{figure}[h]
\begin{center}
\includegraphics[scale=0.35]{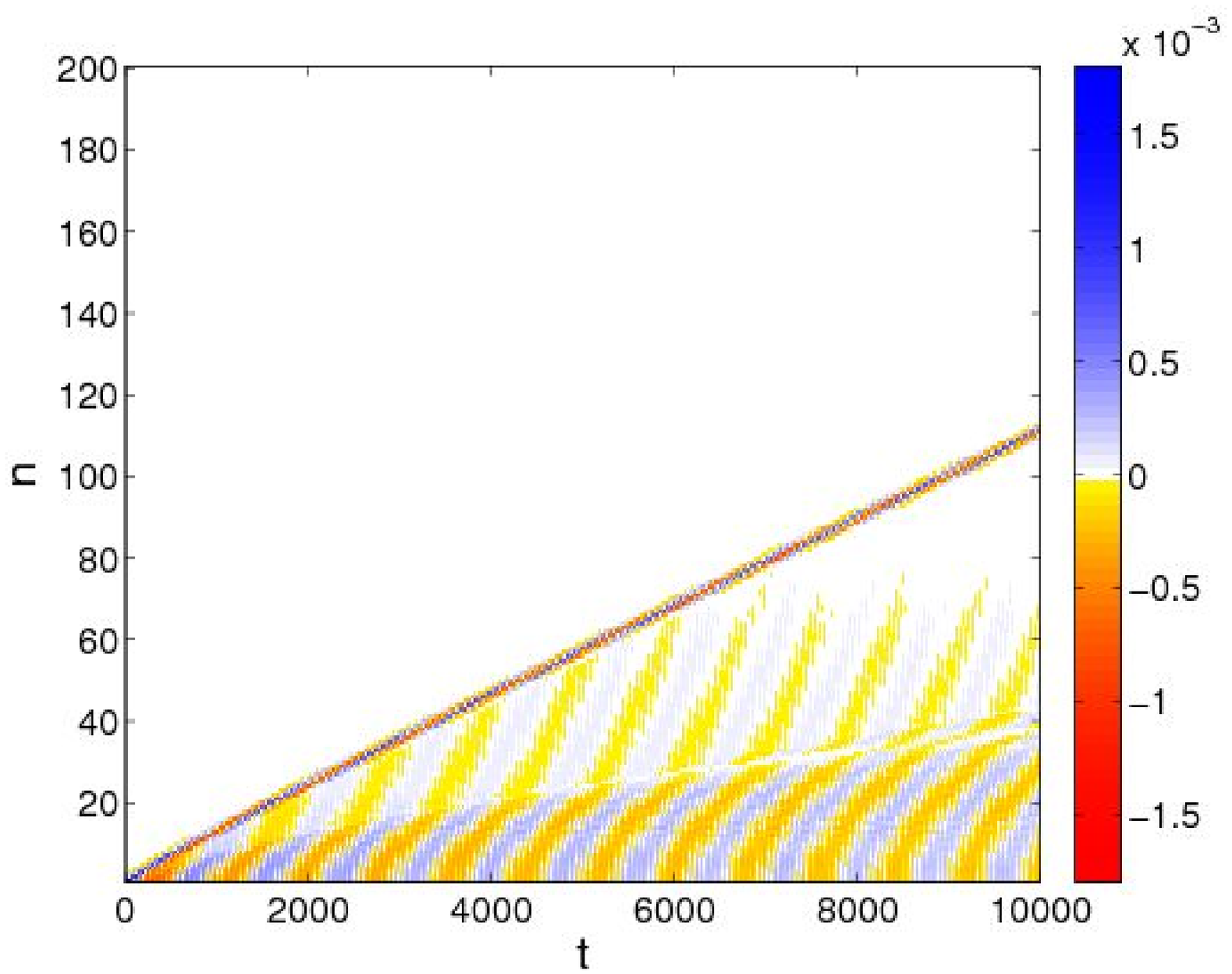}
\includegraphics[scale=0.35]{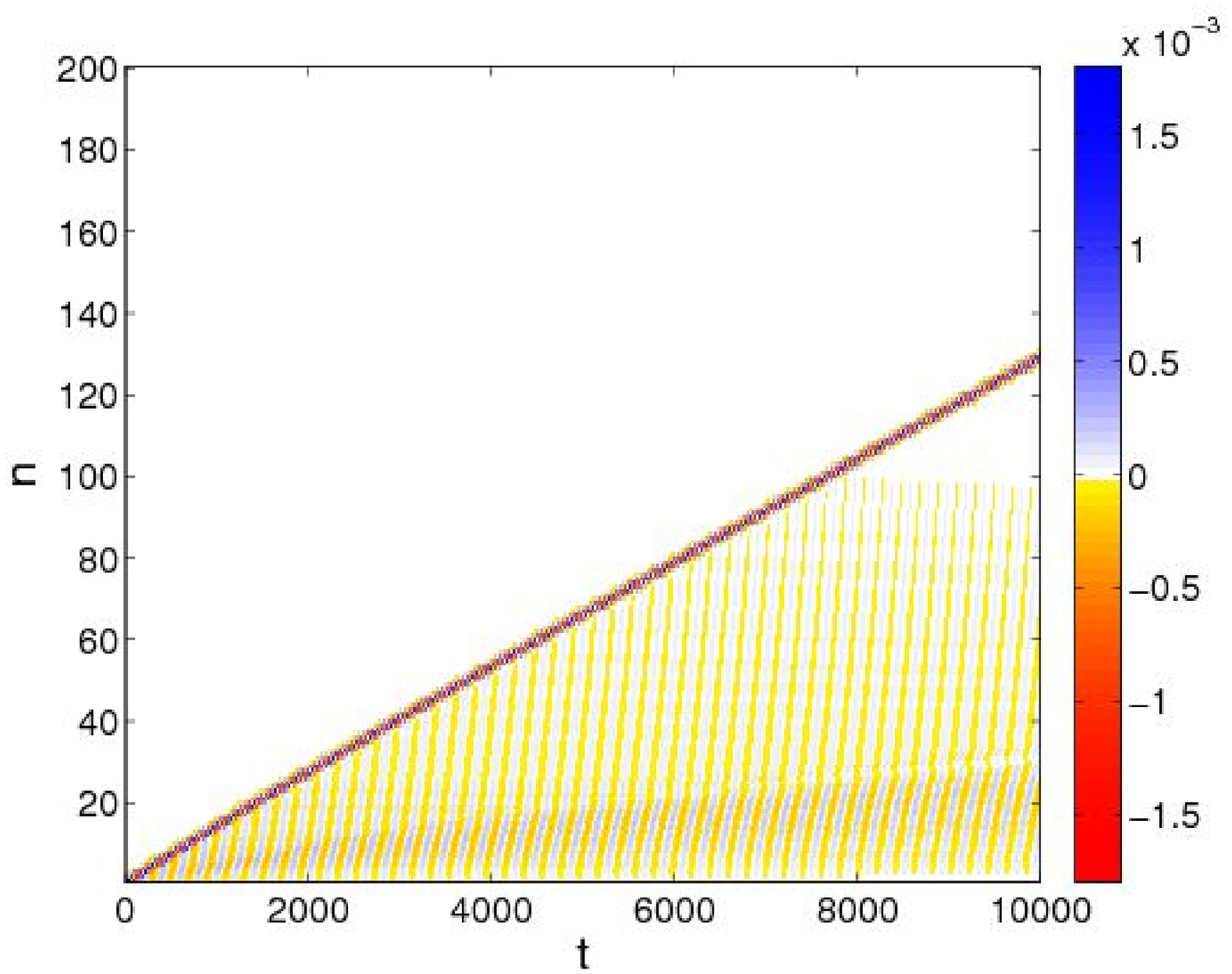}
\end{center}
\caption{\label{scomp2}
Energy density plot showing the comparison between the solution of (\ref{eqm})-(\ref{cic}) (left plot) and
its approximation given by (\ref{dps})-(\ref{approx})-(\ref{ci}) (right plot), for $\epsilon=0.9548 \times 10^{-3}$.
Time $t$ in the bottom plot is related to the slow time $\tau$ through $t=\epsilon^{-1/2}\tau$.
}
\end{figure}

For all $\epsilon >0$ small enough,
the above solution of DpS corresponds to an approximate solution of (\ref{eqm})
given by (\ref{approx}),
satisfying $y_n^{\rm{app}} (0)=0$, $\dot{y}_1^{\rm{app}} (0)=2\, \epsilon + O(\epsilon^{3/2})$,
$\dot{y}_2^{\rm{app}} (0)=O(\epsilon^{3/2})$ and
$\dot{y}_n^{\rm{app}} (0)=0$ for $n \geq 3$.
Figures \ref{scomp} and \ref{scomp2}
compare the corresponding approximate solution (\ref{approx})
and the solution of (\ref{eqm})
with initial condition
\begin{equation}
\label{cic}
y_n(0)=0, \ \ \ \dot{y}_1(0)=2\epsilon, \ \ \ \dot{y}_n(0)=0 \mbox{ for } n \geq 2
\end{equation}
for a small value of $\epsilon$.
One can see that the DpS equation and the full oscillator model give rise to 
similar dynamics,
i.e. the initial impulse splits into a traveling breather and an 
extended wavetrain emitted from
the boundary. The amplitude of the latter is reasonably well reproduced 
by the DpS approximation
over a long transient, while the traveling breather 
amplitude and velocity are overestimated
(e.g., the breather amplitude is approximately $13 \times 10^{-4}$ with
the DpS approximation and $9 \times 10^{-4}$ for the full lattice).
In addition, the traveling breather velocity resulting from the 
DpS approximation is slightly larger than in the full oscillator model.
The same kind of waves are visible up to moderate initial velocities.
The traveling breather
remains highly localized (mainly supported by $7$ lattice sites), and is 
followed by a small oscillatory tail
reminiscent of the periodic traveling waves analyzed in \cite{jamesc}.

\vspace{1ex}

In what follows we analyze the effect of considering the local anharmonic potential (\ref{anpot}).
Due to the smoothness of $W$, the DpS equation remains unchanged with respect to
the harmonic case, as observed in section \ref{dpsrev}. Consequently,
the dynamics of (\ref{eqm})
after the impact is expected to remain unchanged for small excitations, on the
time scales given in section \ref{dpsrev}. However, it is interesting to examine possible additional effects of anharmonicity
occuring on longer time scales or for large amplitude excitations. For example, a
trapping of large amplitude traveling breathers can occur in Klein-Gordon lattices \cite{dauxois,msep}, due to the
Peierls-Nabarro energy barrier separating site-centered
and bond-centered breathers.

\begin{figure}[h]
\begin{center}
\includegraphics[scale=0.35]{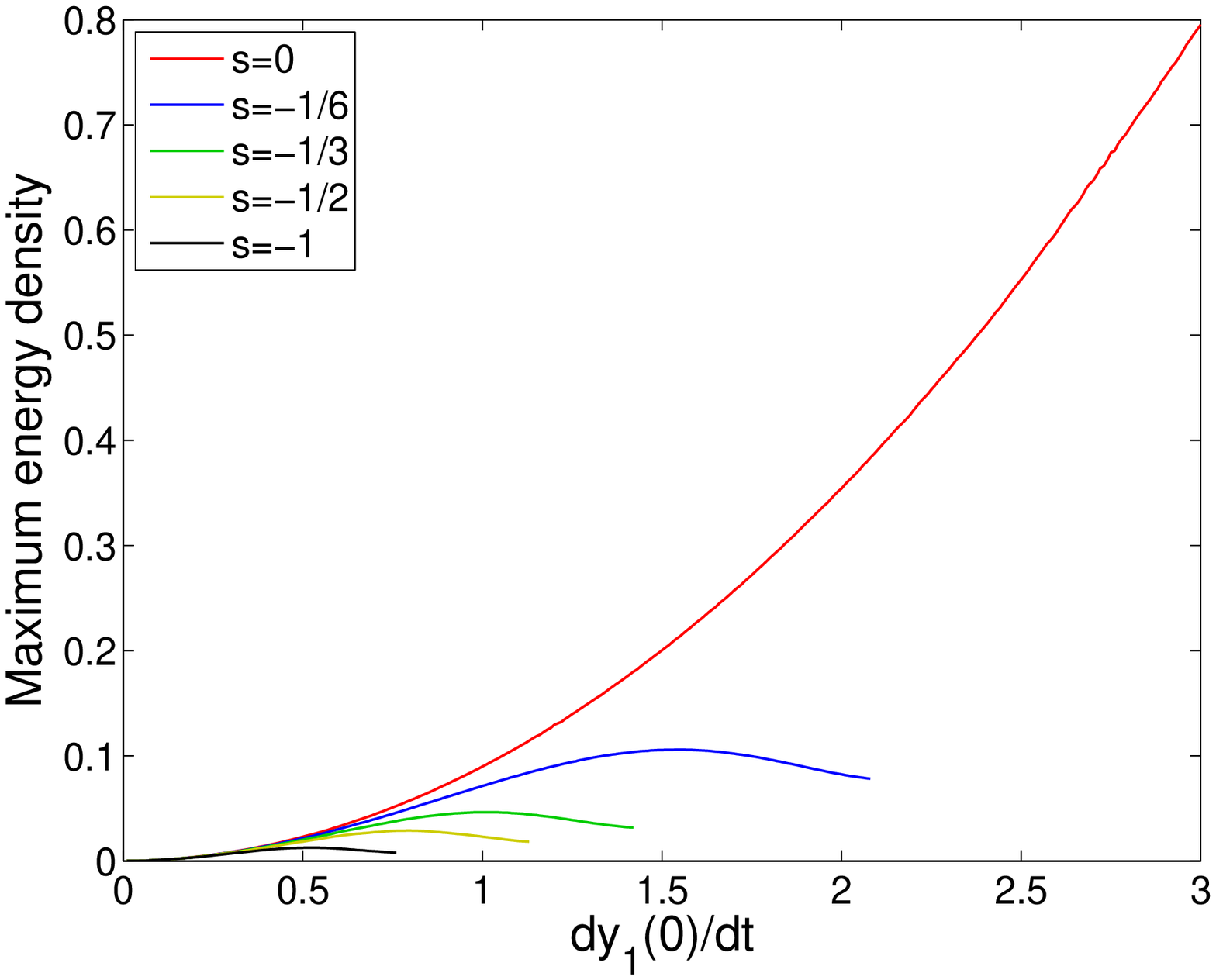}
\includegraphics[scale=0.35]{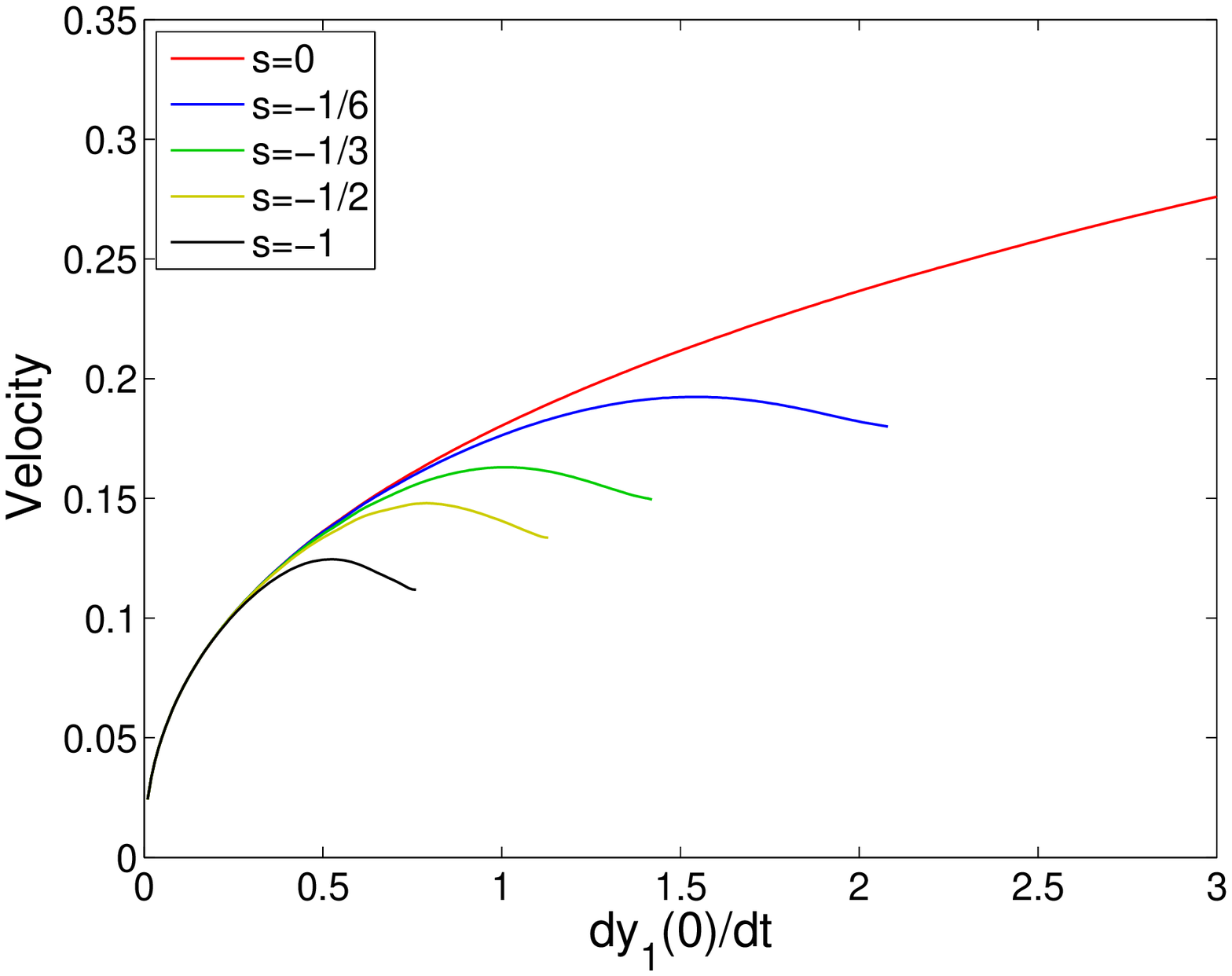}
\end{center}
\caption{\label{ampvel}
Maximum energy density (left plot) and velocity (right plot) of the traveling breather generated in system
(\ref{hamresc})-(\ref{anpot}) with initial condition (\ref{cic}), for several values of $\dot{y}_1(0)$
and anharmonicity parameter $s \leq 0$.}
\end{figure}

\begin{figure}[h]
\begin{center}
\includegraphics[scale=0.35]{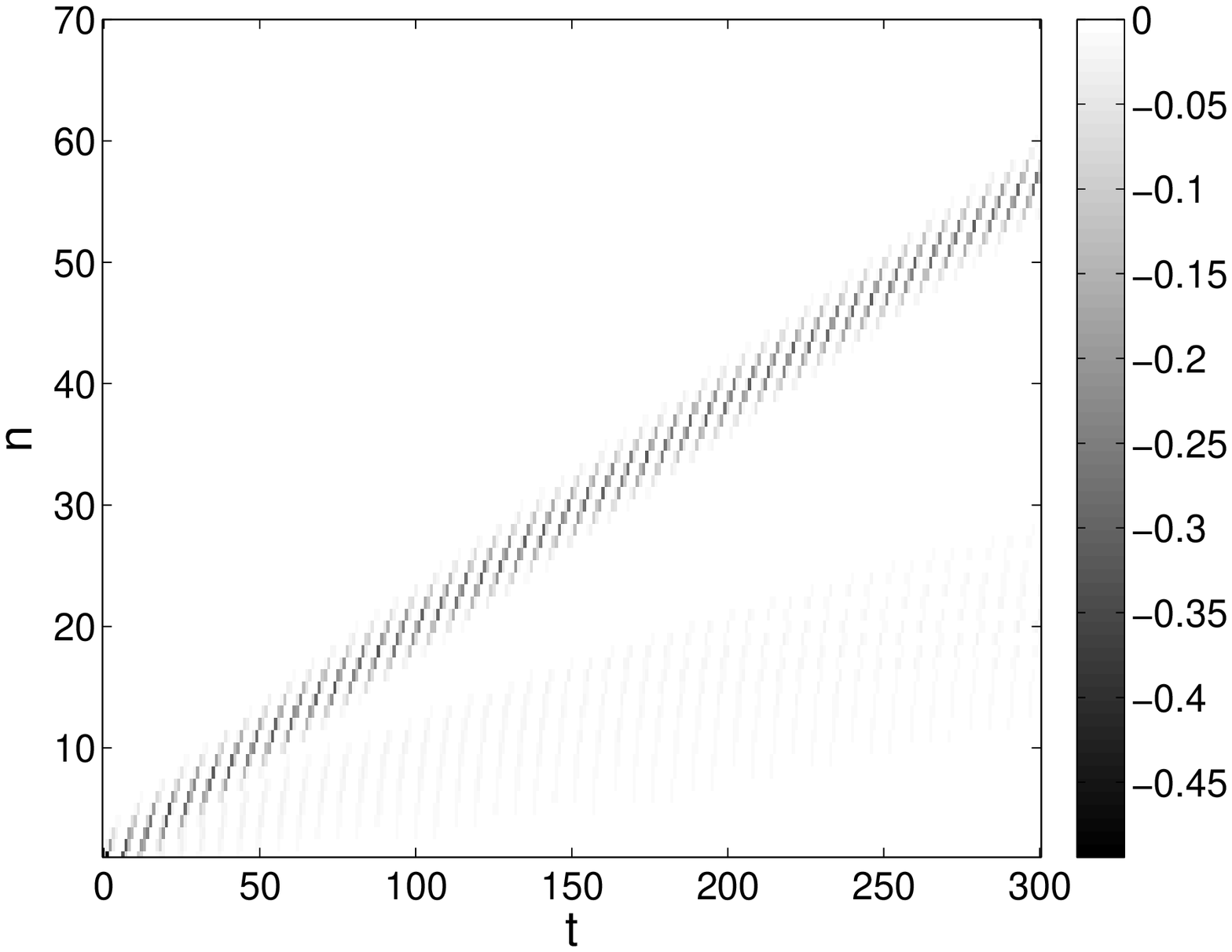}
\includegraphics[scale=0.35]{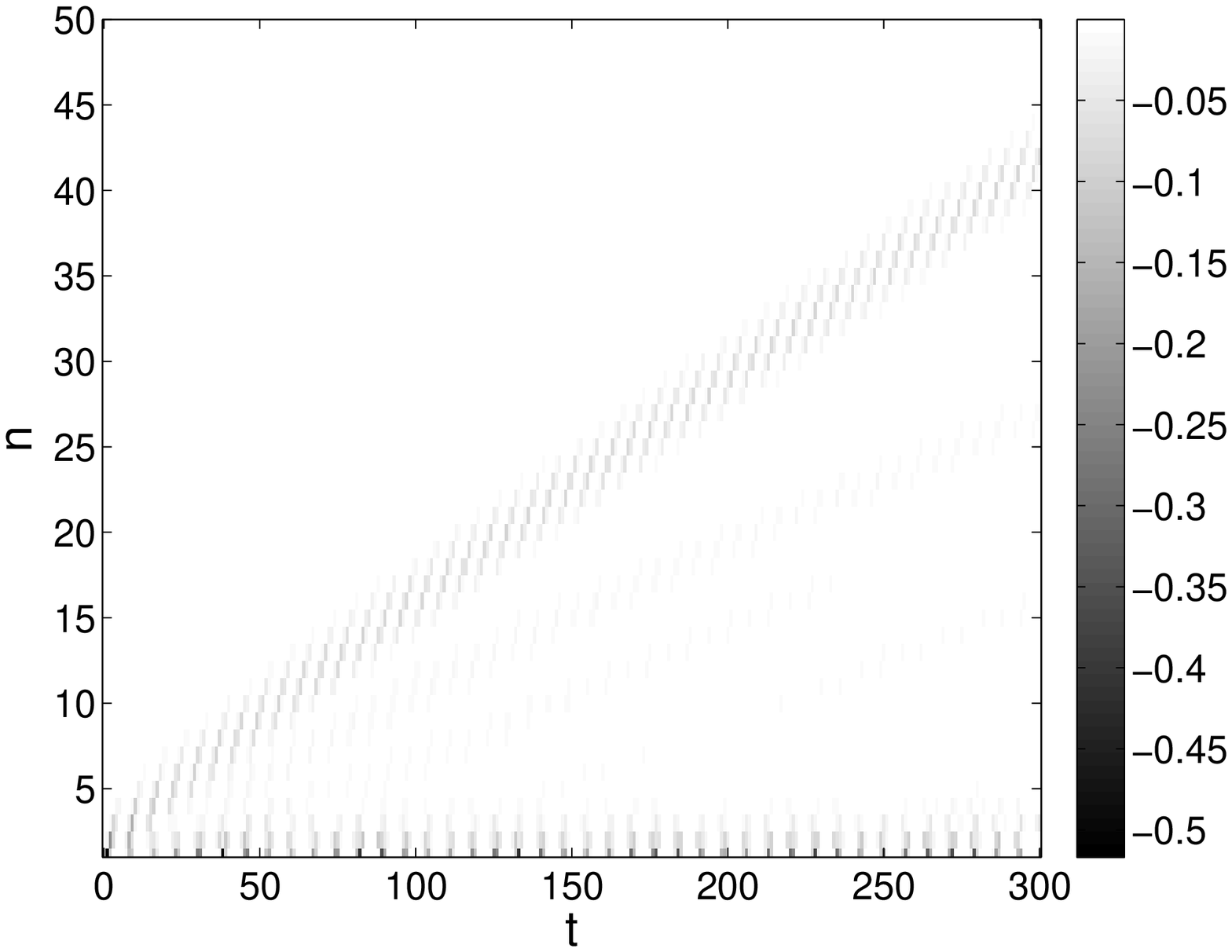}
\end{center}
\caption{\label{surfmode}
Space-time diagrams showing the interaction forces
$f_n = -(y_n - y_{n+1})_+^{3/2}$ in system (\ref{hamresc})-(\ref{anpot})
for the initial condition (\ref{cic}) with $\dot{y}_1(0)= 0.94$.
Forces are represented
in grey levels, white corresponding to vanishing interactions (i.e. beads not in contact)
and black to a minimal negative value of the contact force.
Several values of the anharmonicity parameter are considered~: $s=0$ (left plot),
$s=-0.7$ (right plot).}
\end{figure}

\begin{figure}[h]
\psfrag{n}[0.9]{ $n$}
\psfrag{x}[1][Bl]{ $\dot{y}_n(t) $}
\begin{center}
\includegraphics[scale=0.4]{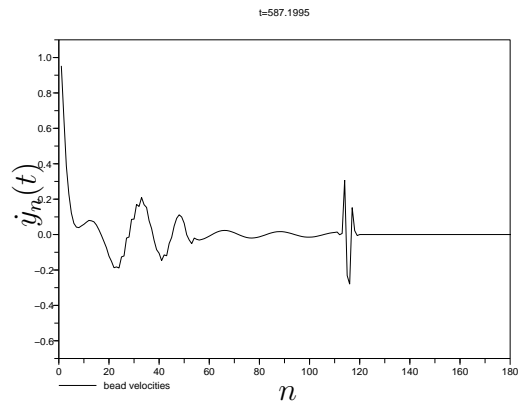}
\end{center}
\caption{\label{smodeandtb}
Snapshot of particle velocities
in system (\ref{hamresc})-(\ref{anpot}) with anharmonicity parameter $s=-1/6$.
The profile is plotted at $t\approx 587$,
for the initial condition (\ref{cic}) with $\dot{y}_1(0) \approx 1.87$.
The excitations of a surface mode
and a traveling breather are clearly visible.
}
\end{figure}

In order to characterize the breather motion we consider the traveling breather energy center $X(t)$ defined by (\ref{encenter}).
The average velocity of the traveling breather is computed as the slope of the linear least squares approximation of the function $X(t)$,
taking only into account the points for which the traveling breather is sufficiently far from the boundary in order to
eliminate boundary effects.
Figure \ref{ampvel} displays the traveling breather velocity and
maximum energy density (computed from (\ref{endens}))
as a function of the initial velocity $\dot{y}_1(0)$, for different values of the parameter $s \leq 0$.
As expected, the different graphs are very close at small initial velocity where the DpS
equation drives the dynamics, but discrepancies appear at larger velocities.
The graphs of figure \ref{ampvel} corresponding to $s<0$ are interrupted above
some critical velocities, because the solution blows-up in finite time when the
initial velocity exceeds some threshold. Below this value, the
anharmonicity of the on-site potential with $s<0$ decreases the breather velocity.
The energy of the traveling breather
(including its kinetic energy) becomes much smaller because a part of the initial energy
remains trapped in the form of a surface mode located near $n=1$.
This phenomenon
is illustrated in figure \ref{surfmode}, which
compares the traveling breather propagation in two chains with $s=0$
(left plot) and $s=-0.7$ (right plot), for
$\dot{y}_1(0)= 0.94$. In the left panel, a highly-localized
traveling breather is clearly visible, while the right panel shows
a surface mode and a lower-energy traveling breather with smaller velocity.
Typical profiles of the surface mode and the traveling breather are displayed in figure
\ref{smodeandtb}.
The possibility of exciting a surface mode by an impact was already pointed out
in reference \cite{dcer}, for a mixed Klein-Gordon - FPU chain with
a sinusoidal local potential, and a Morse interaction potential instead of
the fully-nonlinear Hertzian interactions.

Note that the above-mentioned blow-up phenomenon is due to potential (\ref{anpot})
with $s<0$
and does not occur
for $W(y)=1-\cos{y}$, which corresponds e.g. to the gravitational potential
acting on the usual Newton's cradle.
In the latter case, the dynamics resulting from the impact becomes rather
similar to the phenomena studied in \cite{dcer}. For sufficiently large impact velocities
the traveling breather is replaced by a kink reminiscent of Nesterenko's soliton \cite{neste2},
resulting in the ejection of a finite number of particles at the end of the chain
(result not shown). 
It would be interesting to analyze how the transition between traveling breather
and kink excitations occurs in this system, but this problem lies beyond the scope of the present paper.

\begin{figure}[h]
\begin{center}
\includegraphics[scale=0.4]{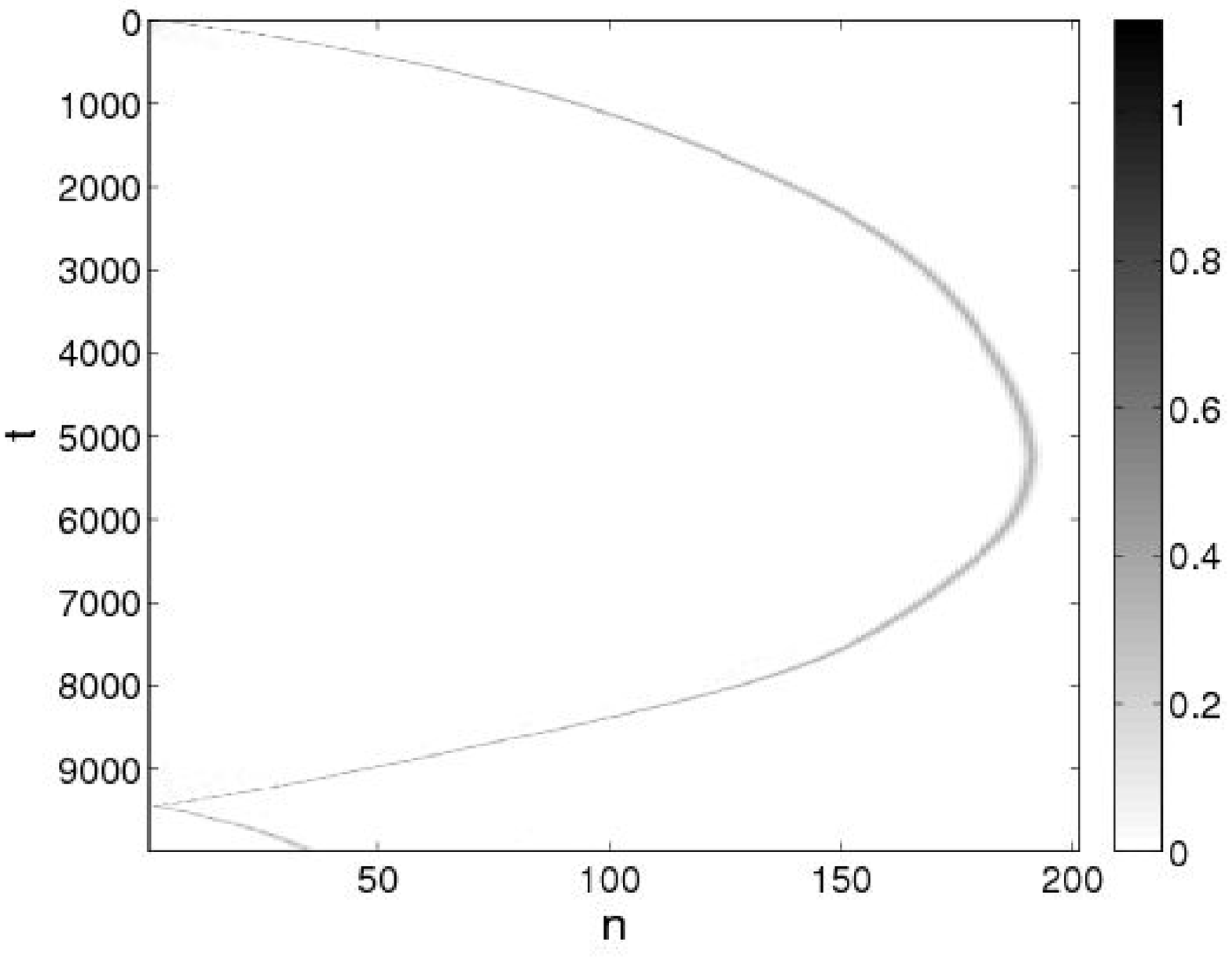}
\includegraphics[scale=0.4]{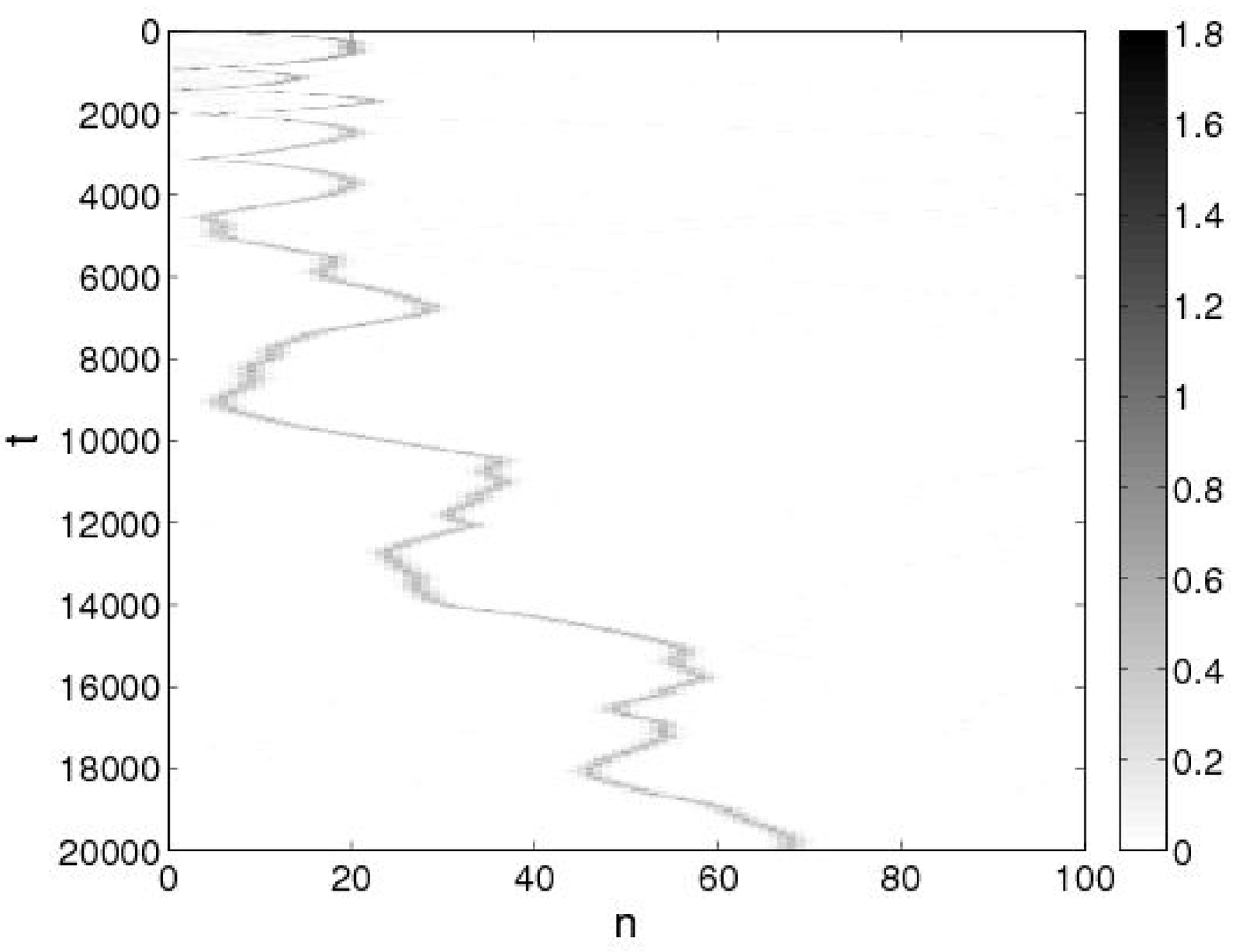}
\includegraphics[scale=0.4]{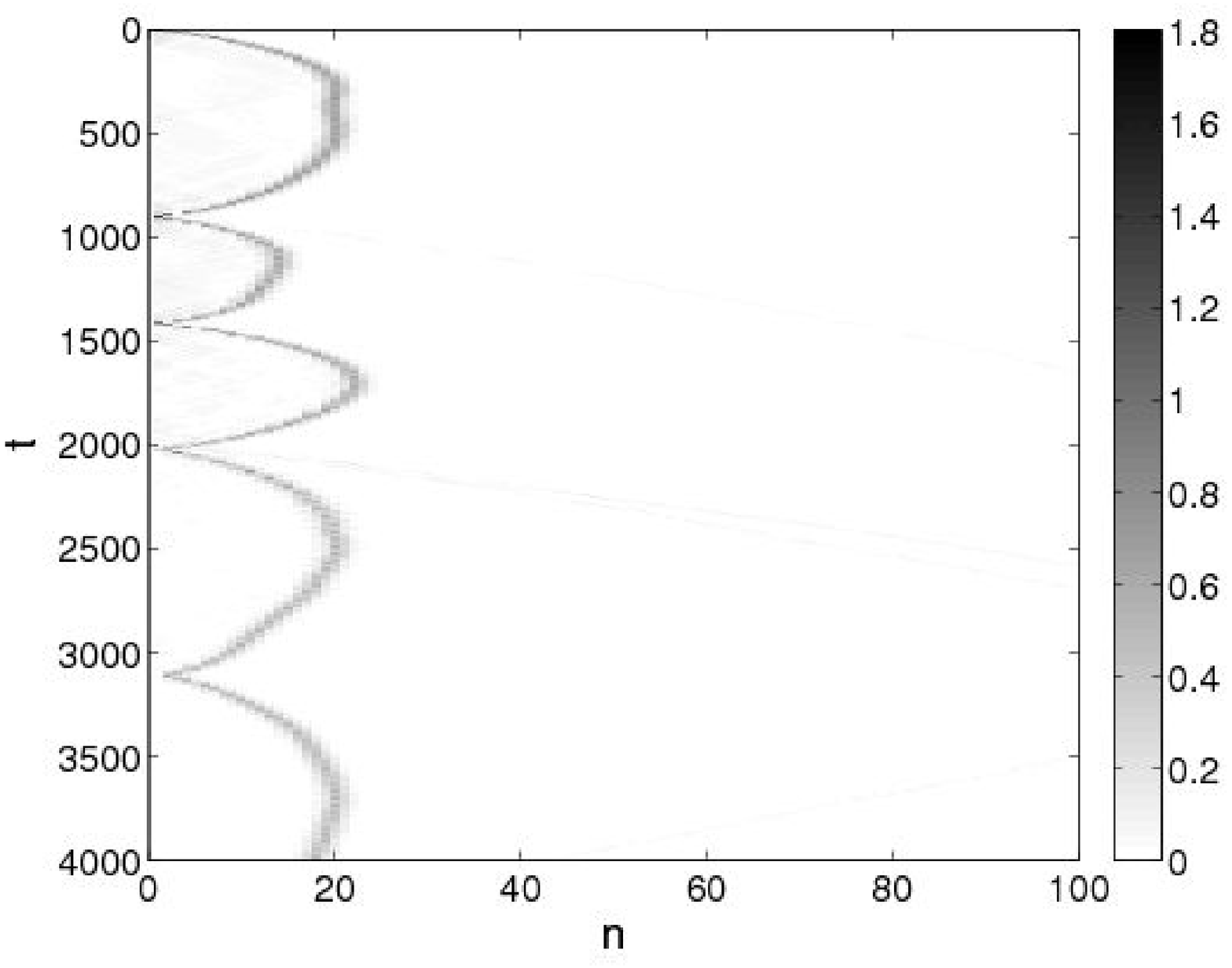}
\end{center}
\caption{\label{impactspos}
Space-time diagram giving the energy density (\ref{endens})
after an initial condition of the form (\ref{cic}),
for the on-site potential (\ref{anpot}) with $s=1$.
The upper plot corresponds to $\dot{y}_1(0)=1.5$
and the middle plot to $\dot{y}_1(0)=1.9$. The bottom plot provides a zoom of the middle one.}
\end{figure}

\begin{figure}
\psfrag{n}[0.9]{ $n$}
\psfrag{x}[1][Bl]{ $\dot{y}_n(t) $}
\begin{center}
\includegraphics[scale=0.33]{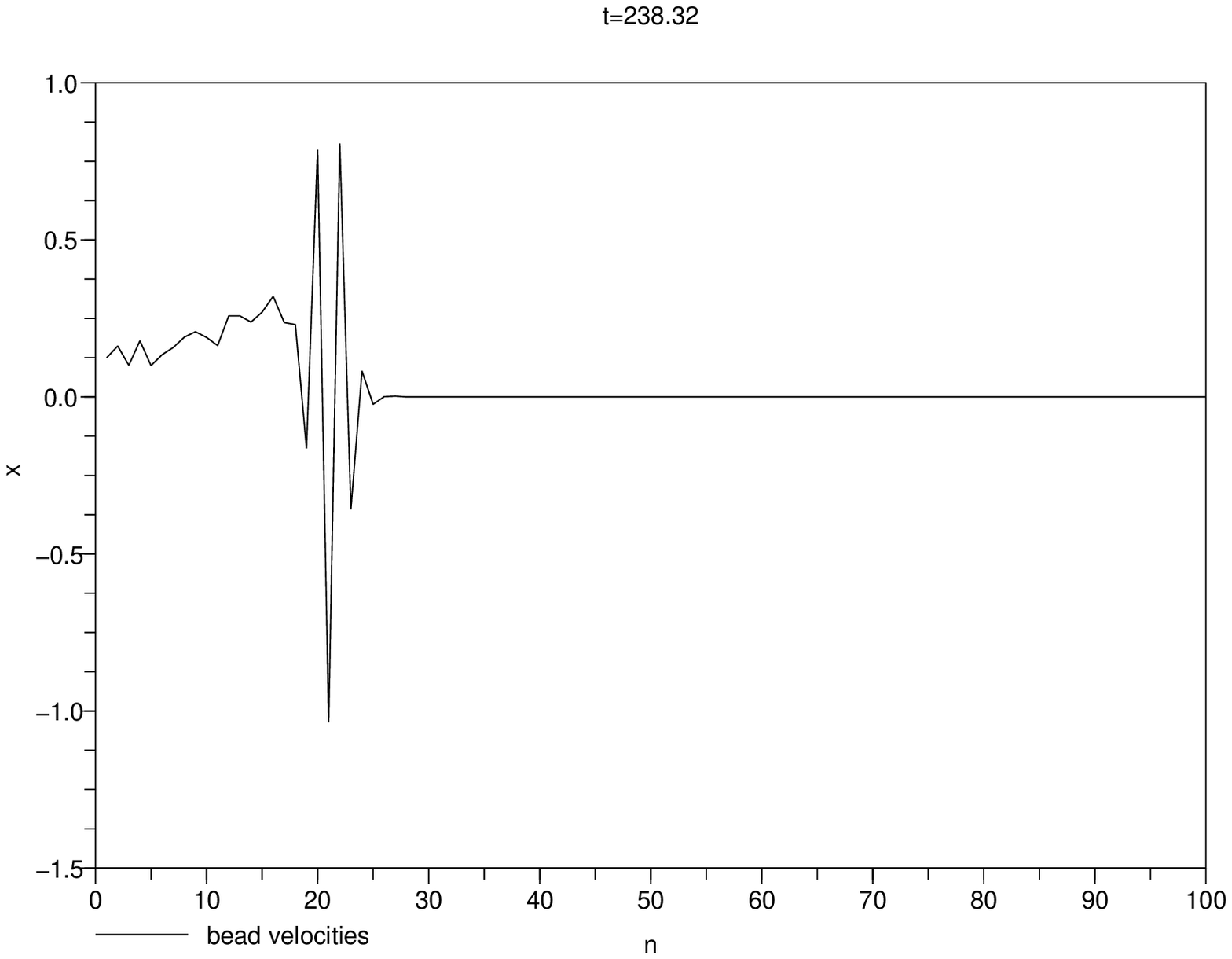}
\includegraphics[scale=0.33]{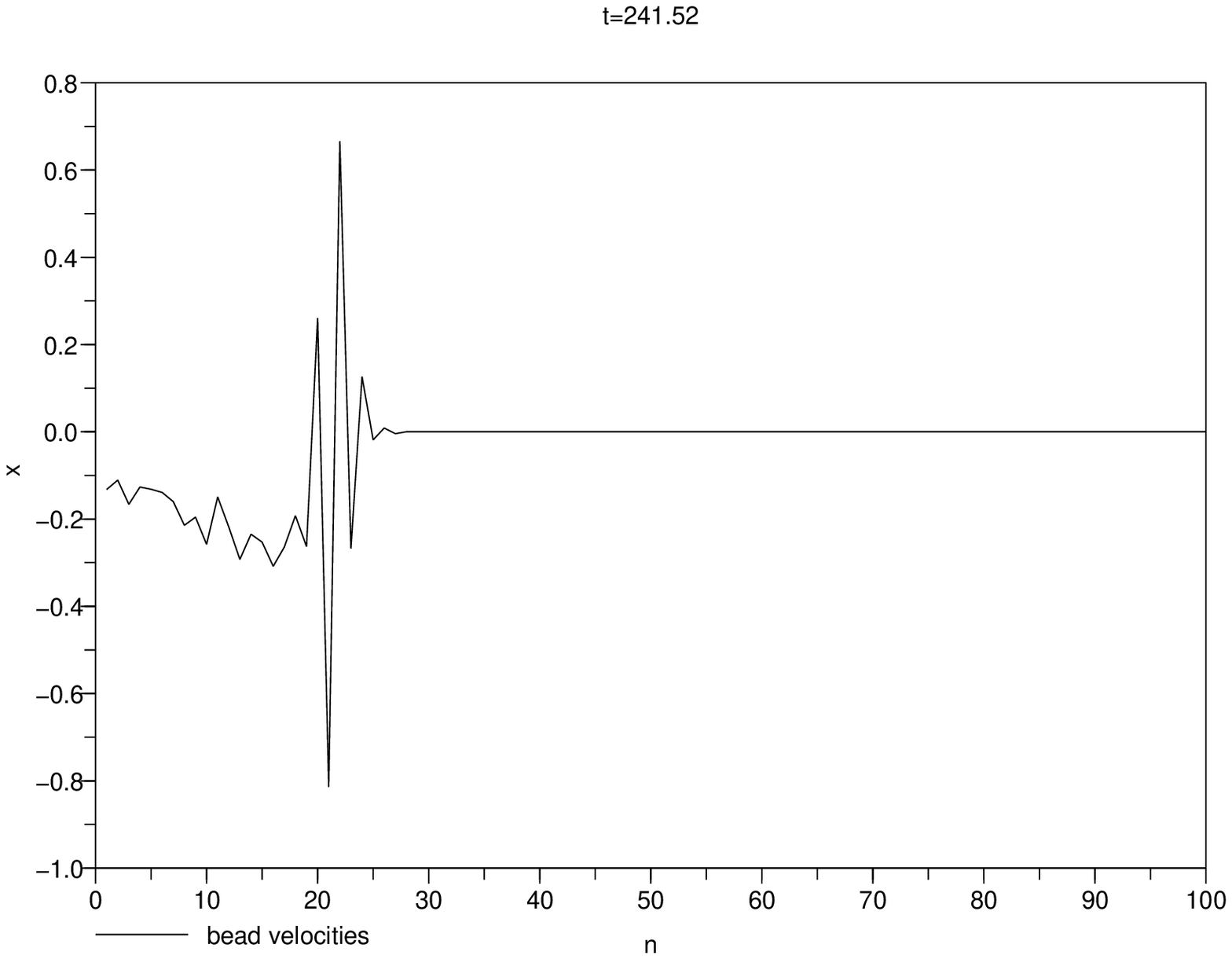}
\end{center}
\caption{\label{dirrevers}
Particle velocities for $s=1$ and
the initial condition (\ref{cic})
with $\dot{y}_1(0) \approx 1.87$.
The traveling breather profile is shown
at two different times close to direction-reversing
(the left panel corresponds to $t\approx 238$, and the right panel
to $t\approx 241$). Nonlinear waves confined between
the traveling breather and the boundary are also visible.}
\end{figure}

In the case $s>0$ of (\ref{anpot}) we observe a different scenario
illustrated by figure \ref{impactspos}.
The traveling breather doesn't move at constant velocity, but instead behaves
like a ``bouncing ball" against the boundary at $n=1$, i.e.
it experiences alternating phases of deceleration,
direction-reversing,
accelerated backward motion towards the boundary, and rebound at the boundary
(bottom panel of figure \ref{impactspos}).
During a few rebounds,
the breather center behaves like a Newtonian particle in an almost constant
effective force field.
This phenomenon seems to take place as soon as $s>0$
(we have checked it for $s \geq 0.005$), but not for $s=0$.
The effective force field both increases with $s$ and
with the imprinted initial velocity. For moderate initial velocities
the traveling breather deceleration is quite slow, as shown by the
top panel of figure \ref{impactspos}.
Figure \ref{dirrevers} displays a traveling
breather profile at the onset of direction-reversing.
These traveling breathers with
direction-reversing motion are
reminiscent of excitations known as ``boomerons'',
consisting of direction-reversing solitons discovered in
different kinds of integrable models
(see \cite{degas} and references therein),
but the link between both phenomena remains quite speculative at this stage.
Although we have no clear explanation of the origin of direction-reversing
for the traveling breather, one possibility might be its interaction with
other nonlinear waves visible in figure \ref{dirrevers}, which are
confined between the traveling breather and the boundary.
In addition, the rebound dynamics can be followed by
phases of intermittent trapping or erratic motion of the breather
(figure \ref{impactspos}, middle panel).

\subsection{\label{rescimpact}Dimensional analysis of the DpS limit}

The DpS limit has been previously described for equation (\ref{eqm}) written
in a normalized form. In this section, we consider a general class of
granular systems with on-site potentials and use suitable scalings to
rewrite the system in the form (\ref{eqm}). Returning to the above impact problem,
this allows us to analyze in which parameter regime the DpS equation
drives the dynamics. As we shall see, this case occurs when
local oscillations are faster than binary collisions.

We consider a chain of identical beads of mass $m$ sitting in local harmonic potentials,
described by the Hamiltonian
\begin{equation}
\label{hamgen}
{\mathcal H}=
\sum_{n}{\frac{m}{2}\, \dot{x}_n^2+\frac{k}{2}\, x_n^2 + \frac{2}{5}\, \gamma (x_n - x_{n+1})_+^{5/2}},
\end{equation}
where $\gamma$ is the nonlinear stiffness of Hertzian interactions
and $k$ the linear stiffness of local potentials.

Let us first consider two interacting beads,
one being initially at rest and the other having an initial velocity $V$,
and temporarily neglect the local restoring force of the on-site potentials.
After collision, their contact time is approximately equal to $2.43\, \tau_h$
with $\tau_h = [m^2/(\gamma^2 V)]^{1/5}$, and their
maximal compression distance is close to $0.76 \,\delta$, where
$\delta = (m V^2 / \gamma )^{2/5} $ \cite{lovett,falcon}.
Moreover, the stiffness constant of
Hertzian interactions linearized at precompression $\delta$ is of the order of
$\kappa_h = \gamma \sqrt{\delta}$.

Including back the local restoring forces,
the displacement $\xi$ at which Hertzian and local forces equilibrate
satisfies $\gamma \, \xi^{3/2}= k \, \xi$ and is given by
$\xi=(k/\gamma)^2$. In addition, the period of local oscillations is
$2\pi \tau_c$ with $\tau_{c}=(m/k)^{1/2}$.

Now we are ready to perform a suitable rescaling of (\ref{hamgen}).
Setting $x_n (t) = \xi\, y_n (t/\tau_c)$, the Hamiltonian (\ref{hamgen}) yields the equations
of motion (\ref{eqm}) in dimensionless form.
Moreover, the initial condition
\begin{equation}
\label{icnrenorm}
x_n (0)=0, \ \ \ \dot{x}_n(0)=V\, \delta_{n,1}
\end{equation}
reads in dimensionless form
\begin{equation}
\label{icrenorm}
y_n(0)=0, \ \ \ \dot{y}_n(0)=\lambda^{5/2}\, \delta_{n,1},
\end{equation}
where
\begin{equation}
\label{deflambda}
\lambda = \kappa_h / k
\end{equation}
measures the relative strengths of the Hertzian interaction at initial velocity $V$ and the
local potential.
Since $\kappa_h = m/\tau_h^2$, we have equivalently
$$
\lambda = \frac{m}{k \tau_h^2},
$$
i.e. $\lambda^{1/2}$ measures the relative duration of local oscillations and binary collisions.

From (\ref{icrenorm}) we deduce that the
DpS regime takes place when $\lambda^{5/2}$ is small.
For example, for a Newton's cradle with strings of length $50\, $cm
and binary collision time $2.43 \,\tau_h = 0.077\, $ms (value taken from \cite{lovett}
for an impact velocity of 
$1.1\, \rm{m.s}^{-1}$), one obtains $\lambda^{5/2}\approx 1.75 \times 10^{19}$,
hence we are extremely far from the DpS regime.
In section \ref{canti} we will introduce a mechanical system for which local oscillators are
much stiffer and the DpS dynamics is relevant.

\vspace{1ex}

\subsection{\label{canti}A lattice model for cantilevers decorated by spherical beads}

Several types of mechanical models have been devised to analyze the properties
of discrete breathers experimentally, see e.g. \cite{ceka,sato,kimura,green}.
In this section we introduce a simplified model of the cantilever system sketched
in figure \ref{boules} (right picture). 
We consider the form (\ref{eqm}) analyzed
previously, but examine the more general situation when the lattice
is spatially inhomogeneous.
With this model, we shall observe that a
moving breather generated by an
impact on the first cantilever
can be almost totally reflected by a localized impurity
corresponding to a moderate increase of the bead radii on a single cantilever.

\vspace{1ex}

We begin by introducing a simplified model of the cantilever system of 
(the right panel of) figure \ref{boules},
where cantilever compression is neglected and bead deformations are treated quasi-statically.
More precisely, each bead is seen as an elastic medium at equilibrium,
clamped at a cantilever at one side, and either free or in contact with one 
bead of a neighboring cantilever
at the opposite side. So any bead deformation is fully determined by two cantilever positions, and can be
approximated by Hertz's contact law.
In addition, each cantilever decorated by two spherical beads is described by
a point-mass model which approximates the dynamics of the slower bending mode,
following a classical approach in the context of atomic force microscope cantilevers \cite{rabe}.
Under these approximations, our
model incorporates a single degree of freedom per cantilever, namely its maximal deflection.

The point-mass model is obtained as follows.
Using a rod model and under the assumption
of small deflection, a cantilever clamped at both ends and bent by a force applied to its mid-point
can be represented by an equivalent linear spring of stiffness $k=192\, E\, I \ell^{-3}$, where
$E$ is the cantilever's Young modulus, $\ell$ its length and
$I=w\, h^3 /12$ its area moment of inertia, $w,h$ denoting the cantilever width and thickness
respectively (see e.g. \cite{ll}, pp. 77 and 81). For a cantilever without attached beads,
the first bending mode frequency satisfies
$\omega_{\rm{min}} \approx 22.4\, [E I / (\rho A)]^{1/2} \ell^{-2}$ (\cite{ll}, p.102)
where $\rho$ denotes the cantilever density and $A=w \, h$ its cross section.
A single cantilever is then represented by an effective mass $m^\ast = k/\omega_{\rm{min}}^2 \approx 0.38\, m_c$,
where $m_c=\rho A \ell$ is the exact cantilever mass. The effective mass of a cantilever
decorated by two beads of masses $m_b$ is then $m=m^\ast + 2m_b$.
For beads of radius $R$ and density $d$ we fix consequently
$m = 0.38 \, m_c + (8/3) \pi d R^3  $.

\vspace{1ex}

Now let us describe the model for a one-dimensinal chain of such cantilevers,
where all beads are made of the same material with
Young's modulus $\mathcal{E}$ and Poisson coefficient $\nu$.
We denote by $R_n = R\, \tilde{R}_n$ the radius of the two beads of the $n$th cantilever
($R$ being a reference value and $\tilde{R}_n$ an adimensional number),
$x_n (t)$ the maximal cantilever deflections and $m_n = 0.38 \, m_c + (8/3) \pi d R_n^3  $ their effective masses.
The array of decorated cantilevers
is then described by the Hamiltonian
\begin{equation}
\label{ham}
{\mathcal H}=
\sum_{n}{\frac{m_n}{2}\, \dot{x}_n^2+\frac{k}{2}\, x_n^2 + \frac{2}{5}\, \gamma_n (x_n - x_{n+1})_+^{5/2}},
\end{equation}
where
$\gamma_n = \gamma \, \eta_n$ is
the nonlinear stiffness constant of Hertzian interactions between two beads on different cantilevers $n$ and $n+1$,
defined by $\gamma = \frac{ E \sqrt{2R}}{3(1-\nu^2)}$ and
$\eta_n = [2 \tilde{R}_n \tilde{R}_{n+1} / (\tilde{R}_n + \tilde{R}_{n+1})]^{1/2}$ (see e.g. \cite{ll}).

Setting $x_n (t) = \xi\, y_n (t/\tau_c)$ as in section \ref{rescimpact}, the Hamiltonian (\ref{ham}) yields the following equations
of motion in dimensionless form
\begin{equation}
\label{eqmnh}
\mu_n \, \ddot{y}_n + y_n = -\eta_n (y_n - y_{n+1})_+^{3/2} + \eta_{n-1} (y_{n-1} - y_{n})_+^{3/2},
\end{equation}
where $\mu_n = m_n /m$. Note that if all beads have radius $R$
(i.e. $\tilde{R}_n =1$) then $\eta_n =\mu_n =1$.

\vspace{1ex}

Our main purpose is to analyze an impact problem in a chain of $N$ cantilevers with free
end boundary conditions, where the first cantilever is hit by a striker at $t=0$.
For this purpose we consider a simpler initial condition where all
cantilevers with index $n \geq 2$ are initially at rest and
the first cantilever has initial velocity $V$ and zero deflection.
This corresponds to fixing the initial condition (\ref{icnrenorm}), which
yields (\ref{icrenorm}) in rescaled form.

Numerical simulations are performed for a chain of $N=200$ stainless steel cantilevers
with $\rho = 8 \times
10^3\,  \rm{kg}.\rm{m}^{-3}$, $E=193$~GPa,
$\ell = 25$~mm, $w=5$~mm, $h=1$~mm, decorated by teflon beads
with $d = 2.2 \times 10^3\,  \rm{kg}.\rm{m}^{-3}$, $\mathcal{E}=1.46$~GPa, $\nu=0.46$ \cite{porter}.
All beads have radius $R=2.38$~mm, except at the middle of the chain where
$\tilde{R}_{100}$ can be tuned.
These values correspond to a cantilever array at the macroscopic scale
(as in reference \cite{kimura}), but extensions to the microscale might be also considered
\cite{sato,yapici}.

We fix the impact velocity $V=1\, \rm{m}.\rm{s}^{-1}$, which
yields $\tau_h \approx 0.047$~ms. Since
$\tau_c \approx 0.025$~ms, we have $\lambda \approx 0.29$ and $\lambda^{5/2}$ is small.
Consequently,
under the above conditions
the DpS approximation is valid in the spatially
homogeneous case, or in sufficiently long homogeneous
segments of a chain including defects.

\vspace{1ex}

The initial impact generates a traveling breather and a fairly extended wavetrain emitted from
the boundary, as previously analyzed in section \ref{impactpb}.
The traveling breather velocity is close to $2030$ sites per second.
Evaluating the traveling breather characteristics at $n=80$, we find
a maximal bead velocity close to $0.5 \, \rm{m}.\rm{s}^{-1}$ (i.e. half the impact velocity),
a maximal cantilever deflection close to $11 \, \mu\rm{m}$
and a maximal interaction force close to $2.8 \, \rm{N}$.
The pulse duration is close to $3.8\, \rm{ms}$ and
the period of internal oscillations close to $0.14 \,\rm{ms}\approx  T_0 / (1.1)$,
$T_0 = 2\pi \tau_c$ being the period of linear local oscillations.

\vspace{1ex}

\begin{figure}[h]
\begin{center}
\includegraphics[scale=0.32]{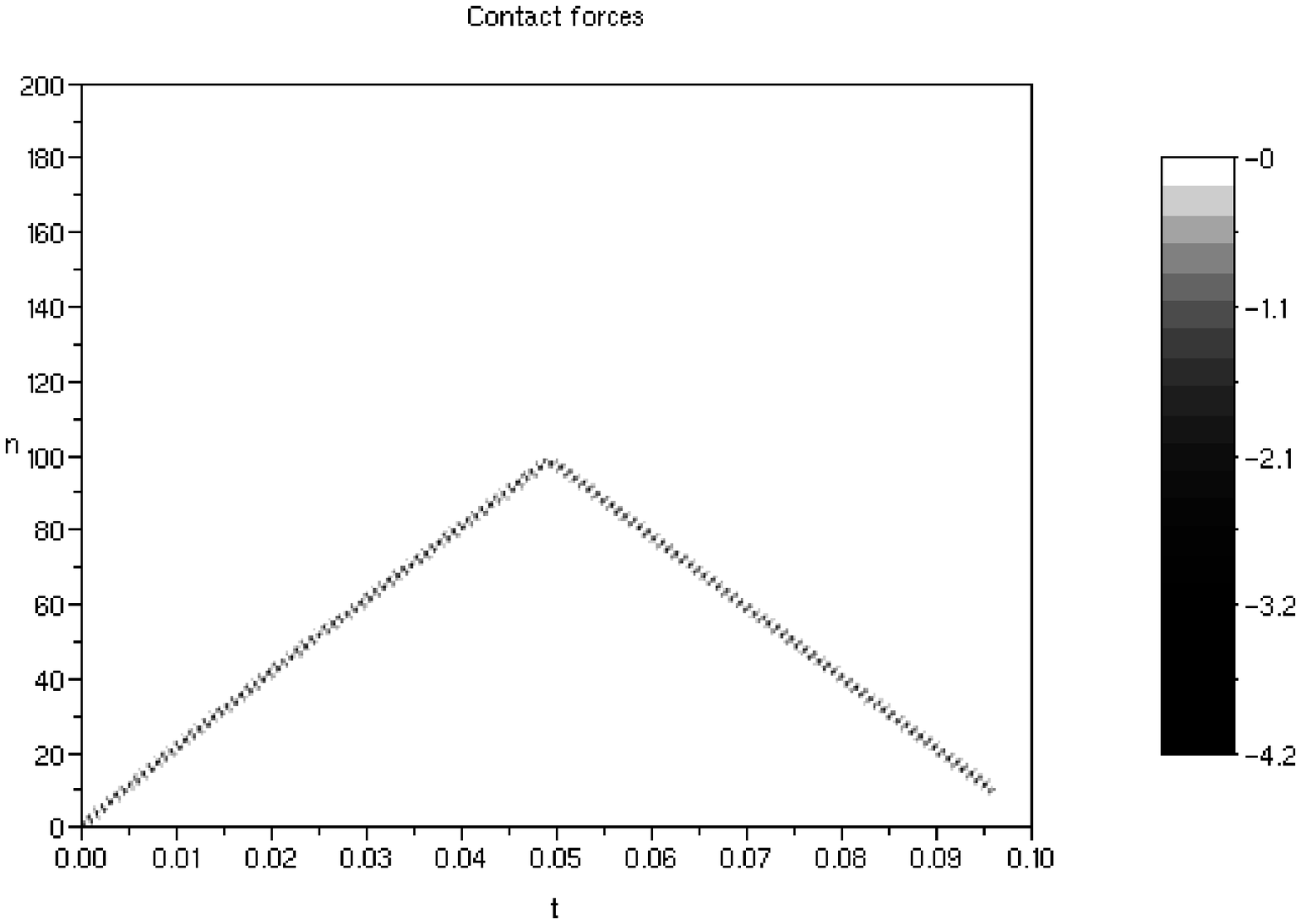}
\includegraphics[scale=0.32]{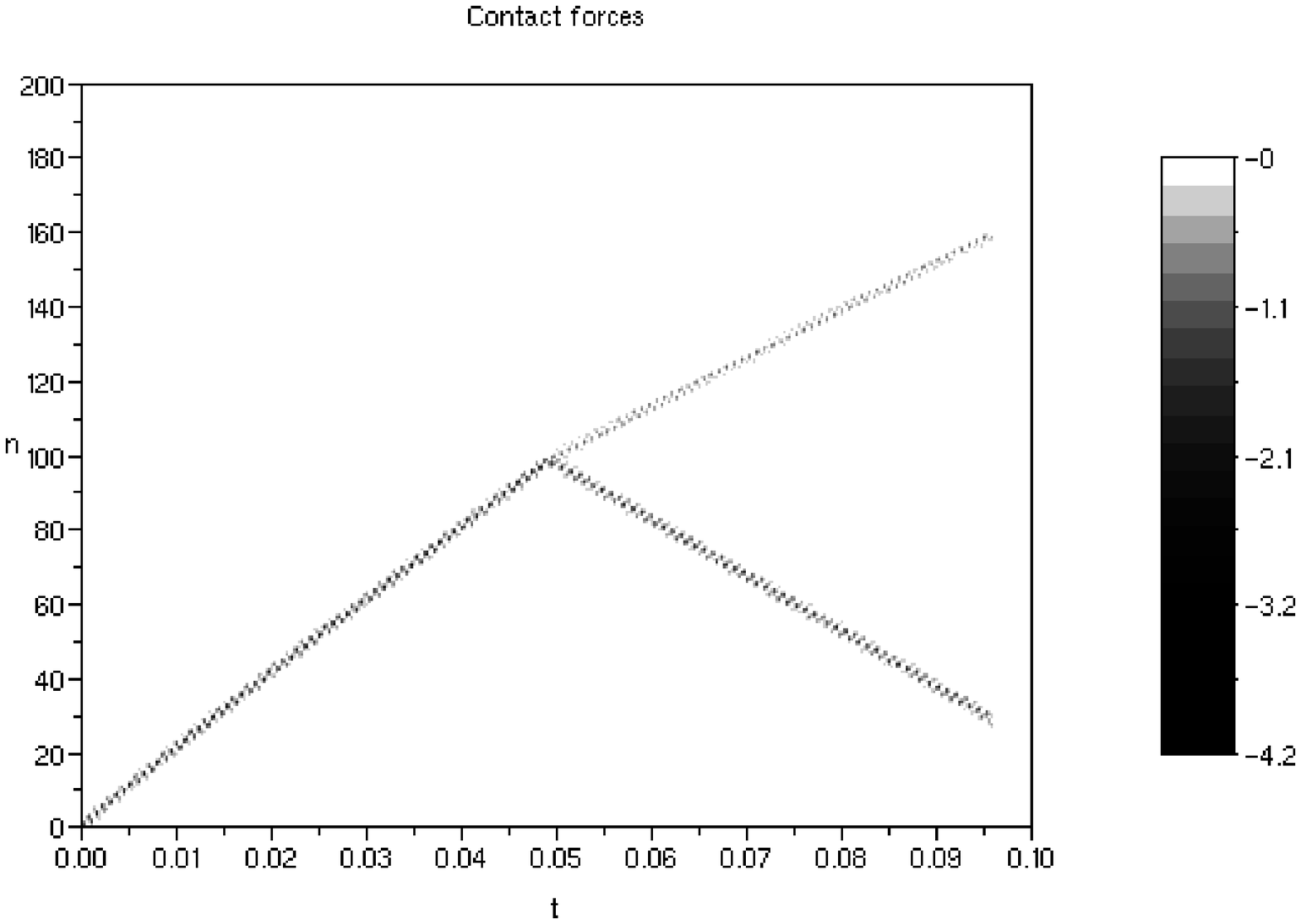}
\end{center}
\caption{\label{forcetotref}
Space-time diagrams showing the interaction forces
$f_n = -\gamma_n (x_n - x_{n+1})_+^{3/2}$ in system (\ref{ham})
for the impact problem described in the text (forces are expressed in $N$).
Forces are represented
in grey levels, white corresponding to vanishing interactions (i.e. beads not in contact)
and black to a minimal negative value of the contact force.
Left plot~:  $\tilde{R}_{100}=1.6$. Right plot~:  $\tilde{R}_{100}=1.1$.
}
\end{figure}

When the breather reaches the defect site,
it appears to be almost totally reflected
for a large enough inhomogeneity, whereas it remains
significantly transmitted for a sufficiently small inhomogeneity.
This phenomenon is illustrated by
figure \ref{forcetotref}, which compares the cases
$\tilde{R}_{100}=1.6$ (almost total reflection)
and $\tilde{R}_{100}=1.1$ (partial reflection). After the breather
reflection by the defect for $\tilde{R}_{100}=1.6$, a small part of the
vibrational energy remains loosely trapped near the defect site.
Such phenomena resulting from breather-defect interactions
have been already numerically observed in different types of
Klein-Gordon lattices \cite{fori,ting,cimpur}.
In the present model, almost total reflection occurs for
physically realistic parameter values, which suggests potential applications of
such systems as shock wave reflectors.

\vspace{1ex}

\section{\label{caseprecomp}Traveling breathers under precompression}

In section \ref{crdps} we have analyzed the properties of discrete breathers
in chains of oscillators coupled by fully nonlinear Hertzian interactions.
We have obtained highly-localized static breathers, which display
a super-exponential spatial decay and have an almost constant width in the small amplitude limit.
Moreover, small perturbations of the static breathers
along a pinning mode generate traveling breathers
propagating at an almost constant velocity with very small dispersion.

These properties are largely due to the fully-nonlinear coupling
between oscillators, which reduces the phonon band to a single frequency.
Intuitively, the absence of linear coupling terms enhances localization,
because linear dispersion tends to disperse localized wave packets.
Though this phenomenon can be compensated by nonlinearity,
breathers in nonlinear lattices with phonon bands
generally have a slow exponential spatial decay
in the limit of vanishing amplitude (see e.g. \cite{jsc,jamesnoble}).
Moreover, due to resonance with phonons, exact traveling breathers
are generally superposed on nondecaying oscillatory tails, a phenomenon
mathematically analyzed in \cite{ioossK,jamessire,sire,ioossj,barashenkov} 
(see also section 4.5 of \cite{flg} for more references); only under
special choices of the speed (or the system parameters) can it then
be the case that the amplitude of these oscillatory tails exactly
vanishes~\cite{melvin1,melvin2}.

Due to these noticeably different breather properties in the presence or absence of phonon band,
it is interesting to consider physical systems possessing a tunable phonon band,
allowing to pass from one situation to the other.
This is the case in particular for granular crystals under tunable precompression,
since the latter
results in a perturbation of the interaction potential inducing an additional
harmonic component. In this section, we incorporate this effect to model (\ref{hamresc}),
formally analyze the existence of discrete breathers through the phenomenon
of modulational instability,
and numerically demonstrate that the existence of a phonon band can drastically
modify the outcome of an initial impact.

\subsection{Granular chain under precompression, and
correspondence to the NLS equation}

We consider the system (\ref{hamresc})
with the modified interaction potential
\begin{equation}
\label{pothertzp}
V(r)=\frac{2}{5}\,  ( d-r )_+^{5/2} +d^{3/2}\, r -\frac{2}{5}\,  d^{5/2},
\end{equation}
where $d > 0$ is a parameter. We have thus for $r\approx 0$
$$
V(r)=v_1\, \frac{r^2}{2}+v_2\, \frac{r^3}{3}+v_3\, \frac{r^4}{4}+O(|r|^5),
$$
with
$$
v_1 = \frac{3}{2}d^{1/2}, \ \ \
v_2 = -\frac{3}{8}d^{-1/2}, \ \ \
v_3 = -\frac{1}{16}d^{-3/2}.
$$
This modified potential possesses
a harmonic component of size $d^{1/2}$ in the neighborhood of the origin,
and it becomes linear for $r \geq d$.
The first term of (\ref{pothertzp}) corresponds to the classical Hertzian potential
including a precompression effect. For example, this type of interaction
can be achieved in the cantilever system of figure \ref{boules}
in the case of a sufficiently long chain. This can be done
by applying a force at both ends of the system when the
cantilevers are unclamped, which results in a compression of all
the beads by a distance $d$ (compression becomes uniform
for an infinite system), and by clamping the cantilevers
at this new equilibrium state.
The second and third terms of (\ref{pothertzp}) do not modify the equations of motion,
and just aim at putting the modified Hertz potential in a standard form
with $V(0)=0$, $V^\prime (0)=0$.

\vspace{1ex}

System (\ref{hamresc})-(\ref{anpot})-(\ref{pothertzp}) consists of a mixed Klein-Gordon - FPU lattice,
which admits a phonon band with a finite width (of order $O(d^{1/2})$ when $d\approx 0$). The
phonons $y_n (t)=A\, e^{i\, (q n -\omega t)}+\mbox{c.c.}$
of the system linearized
at $y_n=0$ obey the dispersion relation
\begin{equation}
\label{dispers}
\omega^2 (q)= 1+2 v_1 (1-\cos{q}),
\end{equation}
where $q \in [0,\pi ]$ denotes the wavenumber and $\omega$ the phonon frequency.

For this class of systems combining anharmonic local and interaction potentials,
the modulational instability (MI) of small amplitude periodic and standing waves
has been studied in a number of references \cite{flachtan,dorignac,johan,gian,gian3}.
This phenomenon has been analyzed in \cite{gian,gian3} through the
continuum nonlinear Schr\"odinger (NLS) equation, which describes the slow spatio-temporal
modulation of small amplitude phonons under the effects of nonlinearity and dispersion
(see also the basic papers \cite{remoiss,kono}).
From the general results of \cite{gian,gian3},
system (\ref{hamresc})-(\ref{anpot})-(\ref{pothertzp})
admits solutions of the form
\begin{equation}
y_{n}(t)=\epsilon \,A[\epsilon ^{2}t,\epsilon (n-c\,t)]\,e^{i(qn-\omega
t)}+c.c.+O(\epsilon ^{3/2})  \label{ansatz}
\end{equation}
on time intervals of length $O(\epsilon^{-2})$ ($\epsilon$ being a small parameter),
where $A(\tau ,\xi )$ satisfies the NLS equation
\begin{equation}
i\,\partial _{\tau}A=-\frac{1}{2}\omega^{\prime \prime }(q)\,\partial _{\xi
}^{2}A+h|A|^{2}A.  \label{nls}
\end{equation}
In the above expressions, $\omega$ is given by (\ref{dispers}),
$c=\omega^{\prime }(q)$ is the group velocity,
$h=\beta / \omega$, $\omega^{\prime \prime}=v_1 \omega_2 / \omega$ and
$$
\beta = \frac{16 v_2^2 (\sin{q})^2 (1-\cos{q})^2}{4 v_1 (1-\cos{q})^2 +3}+\frac{3}{2}[\, 4v_3 (1-\cos{q})^2+s   \, ] ,
$$
$$
\omega_2 = \cos{q}-\frac{v_1}{\omega^2}\, (\sin{q})^2
$$
(see \cite{gian}, equation (2.12) p. 557).

The so-called focusing case of the NLS equation occurs for $\omega^{\prime \prime}(q)\,h<0$,
i.e. under the condition
\begin{equation}
\label{focus}
\Phi := -  \beta\,  \omega_2 >0.
\end{equation}
In that case the spatially homogeneous solutions of (\ref{nls}) are unstable,
and (\ref{nls}) admits {\em sech}-shaped solutions corresponding
(at least on long transients) to small amplitude traveling breather solutions
taking the form
\begin{equation}
y_{n}(t)=\epsilon\, M\,
\frac{e^{i[qn-(\omega -\epsilon^{2} \omega^{\prime \prime} /2 )  t+\varphi ]}}{\cosh {[\epsilon
(n-c\,t)]%
}}+c.c.+O(\epsilon ^{3/2}) ,  \label{tbs}
\end{equation}%
where $M=(-\omega^{\prime \prime}/h)^{1/2} $.
These solutions decay exponentially in space and broaden in the small amplitude
limit $\epsilon \rightarrow 0$, in contrast with the traveling breathers
numerically obtained in section \ref{impactpb} in the absence
of precompression.
For Klein-Gordon lattices (i.e. for harmonic interaction potentials),
the existence of exact traveling breather solutions close to (\ref{tbs})
has been proved in special cases in \cite{ioossK,jamessire,sire}
(breathers are superposed on a nondecaying oscillatory tail, exponentially small in $\epsilon$).
However these results do not directly apply to our model having anharmonic interaction potentials.

The frequency $\omega (q)$ defined by (\ref{dispers}) admits a unique
inflection point in the interval $(0,\pi )$, at the wavenumber $q=q_c \in (0, \pi/2)$ satisfying
$\cos{q_c}=v_1\, (1-\cos{q_c})^2$.
In the generic case when $\beta (q_c) \neq 0$, it follows that $\Phi$
changes sign at $q=q_c$ (since $w_2$ changes sign).
Consequently, MI generically occurs for wavenumbers in some interval
lying at one side of $q_c$.  This interval may extend
or not up to one edge of the phonon band, depending on parameter values.
For $q=0$ (in-phase mode) the condition $\Phi>0$ reduces to $s<0$,
and for $q=\pi$ (out-of-phase mode) it reduces to $16 v_3 + s >0$.
These conditions have been also obtained in \cite{flachtan} through a Hill's type analysis.

In the next section, we numerically check that MI can lead to the formation of
traveling breathers on long transients, and revisit the impact problem of section \ref{impactpb}.

\subsection{Excitation of traveling breathers}

\begin{figure}[h]
\psfrag{q}[0.9]{ $q$}
\psfrag{phi}[1][Bl]{ $\Phi$}
\begin{center}
\includegraphics[scale=0.4]{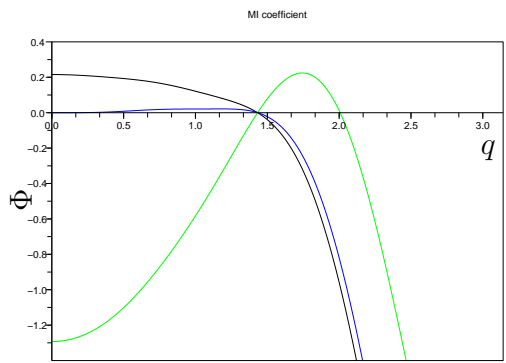}
\end{center}
\caption{\label{graphmi}
Graphs of the MI coefficient $\Phi$ defined by (\ref{focus}) as a function of wavenumber $q$, for
$s=-1/6$ (black curve), $s=0$ (blue curve), $s=1$ (green curve). Modulational instability
occurs in the bands where $\Phi >0$.
}
\end{figure}

In this section we fix $d=1/2$, so that
$v_1\approx 1.06$, $v_2 \approx - 0.53$ and $v_3 \approx - 0.17$, and
consider different values of the anharmonicity parameter
$s=1$, $s=0$ and $s=-1/6$.
For all these values, there exists a band of unstable phonon modes
characterized by $\Phi (q) >0$ (see figure \ref{graphmi}).
One can notice that $\Phi$ is much smaller for $s=0$ inside the
band of unstable modes, due to the smallness of $h$. In that case, slower MI
should occur according to the NLS approximation, but
at the same time the applicability of the latter should be restricted to
smaller values of $\epsilon$.

\begin{figure}
\psfrag{n}[0.9]{$n$}
\psfrag{x}[1][Bl]{ ${y}_n(t) $}
\begin{center}
\includegraphics[scale=0.34]{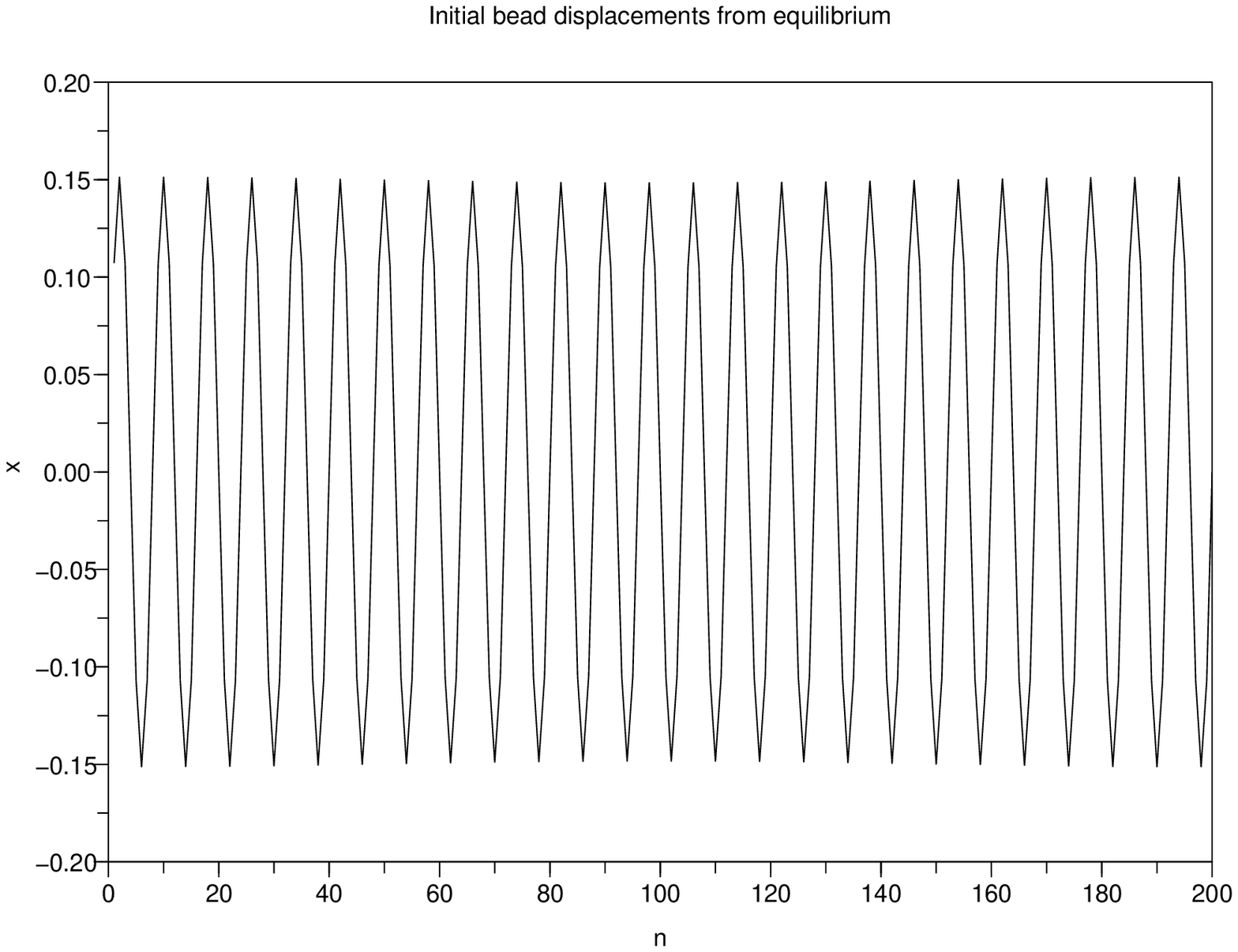}
\includegraphics[scale=0.34]{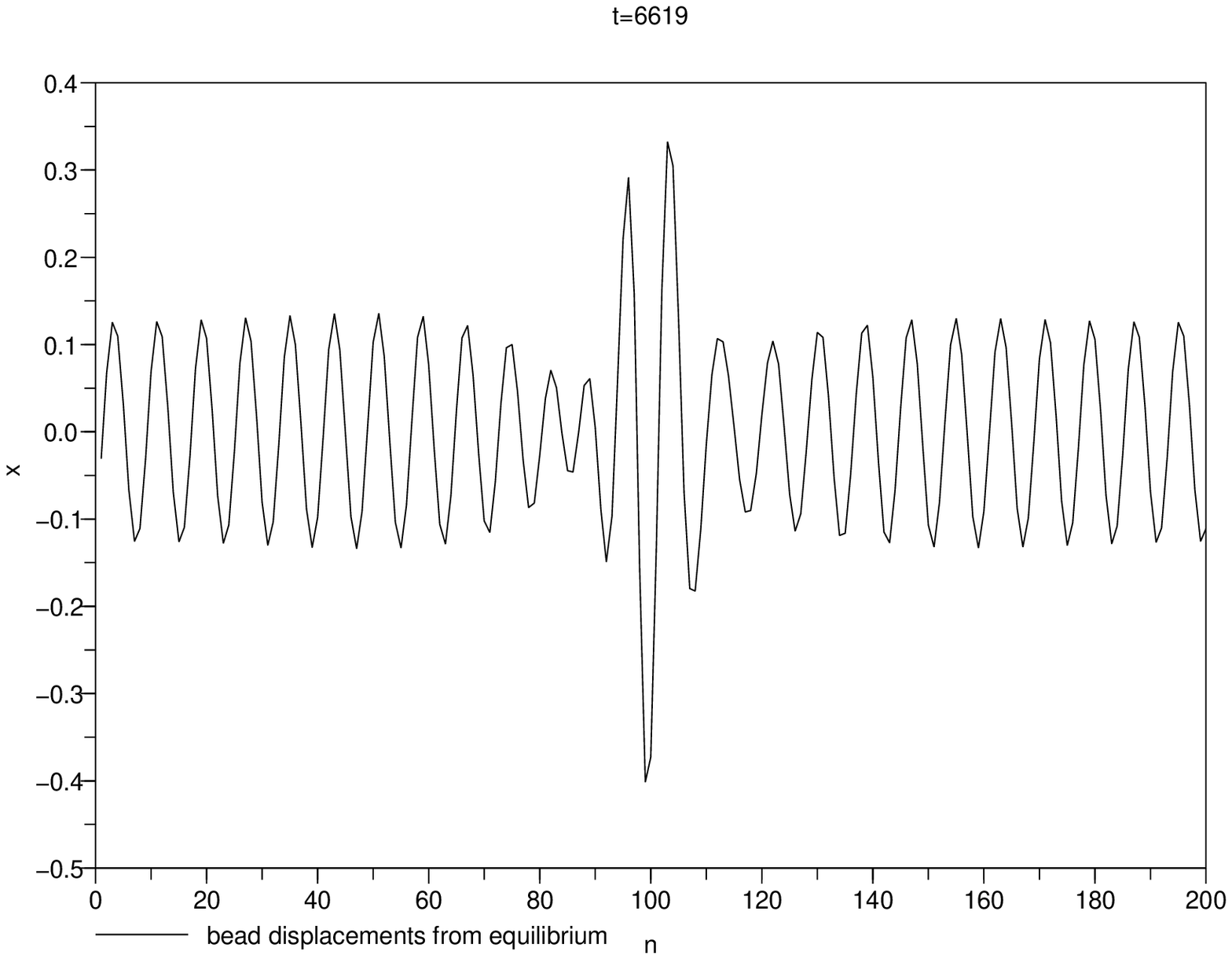}
\end{center}
\caption{\label{mism1o6}
Evolution of particle positions
in the system (\ref{hamresc})-(\ref{anpot})-(\ref{pothertzp}) with 
periodic boundary
conditions ($N=200$ particles). We consider the case $s=-1/6$
with precompression $d=0.5$. The initial condition (\ref{cimi})
is plotted in the left panel (case $q=\pi /4$).
Particle displacements plotted at time $t= 6619$ (right panel)
reveal the formation of a traveling breather resulting from a modulational instability
(the envelope propagates rightwise).}
\end{figure}

To illustrate the MI phenomenon, we integrate (\ref{hamresc})-(\ref{anpot})-(\ref{pothertzp})
numerically for initial conditions
\begin{eqnarray}
\label{cimi}
x_n(0)&=&a\, \sin{(q n)}\, (1+b\, \cos{(2n\pi /N)}), \\
\nonumber
\dot{x}_n(0)&=&-a\, \omega\, \cos{(q n)}\, (1+b\, \cos{(2n\pi /N)})
\end{eqnarray}
corresponding to slowly modulated phonons, with
$a=0.15$, $b=0.01$, a wavenumber $q$ in the band of unstable modes (see fig. \ref{graphmi}),
and $\omega$ determined by (\ref{dispers}).
We consider a chain of $N$ particles with
periodic boundary conditions.
Figure \ref{mism1o6} displays the results for $s=-1/6$, $q=\pi /4$ and $N=200$.
The initial perturbation generates a traveling breather over a long 
transient
(at the end of which a splitting of the pulse occurs).
The same phenomenon occurs for $s=1$ and $s=0$, albeit the latter case
results in slower instabilities and less localized traveling breathers (results not shown).

According to the previous computations, traveling breathers with profiles
reminiscent of (\ref{tbs}) can be generated from slow modulations of
small amplitude unstable phonons. This raises
the question of the nucleation of
traveling breather from other types of
initial conditions, in particular for a localized impact.
In section \ref{impactpb} we observed that
this type of excitation systematically generates traveling breathers.
By extending this study to the case of potential (\ref{pothertzp}), we will
show in which way linear dispersion may modify the impact dynamics.

In what follows we keep the same values of
parameters $d, s$ and
integrate (\ref{hamresc})-(\ref{anpot})-(\ref{pothertzp}) numerically (for free end boundary conditions),
starting from the initial condition (\ref{cic}) with $\dot{y}_1(0) \approx 1.87$.
Depending on the value of $s$, the initial excitation may lead to
different dynamical phenomena, and notable differences
with respect to the case without precompression are observed.

\begin{figure}[!b]
\psfrag{n}[0.9]{ $n$}
\psfrag{x}[1][Bl]{ $\dot{y}_n(t) $}
\begin{center}
\includegraphics[scale=0.4]{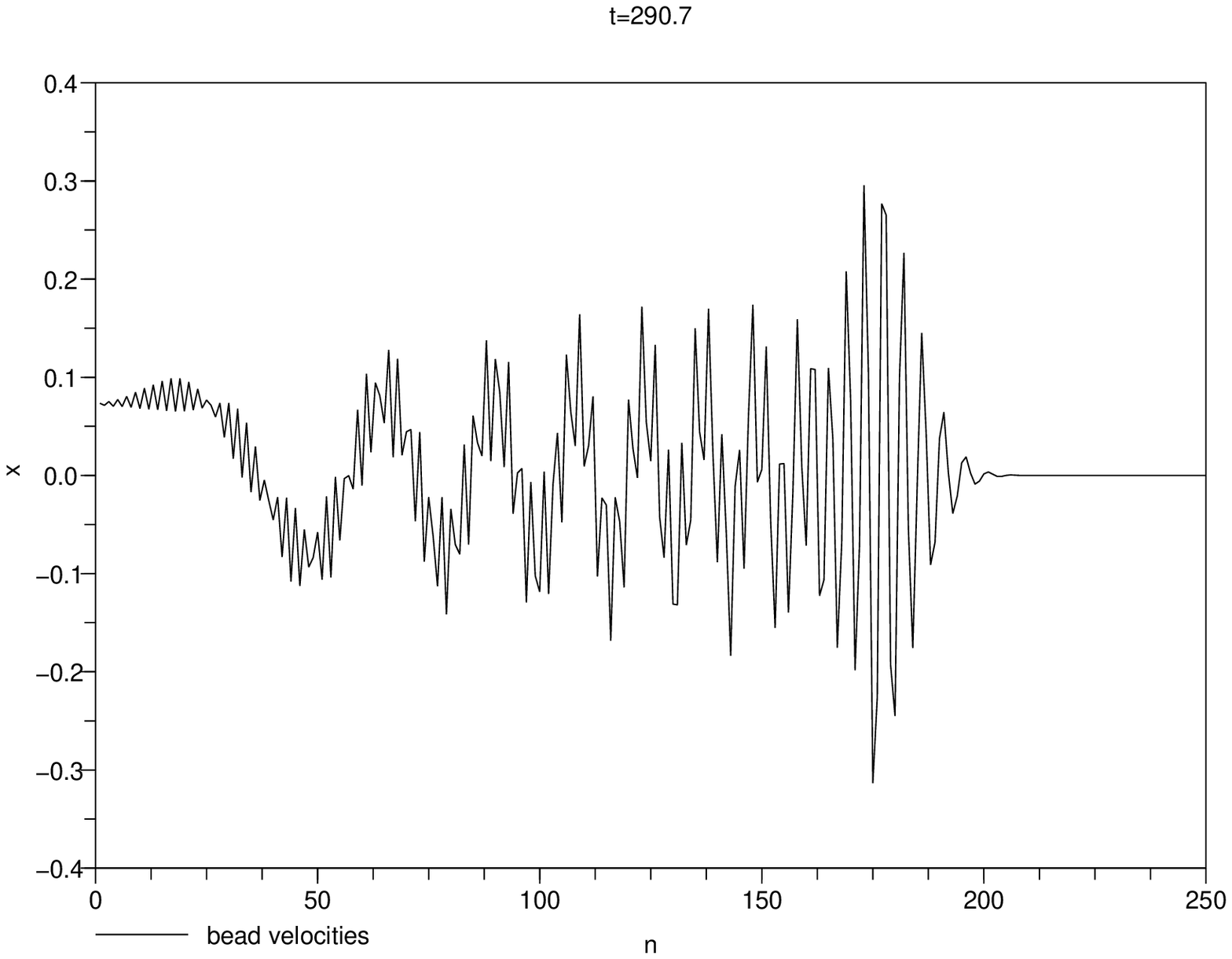}
\includegraphics[scale=0.4]{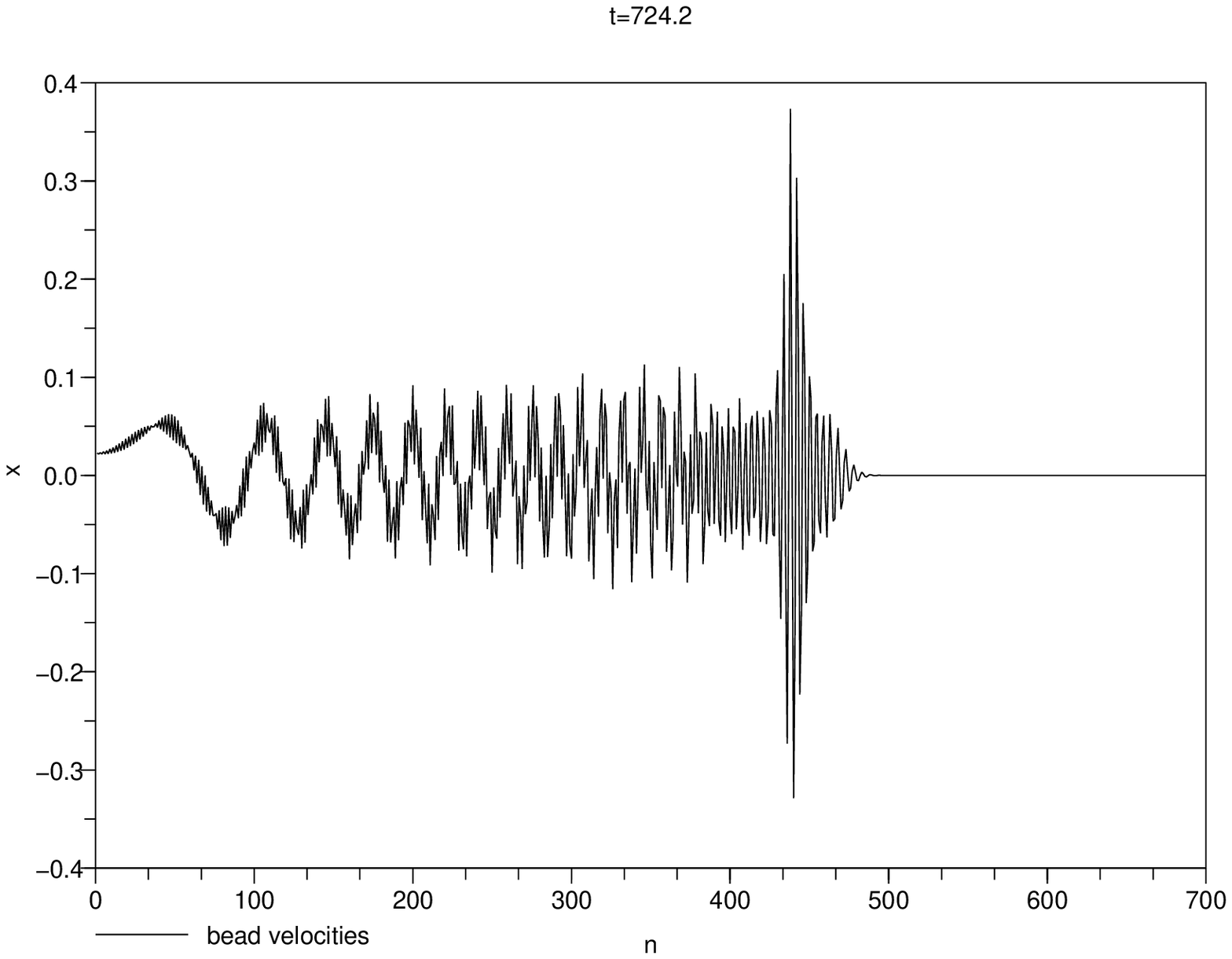}
\includegraphics[scale=0.4]{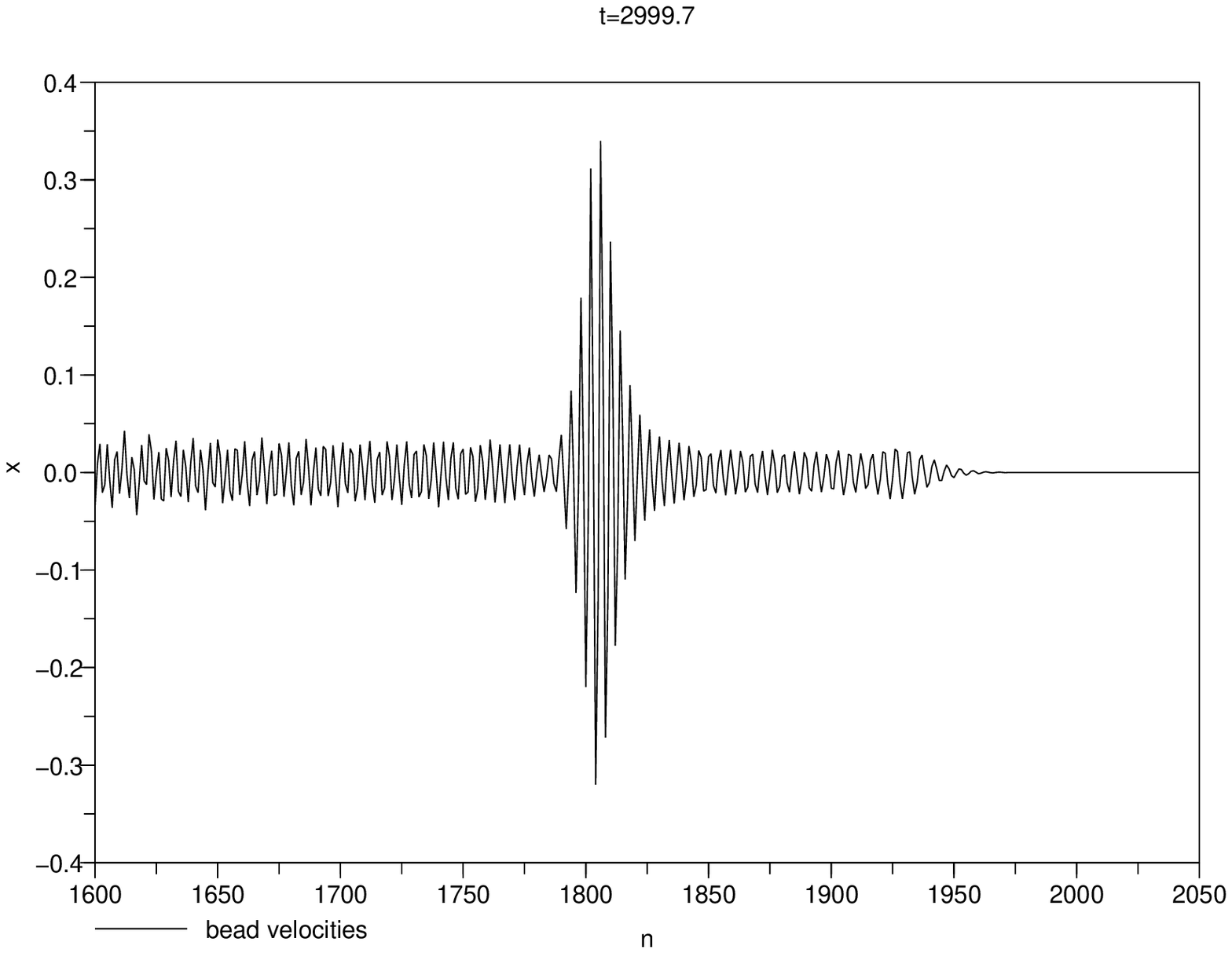}
\end{center}
\caption{\label{exmips1}
Snapshot of particle velocities
in system (\ref{eqm})-(\ref{anpot})-(\ref{pothertzp}) with anharmonicity parameter $s=1$
and precompression $d=0.5$, for
the initial condition (\ref{cic}) with $\dot{y}_1(0) \approx 1.87$.
The profile is plotted at three different times $t\approx 291$ (top panel),
$t\approx 724$ (middle panel) and
$t\approx 3000$ (bottom panel),
showing the formation of a traveling breather that coexists with
a sizeable dispersive wavetrain.}
\end{figure}

The case $s=1$ is described in figure \ref{exmips1}, which shows
the particle velocity profiles at different times.
A traveling breather reminiscent of the {\em sech}-type envelope solitons (\ref{tbs})
appears after the impact. It forms around $t=290$ (top plot),
and remains much less localized than the
traveling breathers previously obtained without precompression
(compare figures \ref{exmips1} and \ref{dirrevers}).
The breather propagates away from the boundary (middle and bottom plots)
and the ``boomerang effect" that occurs without precompression
disappears. In addition, the initial perturbation generates a
dispersive wavetrain of substantial amplitude,
and the traveling breather becomes ultimately superposed on an oscillatory tail at both sides of the
central pulse (bottom plot), which yields a traveling breather
profile reminiscent of the waves computed in \cite{sirenum} 
(see also~\cite{melvin2}).

\begin{figure}[h]
\psfrag{n}[0.9]{ $n$}
\psfrag{x}[1][Bl]{ $\dot{y}_n(t) $}
\begin{center}
\includegraphics[scale=0.4]{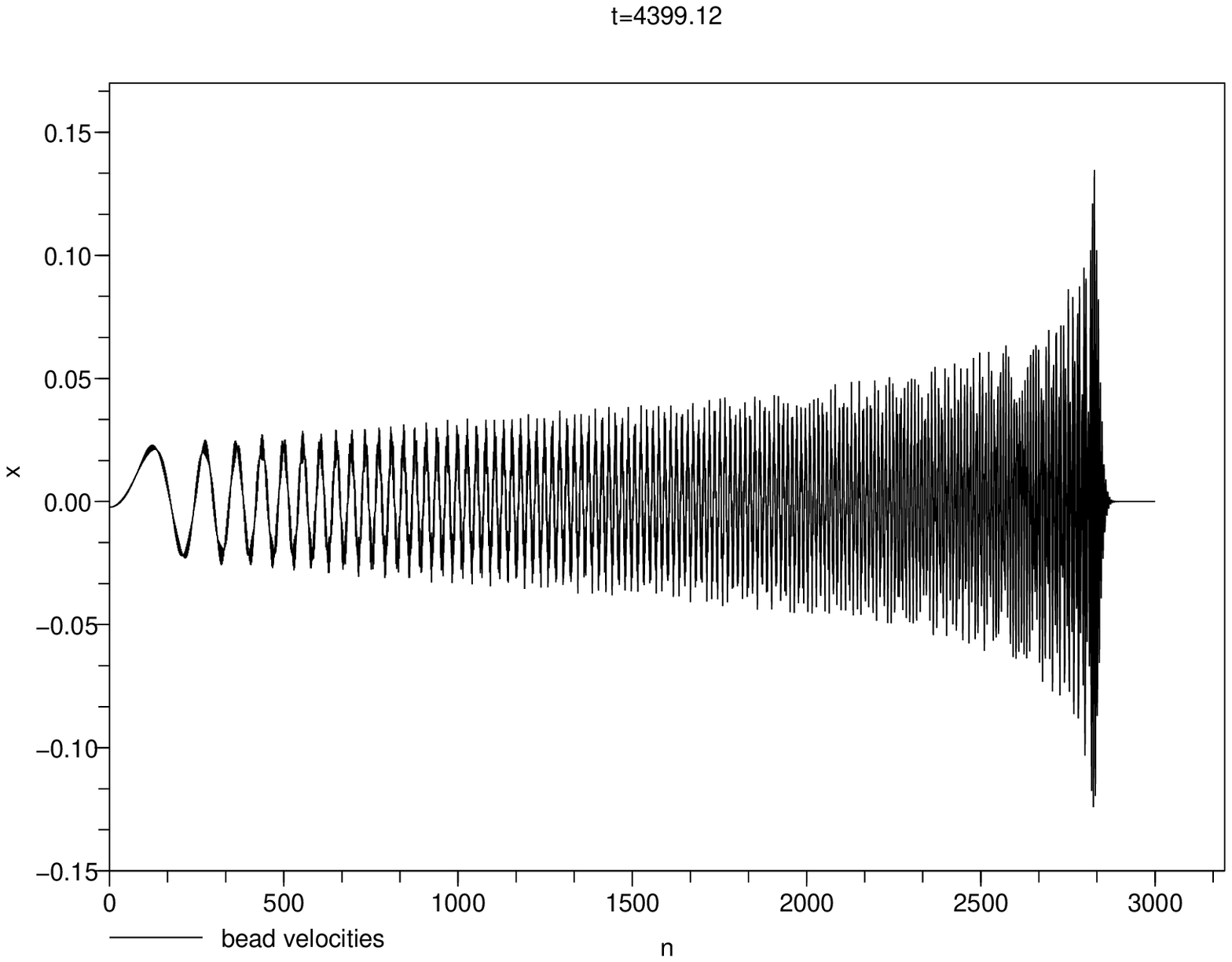}
\includegraphics[scale=0.4]{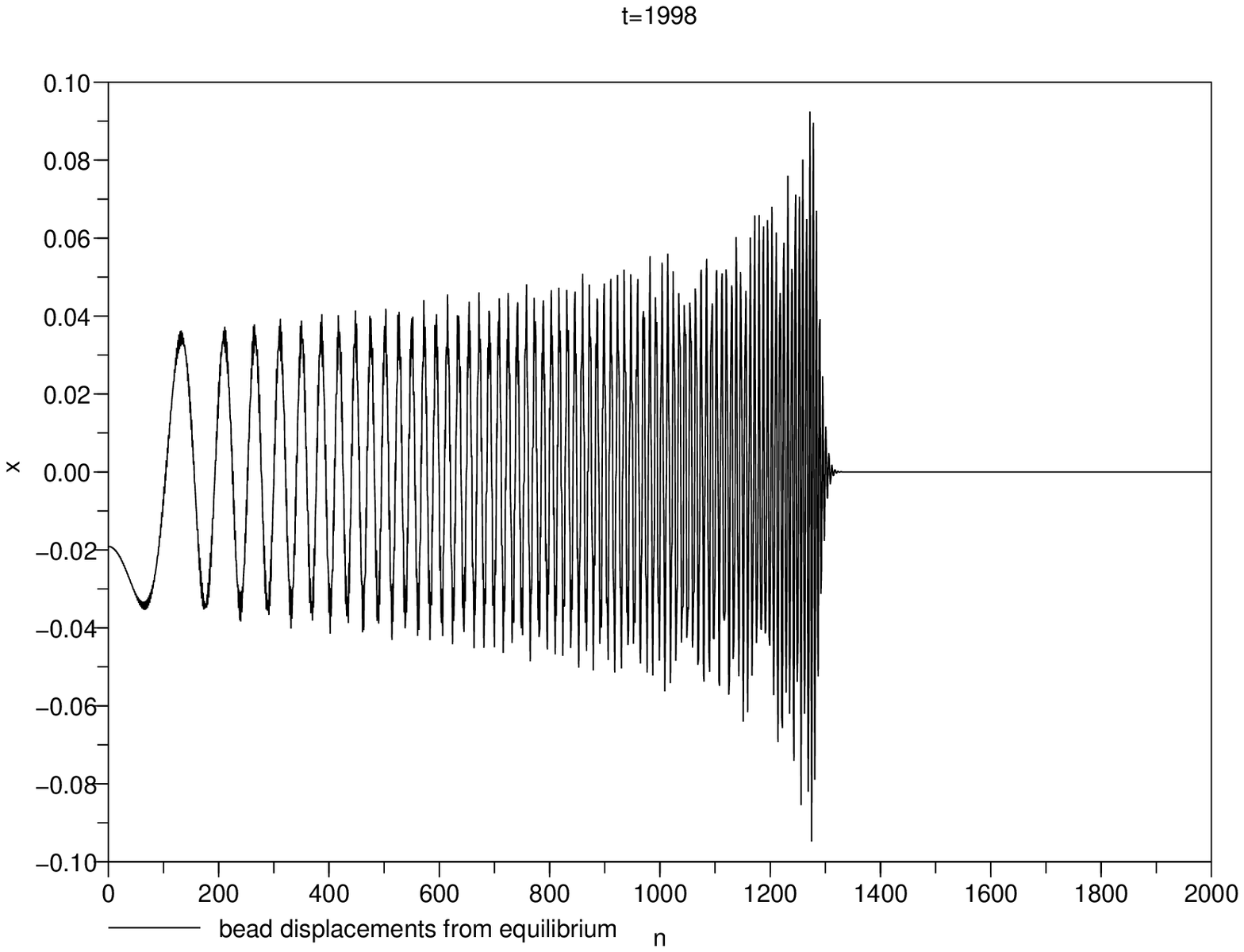}
\end{center}
\caption{\label{exmip}
Same as in figure \ref{exmips1}, for $s=0$ (top panel, particle velocities at $t\approx 4399$) and
$s=-1/6$ (bottom panel, particle velocities at $t\approx 1998$).}
\end{figure}

The cases $s=-1/6$ and $s=0$ yield a different situation described in figure \ref{exmip}.
The difference with the case without precompression is striking
(compare figure \ref{exmip} with figure \ref{smodeandtb}).
The initial localized perturbation generates an important dispersive wavetrain,
and no traveling breather is excited, at least on the timescales of the simulation.
Moreover, in the present case we do not observe the formation of a surface mode.

As a conclusion, according to our results,
introducing a precompression
attenuates spatial localization and enhances dispersive effects,
a phenomenon linked with an additional linear component
embedded within the Hertzian interactions.

\vspace{1ex}

\section{\label{conclu}Conclusion}

We have analyzed the properties of discrete breathers in FPU lattices
and mixed FPU-Klein-Gordon lattices with Hertzian interactions.
While static breathers don't exist in the absence of
precompression and of onsite potentials, the addition of 
the latter creates highly localized breathers, which
display a particularly strong mobility, a phenomenon
well-described by the DpS equation in the small amplitude regime
and associated with the spectral properties (i.e., the pinning mode)
of such states. 
Beyond the DpS limit, we have identified different phenomena
depending on the softening or hardening character of the local potential,
namely the generation of a surface mode after an impact
or the existence of direction-reversing traveling breathers. 
Importantly also the stability of both the on-site and
inter-site breather states obtained was critically dependent
on the strength (and sign) of the anharmonicity.

We have also introduced a mechanical model consisting of
a chain of stiff cantilevers decorated by spherical beads, which
may allow to realize the above localized excitations.
According to our study, an impact at one end of the cantilever chain
should generate a highly-localized traveling breather.
In this regime, contrary to what is the case for a regular
cradle under gravity, the ranges of parameters of the
problem (e.g., beads of about 1cm diameter, loads of about 1N, and
cantilever width of about 1cm) 
are deemed relevant for the observation of such breathers and for 
the description of the system by the DpS approximation examined herein.

Obviously, one has to stress that the lattice model (\ref{eqmnh})
is simplified and important corrections may apply.
For example, a finite-element modeling would be helpful to validate the model
or improve its calibration. In addition, it would be important to take
dissipation into account, following e.g. the approach of \cite{ricardo}.
Since many sources of dissipation are present (friction, plasticity effects,
transmission of vibrational energy through the walls),
one can wonder if dissipation may overdamp the dynamics and completely destroy the breathers.
However, recent experimental results \cite{boe} have demonstated that
static breathers with lifetimes of the order of $10$ ms could be generated
in diatomic granular chains. During this time,
the moving breather computed in section \ref{canti} would travel 
over approximately
$20$ sites (performing roughly $70$ internal oscillations), which would
allow for an experimental detection, provided
this excitation persists in the presence of dissipation,
with moderate changes in velocity and frequency.
Although the setting of decorated cantilevers proposed herein would
have the additional source of dissipation through radiating energy
into the ground (through the clamping of the cantilevers), it is
certainly deemed worthwhile to consider such experimental setups and to 
examine systematically the resulting dynamics.

\vspace{1ex}

A different approach which may allow to generate static breathers
is linked with modulational instability. Indeed,
static breathers have been
excited by modulational instabilities
in experiments performed on diatomic granular chains \cite{boe},
a phenomenon also numerically illustrated in the Newton's cradle
\cite{jamesc}. In this respect,
an extensive study of MI in the cradle model
(with the help of the DpS equation) would be of interest. 
A related aspect concerns the actuation of the system through
the driving of a bead with a particular frequency. In fact,
the experiments of~\cite{boe} were realized based on such
actuation of the chain at modulationally unstable frequencies
rather than the generation of suitable spatially extended,
modulationally unstable states. In that regard, it should be noted
that it is not straightforward
to experimentally
initialize desired spatial profiles throughout the lattice in this
system.

As we have seen, static breathers may be deformed by
weak instabilities resulting in a translational motion and traveling
counterparts thereof. However,
in an experimental context, these weak instabilities are likely to
be irrelevant due to dissipation. To fix the ideas, let us assume
a breather lifetime of the order of $10$ ms in the presence of dissipation,
as in the experiments of \cite{boe}.
In the computations of
section \ref{canti}, the breather periods at small amplitude were 
(roughly) close to $0.15$ ms, therefore
unstable Floquet eigenvalues
$1+\epsilon$ would have an effect over times of order $0.15 \epsilon^{-1}$ ms.
Consequently, dissipation
should destroy the breather well before the instability becomes observable
as soon as $\epsilon < 0.015$, and thus the instabilities identified in section
\ref{stabmob} (where $\epsilon < 10^{-3}$) would be largely dominated by 
dissipative effects.

\vspace{1ex}

From a numerical point of view, an interesting open problem concerns
the computation of traveling breathers.
In the above computations, 
approximate traveling breathers were generated by the
dynamics after an impact at one end of the chain. It would be interesting
to compute exact traveling breather solutions using the Newton 
method, as in references
\cite{aubryC,sirenum,melvin2}. Moreover, the existence (and physical
explanation) of direction-reversing 
traveling breathers remains to be elucidated. Furthermore, it would
be relevant to understand in more detail the nature of interactions
of these traveling breather with static defects. Studies in these
directions are currently in progress and will be reported in future
publications.

\vspace{1ex}

\textbf{Acknowledgements.}
{Part of this work has been carried out during
a visit of P.G.K. to laboratoire Jean Kuntzmann,
supported by the CNRS, France, to which P.G.K. is grateful
for the hospitality.
The authors are grateful to Nicholas Boechler for helpful inputs
on experimental issues.
G.J. acknowledges stimulating discussions with
E. Dumas, B. Bidegaray, M. Peyrard and G. Theocharis. J.C. acknowledges financial support from the MICINN project FIS2008-04848.
P.G.K. also acknowledges support from the US National Science Foundation
through grant CMMI-1000337 and also from the Alexander S. Onassis 
Public Benefit Foundation through the grant RZG 003/2010-2011.}

\end{document}